\documentclass[a4paper,11pt]{article}
\usepackage{jcappub}
\usepackage{graphicx}
\usepackage{amsmath,amssymb,amsfonts}
\usepackage{xspace}
\usepackage{subfig}
\usepackage{float}
\usepackage{color,bm}
\usepackage{xcolor}
\usepackage{bbding}
\usepackage[utf8]{inputenc}
\usepackage{commath}
\usepackage{mathrsfs}
\usepackage{slashed}
\usepackage{fancyvrb}
\usepackage[utf8]{inputenc}
\usepackage[english]{babel}
\usepackage{orcidlink}
\usepackage{cancel}
\usepackage{hyperref}
\usepackage{tabularray}
\UseTblrLibrary{booktabs}
\usepackage{gensymb}
\usepackage{scalerel}
\usepackage[titletoc]{appendix}

\usepackage{multirow}

\usepackage{xcolor}

\title{Inelastic Self-interacting Dark Matter and LUX-ZEPLIN 248 keV Event in a Dirac Modular Inverse Seesaw} 

\author[a]{Pritam Das$^{\orcidlink{0000-0001-6195-6944}}$,}
\emailAdd{prtmdas9@gmail.com}
\affiliation[a]{Department of Physics, Salbari College, Baksa, Assam 781318, India}

\author[b]{Biswajit Karmakar$^{\orcidlink{0000-0003-0259-8061}}$,}
\emailAdd{biswajit.karmakar@us.edu.pl}
\affiliation[b]{Institute of Physics, University of Silesia,  Katowice, Poland}

\author[c]{Satyabrata Mahapatra$^{\orcidlink{0000-0002-4000-5071}}$,}
\emailAdd{satyabrata@iitgoa.ac.in}
\affiliation[c]{School of Physical Sciences, Indian Institute of Technology Goa, Ponda-403401, Goa, India.}
\author[d]{and Partha Kumar Paul$^{\orcidlink{0000-0002-9107-5635}}$}
\emailAdd{ph22resch11012@iith.ac.in}
\affiliation[d]{Department of Physics, Indian Institute of Technology Hyderabad, Kandi, Telangana 502285, India}
\abstract{We propose a novel framework that simultaneously addresses the origin of Dirac neutrino masses and the nature of self-interacting dark matter (SIDM). The model is based on an $A_{4}$ modular symmetry to ensure the Diracness of neutrinos as well as the stability of the DM. The neutrino sector realizes a Dirac Inverse Seesaw mechanism where the smallness of the neutrino mass is governed by the vacuum expectation value (VEV) of a singlet scalar $\phi$. This same scalar couples to a vector-like fermion DM candidate, inducing a tiny Majorana mass splitting that renders the DM pseudo-Dirac and inelastic. Crucially, the scalar also acts as a light mediator for DM self-interactions, potentially solving the small-scale structure problems of Cold DM. In light of the recent 248 keV nuclear-recoil event, LZ230616, observed by LUX-ZEPLIN (LZ) DM direct detection experiment, we demonstrate that our inelastic SIDM parameter space naturally accommodates this signal via endothermic scattering kinematics. Furthermore, the spontaneous breaking of the dark parity required for this inelasticity produces a network of cosmological domain walls. We show that the explicit symmetry breaking needed to safely annihilate these walls generates a stochastic gravitational wave background. The non-holomorphic modular symmetry reduces the free parameters, correlating neutrino observables, addressing DM phenomenology, $\Delta N_{\rm eff}$, and gravitational-wave signatures.}

\begin{document}

\maketitle
\flushbottom

\section{Introduction}

The discovery of neutrino oscillations provides compelling evidence for physics beyond the Standard Model (BSM), establishing that neutrinos are massive \cite{ParticleDataGroup:2018ovx}. Nevertheless, the origin and nature of these tiny masses remain open questions. In particular, whether neutrinos are Dirac or Majorana fermions is still unknown. The absence of a confirmed signal in neutrinoless double-beta decay ($0\nu\beta\beta$) keeps the Dirac possibility viable and motivates mechanisms capable of generating sub-eV Dirac masses without invoking unnaturally tiny Yukawa couplings. In a conventional Dirac framework, reproducing $m_\nu\lesssim 0.1$ eV requires Yukawa couplings of order $10^{-12}$. A Dirac inverse seesaw mechanism provides an attractive alternative, in which the small neutrino masses arise from the suppression associated with a heavy messenger sector and a small symmetry-breaking scale.

At the same time, the flavor structure of the lepton sector remains unexplained within the Standard Model (SM). Non-Abelian discrete flavor symmetries, in particular $A_4$, have been extensively employed to account for the observed neutrino mixing pattern  \cite{Altarelli:2010gt, King:2013eh,Petcov:2017ggy, Chauhan:2023faf, Borah:2024gql,Borah:2017dmk, Das:2018qyt}. Conventional implementations, however, generally require flavon fields whose vacuum alignment introduces additional parameters and dynamics. Modular flavor symmetry { proposed by Feruglio \cite{Feruglio:2017spp}} offers an economical alternative in which the vacuum expectation value of a complex modulus $\tau$ controls the breaking of the flavor symmetry, eliminating the need for flavons. While the standard modular framework is usually formulated in terms of holomorphic modular forms \cite{Feruglio:2021dte, Ding:2023htn, Nomura:2019xsb,Asaka:2019vev, Nomura:2019lnr,Chen:2025ruj,Granelli:2025lds,Pathak:2024sei}, non-supersymmetric constructions based on non-holomorphic modular forms provide a broader class of modular-invariant structures, including negative modular weights \cite{Qu:2024rns}. In this work we employ polyharmonic Maa{\ss} forms of level three, realizing the finite modular group $\Gamma_3\simeq A_4$, to construct a predictive framework for the lepton sector \cite{Gao:2025jlw,Tapender:2026ets,Nasri:2026nbf,Okada:2025nap,Zhang:2025dsa,Qu:2025ddz,Nomura:2024vus,Kobayashi:2025hnc,Cheshta:2026fls,Behera:2026tmo,CentellesChulia:2023osj,Wang:2020dbp,Singh:2024imk}.

The dark matter (DM) problem provides a second major motivation for extending the Standard Model. Although collisionless cold DM successfully describes the formation of structure on cosmological scales, tensions have been discussed at galactic and sub-galactic scales \cite{Spergel:1999mh, Tulin:2017ara, Bullock:2017xww}. Self-interacting dark matter (SIDM), mediated by a light scalar or vector state with self-interaction cross-section per unit DM mass $\sigma/m \sim 1 \; {\rm cm}^2/{\rm g} \approx 2 \times 10^{-24} \; {\rm cm}^2/{\rm GeV}$, can modify the inner structure of DM halos while retaining the successful large-scale predictions of cold DM \cite{Buckley:2009in, Feng:2009hw, Feng:2009mn, Loeb:2010gj, Bringmann:2016din, Kaplinghat:2015aga, vandenAarssen:2012vpm, Tulin:2013teo, Borah:2022ask,Borah:2024wos}. However, a light mediator that generates sizable DM  can also induce appreciable elastic scattering of DM with ordinary matter, leading to strong constraints from direct-detection experiments~\cite{LZ:2024zvo, XENON:2024wpa, PandaX:2025rrz, Choi:2026kxe}. This tension motivates inelastic dark matter (iDM), in which the dark sector contains two nearly degenerate states and scattering off a nucleus proceeds predominantly through a transition between them~\cite{Tucker-Smith:2001myb, Cui:2009xq, He:2020sat, Borah:2020smw, Cho:2024lhp, Borah:2020jzi}. If the mass splitting is sufficiently large compared with the typical kinetic energy of halo DM, low-energy elastic scattering is suppressed while dark-sector self-interactions can remain efficient.

This mechanism has acquired renewed phenomenological interest following the recent result from the LUX-ZEPLIN (LZ) experiment \cite{LZ:2026axp}. Using an exposure of $2.84$ tonne-years and an extended nuclear-recoil energy window reaching approximately $270$ keV, the LZ collaboration reported a single event consistent with a nuclear recoil of
$
E_R = 248\pm23_{\rm stat}\pm23_{\rm sys}\ {\rm keV},
$
in a region with a low expected background. The background-only hypothesis is disfavored at a global significance of $2.6\sigma$, with a maximum local significance of $3.4\sigma$ across the models considered. Importantly, the extended recoil window permits sensitivity to interactions whose spectra are harder than those of conventional spin-independent elastic scattering, including inelastic DM scenarios. The high recoil energy is particularly suggestive of an inelastic transition because the required incoming velocity depends directly on the mass splitting. For a transition between the dark sector partners $X_1\rightarrow X_2$ with splitting $\delta$, the minimum velocity required to produce a nuclear recoil $E_R$ is
\begin{equation}
v_{\rm min}(E_R)=
\frac{1}{\sqrt{2m_NE_R}}
\left(
\frac{m_NE_R}{\mu_{_{XN}}}+\delta
\right),
\end{equation}
where $\mu_{_{XN}}$ is the DM--nucleus reduced mass. The recoil energy at which $v_{\rm min}$ is minimized is
$
E_R^{\rm min}=({\mu_{_{XN}}}/{m_N})\,\delta,
$
which approaches $\delta$ for $m_X\gg m_N$. Hence, a recoil at a few hundred keV naturally points towards an inelastic mass splitting of the same order. 
Consequently, a splitting of several hundred keV naturally shifts the recoil spectrum towards the high-energy region probed by LZ. Some very recent studies to provide a possible interpretation of the observed event are~\cite{Fan:2026kxx, Lou:2026idn, Freese:2026sga, Su:2026rwz, Chattopadhyay:2026ryw, Yamashita:2026ump, Visinelli:2026kgt, DiMauro:2026ldr, Rodd:2026tyn, Jeesun:2026vzo, McCabe:2026crm, Unwin:2026rdp, Smirnov:2026aqk, Du:2026guj}. These observations motivate a systematic investigation of whether a theoretically motivated inelastic DM framework can realize the same kinematic regime while simultaneously satisfying cosmological and astrophysical requirements.

{Building on these considerations,} we construct such a framework by connecting the inelastic dark sector to the origin of Dirac neutrino masses through a common scalar singlet $\phi$. The neutrino sector contains vector-like fermions $N_{L,R}$ and right-handed neutrinos $\nu_R$. After symmetry breaking, integrating out the heavy vector-like states generates an effective Dirac mass,
$
m_\nu^{\rm eff}\sim
(Y_Hv_h){M_N}^{-1}(Y_\phi v_\phi),
$
where $v_\phi$ denotes the VEV of $\phi$. Thus, the smallness of the neutrino masses is controlled by the same scalar VEV that enters the dark sector. The DM candidate is a vector-like fermion $\Psi$ {($=\Psi_L+\Psi_R$)} whose Dirac mass is accompanied by small Majorana masses induced by the $\phi$ VEV. The resulting spectrum consists of two nearly degenerate Majorana states, $\chi_1$ and $\chi_2$, with a mass splitting determined by the symmetry-breaking structure.

A central feature of the construction is an approximate left-right interchange symmetry, denoted by $\mathcal{Z}_2^{LR}$. Under this symmetry, the two chiral components of the dark fermion are interchanged while $\phi$ changes sign. In the exact symmetry limit, the Yukawa couplings satisfy $Y_R=-Y_L$. This relation removes the diagonal coupling of the scalar mediator to the DM mass eigenstates and consequently suppresses elastic DM--nucleus scattering. However, the exact symmetry also forces the two Majorana states to remain degenerate and, because $\langle\phi\rangle\neq0$ spontaneously breaks $\mathcal{Z}_2^{LR}$, leads to a cosmologically problematic network of domain walls~\cite{Vilenkin:2000jqa,Gelmini:1988sf,Larsson:1996sp,Saikawa:2017hiv,Nakayama:2016gxi,Paul:2024iie,Ma:2025bjf,Borah:2026kfo}.

To evade these issues, we therefore introduce a small explicit breaking of $\mathcal{Z}_2^{LR}$, parameterized by $
Y_R=-Y_L(1-\epsilon)$ with $\epsilon\ll1$. 
The resulting Majorana mass splitting is $
\delta\simeq\sqrt{2}\epsilon Y_Lv_\phi$, while the diagonal scalar coupling remains suppressed by $\epsilon$. At the same time, the explicit breaking lifts the degeneracy between the two $\mathcal{Z}_2^{LR}$ vacua and generates a pressure difference across the domain walls. The domain-wall network therefore becomes unstable and annihilates in the early Universe. The small parameter $\epsilon$ thus plays a particularly important role in the framework: it simultaneously controls the inelastic DM mass splitting, the residual elastic coupling, and the same explicit breaking leads to the domain wall network dynamics. The explicit breaking of $\mathcal{Z}_2^{LR}$ gives rise to a second observable consequence. The scalar potential possesses two approximately degenerate vacua connected by domain walls, while the small symmetry-breaking term introduces a vacuum-energy bias. The resulting domain-wall annihilation produces a stochastic gravitational-wave background whose peak frequency and amplitude are determined by the wall tension and annihilation temperature. Since both the wall tension and the DM mass splitting depend on the same symmetry-breaking scale and parameters, the gravitational-wave spectrum is correlated with the neutrino and DM phenomenology. We also explore the constraints on another cosmological observable $\Delta N_{\text{eff}}$ arising from the light mediator’s coupling to right-handed neutrinos which is tightly correlated with the neutrino phenomenology. Thus the model therefore provides a multi-messenger connection between LZ event, DM and neutrino phenomenology and cosmological observables like gravitational waves and $\Delta N_{\rm eff}$.

{To the best of our knowledge, this framework provides a novel realization of inelastic DM within a Dirac inverse-seesaw construction in which the DM mass splitting is directly connected to the vacuum expectation value of the scalar field that participates in the neutrino-mass mechanism, with the underlying flavor structure controlled by non-holomorphic modular $A_4$ symmetry. The resulting framework therefore connects Dirac neutrino masses, inelastic self-interacting DM, and the cosmological consequences of dark-sector symmetry breaking within a common construction while providing a viable explanation for the recently observed LZ230616 nuclear recoil event.}

We first construct the modular Dirac Inverse Seesaw framework in section~\ref{sec:model} and identify the neutrino-compatible parameter space and evaluate the model prediction for $\Delta N_{\rm eff}$ in section~\ref{sec:nupheno}. We then analyze the pseudo-Dirac DM phenomenology, including self-interactions, relic abundance, and detection constraints in section~\ref{sec:DM}. We subsequently confront the model with the LZ230616 event and identify the region capable of producing the observed high-energy recoil in section \ref{sec:lzevent}. Finally, we study the annihilation of the $\mathcal{Z}_2^{LR}$ domain-wall network and the resulting stochastic gravitational-wave spectrum in section~\ref{sec:DW}, highlighting the correlations between the LZ-compatible DM parameter space and the gravitational-wave signal. Finally, we conclude our work in section \ref{sec:conclusion}.

\section{Model Framework}
\label{sec:model}
We consider a framework that simultaneously addresses the origin of Dirac neutrino masses and the phenomenology of self-interacting inelastic dark matter (SIiDM), within a non-holomorphic modular $A_4$ flavor symmetry. The light neutrino masses are generated through a Dirac inverse seesaw mechanism, while the dark sector contains a Dirac fermion that becomes pseudo-Dirac after the spontaneous breaking of the $A_4$ discrete symmetry. 

In addition to the SM field content, we introduce two sets of heavy fermions ($N_\alpha, \psi_\alpha$), right-handed neutrino (RHN) $\nu_R$ and a scalar singlet $\phi$. The field content with their corresponding charge under $A_4$ discrete symmetry and modular weights ($k_I$) along with the Yukawa couplings are shown in Table \ref{tab:fiels}. The charged-lepton mass matrix is diagonal for the chosen charge assignments. In particular, the direct Yukawa interaction $\overline{L}\widetilde{H}\nu_R$ is forbidden by the modular $A_4$ structure, thereby avoiding the need to introduce unnaturally small fundamental Dirac Yukawa couplings.  The $A_4$ product rules and the construction of the non-holomorphic modular forms employed in this work are summarized in Appendices~\ref{appendA4} and~\ref{appenModular}, respectively.

\begin{table}[h]
    \centering
    \resizebox{\textwidth}{!}{
    \begin{tabular}{|c||ccccccccc||cccc|}
         \hline
         Charges& &&&Fields&&&&&&&&Couplings&\\
         \hline
         &$\bar{L}$&$H$&$e_R$&$N_R$&$\overline{N_L}$&$\nu_R$&$\phi$&$\Psi_L$&$\Psi_R$  &$Y_H$&$Y_\phi$&$Y_{L,R}$&$\lambda_{H\phi}$\\
         \hline
         $SU(2)$&2&2&1&1&1&1&1&1&1&-&-&-&-\\
$A_4$&1,1$^{\prime\prime}$,1$^\prime$&1&1,1$^\prime$,1$^{\prime\prime}$&3&3&3&1&1&1& 3&3&1&1\\
         $k_I$&-2&0&0&4&-4&6&-6&-3&-3&-2&-4&0&-4\\
\hline
    \end{tabular}}
    \caption{Charge assignments for respective field and couplings}
    \label{tab:fiels}
\end{table}

\subsection{Neutrino sector}
\label{subsec:neutrino_model}
The relevant neutrino-sector interactions are described by the Lagrangian:
\begin{equation}\label{lagNu}
\mathcal{L_\nu}=Y^{(2)}_H\bar{L}\tilde{H}N_R+Y^{(-4)}_\phi\overline{N_L}\phi\nu_R+M_N\overline{N_L}N_R+h.c.
\end{equation}
where the superscripts indicate the modular weights of the corresponding modular forms. The heavy fermions $N_L$ and $N_R$ form vector-like states with a common mass scale $M_N$. For the real scalar $\phi$, the scalar potential relevant for symmetry breaking is
\begin{equation}
\label{scalarPotential}
V(H,\phi)=-\mu_H^2 H^\dagger H
+\lambda_H(H^\dagger H)^2
-\frac{\mu_\phi^2}{2}\phi^2
+\frac{\lambda_\phi}{4}\phi^4
+\frac{\lambda_{H\phi}}{2}(H^\dagger H)\,\phi^2.
\end{equation}

Using the $A_4$ product rules, the neutrino-sector Lagrangian can be explicitly expanded as:
\begin{eqnarray}\label{lagEx}
 \nonumber   \mathcal{L}_\nu=&& \alpha \bar L_1 \tilde{H} \left(Y_{H1} N_{R1} + Y_{H2} N_{R2} + Y_{H3} N_{R3} \right)+ \beta \bar L_2 \tilde{H} \left(Y_{H1} N_{R1} + \omega^2Y_{H2} N_{R2} + \omega Y_{H3} N_{R3} \right)\\&&+ \nonumber \gamma \bar L_3 \tilde{H} \left(Y_{H1} N_{R1} +\omega Y_{H2} N_{R2} + \omega^2 Y_{H3} N_{R3} \right) + \alpha_\phi \Big[
(Y_{\phi2}N_{L3} + Y_{\phi3}N_{L2})\phi\nu_{R1} \\&&+ (Y_{\phi3}N_{L1} + Y_{\phi1}N_{L3})\phi\nu_{R2} + (Y_{\phi1}N_{L2} + Y_{\phi2}N_{L1})\phi\nu_{R3}
\Big]_{sym}\\&& \nonumber+\beta_\phi \Big[
(Y_{\phi2}N_{L3} - Y_{\phi3}N_{L2})\phi\nu_{R1} + (Y_{\phi3}N_{L1} - Y_{\phi1}N_{L3})\phi\nu_{R2}+ (Y_{\phi1}N_{L2} - Y_{\phi2}N_{L1})\phi\nu_{R3}
\Big]_{Asym}\\&& + M_N \left( \bar N_{L1} N_{R1} + \bar N_{L2} N_{R2} + \bar N_{L3} N_{R3} \right) \nonumber\label{lagr1}
\end{eqnarray}
While writing the Lagrangian in Eq. \eqref{lagNu}, we have rephrased the Yukawa couplings without introducing the free parameters. In Eq.\eqref{lagEx}, we explicitly expanded the interaction terms and introduced the free parameters. 

After spontaneous symmetry breaking, the scalars get VEV as $\langle H\rangle=v_h,
\langle\phi\rangle=v_\phi$, and the matrices associated with the two Yukawa interactions and the heavy vector-like mass term can consequently be written as:
\begin{eqnarray}
&&\nonumber M_D^{(H)} =
v_h
\begin{pmatrix}
\alpha Y_1 & \alpha Y_2 & \alpha Y_3 \\
\beta Y_1 & \omega^2\beta Y_2 & \omega\beta Y_3 \\
\gamma Y_1 & \omega\gamma Y_2 & \omega^2 \gamma Y_3
\end{pmatrix}\,, \quad M_R =
M_N
\begin{pmatrix}
1 & 0 & 0 \\
0 & 1 & 0 \\
0 & 0 & 1
\end{pmatrix},\\&& M_D'=v_\phi \begin{pmatrix}
   0&Y_{\phi3}(\alpha_\phi-\beta_\phi)&Y_{\phi2}(\alpha_\phi+\beta_\phi)\\Y_{\phi3}(\alpha_\phi+\beta_\phi)&0&Y_{\phi1}(\alpha_\phi-\beta_\phi)\\Y_{\phi2}(\alpha_\phi-\beta_\phi)&Y_{\phi1}(\alpha_\phi+\beta_\phi)&0
\end{pmatrix}\,.
\end{eqnarray}

The neutral-fermion mass matrix, written in the basis
$
\psi_\nu=
(\nu_L,N_R^c,N_L,\nu_R^c)^T$
takes the form:
\begin{eqnarray}
 \mathcal{M} = \begin{pmatrix} 0 & Y_H v_h & 0 & 0 \\ Y_H^T v_h & 0 & M_N^T & 0 \\ 0 & M_N & 0 & Y_\phi v_\phi \\ 0 & 0 & Y_\phi^T v_\phi & 0 \end{pmatrix}  . 
\end{eqnarray}

For the hierarchy
$
M_N\gg v_\phi \gg v_h,
$
the heavy states can be integrated out, yielding the effective light Dirac neutrino mass:
\begin{eqnarray}\label{eq:mnuEff}
  m_{\rm \nu}^{\rm eff} \approx (Y_H v_h) \frac{1}{M_N} (Y_\phi v_\phi)  .
\end{eqnarray}

Thus, the smallness of the observed neutrino masses follows from the inverse-seesaw-like suppression associated with the heavy messenger scale, without requiring extremely small fundamental Yukawa couplings. At the effective level, the corresponding operator is \cite{CentellesChulia:2020dfh}
\begin{equation}
\label{DiracOperator}
\mathcal{O}_{\rm Dirac}
=
\frac{1}{\Lambda}
(\overline{L}\widetilde{H})\phi\nu_R,
\qquad
\Lambda\sim
\frac{M_N}{Y_HY_\phi}.
\end{equation}

It is to be noted that the same scalar vacuum expectation value $v_\phi$ that enters Eq.~\eqref{eq:mnuEff} will play a crucial role in the dark sector, providing the origin of the small Majorana mass terms that split the dark Dirac fermion into a pseudo-Dirac pair.

Throughout the analysis, generalized CP (gCP) symmetry is imposed so that the coefficients multiplying the modular forms can be chosen real. In the absence of gCP, these coefficients would in general be complex and would introduce additional independent phases. The implementation of gCP in the non-holomorphic modular framework is discussed in Appendix~\ref{nonholo}.

\subsection{Dark sector}
\label{subsec:dark_model}
The dark sector consists of the Dirac fermion $\Psi$ and the scalar $\phi$. The relevant DM Lagrangian is: 
\begin{equation}\label{lagDM}
    \mathcal{L}_{Dark} =i\bar{\Psi}\gamma^\mu\partial_\mu\Psi-M_\psi\bar{\Psi}\Psi-Y^{(0)}_L\overline{(\Psi_L)^C}\phi\Psi_L-Y^{(0)}_R\overline{(\Psi_R)^C}\phi\Psi_R.
\end{equation}
The scalar field also participates in the neutrino sector through the interaction in Eq.~\eqref{lagNu}. Consequently, $\phi$ provides a common connection between the neutrino and dark sectors. The dark fermion mass structure becomes particularly interesting after $\phi$ develops a vacuum expectation value. The Yukawa interactions in Eq.~\eqref{lagDM} generate Majorana mass terms $m_{L,R}(\equiv \sqrt{2}Y_{L,R}v_\phi)$ for the two chiral components. Together with the bare Dirac mass $M_\psi$, the dark-fermion mass matrix in the Weyl basis $(\psi_L,\psi_R^c)$ is therefore
\begin{equation}
\label{darkMassMatrix}
\mathcal{M}_\psi=
\begin{pmatrix}
m_L & M_\psi\\
M_\psi & m_R
\end{pmatrix}.
\end{equation}

For $M_\psi\gg m_{L,R}$, the two Majorana eigenstates form a pseudo-Dirac pair. The detailed diagonalization, the resulting mass splitting, and the scalar couplings in the physical basis are presented in Sec.~\ref{sec:DM}.

A central issue is that sizeable DM self-interactions generally require appreciable values of the scalar Yukawa couplings $Y_L$ and $Y_R$. Without an additional symmetry, such couplings also generate diagonal interactions between the scalar mediator and the DM mass eigenstates. Through the Higgs portal, these interactions can induce elastic scattering of DM on nuclei and can therefore lead to stringent direct-detection constraints \cite{LZ:2024zvo,XENON:2024wpa,PandaX:2025rrz}.

This motivates a discrete left-right interchange symmetry in the dark fermion sector, which we denote by $\mathcal{Z}_2^{LR}$. Its action is defined as
\begin{equation}
\label{Z2LR}
\Psi_L\leftrightarrow\Psi_R,
\qquad
\phi\rightarrow-\phi,
\qquad
\nu_R\rightarrow-\nu_R.
\end{equation}
This should be regarded as a dark-sector left-right interchange symmetry rather than the conventional left-right gauge symmetry. The transformation of $\nu_R$ is included so that the neutrino-sector interaction $\overline{N_L}\phi\nu_R$ remains invariant. Hence, the symmetry does not introduce or generate a Majorana mass for $\nu_R$ and does not spoil the Dirac nature of the light neutrinos.

Under Eq.~\eqref{Z2LR}, the dark-sector Yukawa terms transform as
\begin{equation}
Y_L\overline{(\Psi_L)^C}\phi\Psi_L
+
Y_R\overline{(\Psi_R)^C}\phi\Psi_R
\rightarrow
-Y_L\overline{(\Psi_R)^C}\phi\Psi_R
-
Y_R\overline{(\Psi_L)^C}\phi\Psi_L.
\end{equation}
An exact $\mathcal{Z}_2^{LR}$ therefore requires
\begin{equation}
\label{exactZ2}
Y_R=-Y_L.
\end{equation}
This relation has an important physical consequence such as it removes the leading diagonal scalar coupling in the pseudo-Dirac limit and thereby naturally suppresses elastic DM scattering. However, an exact $\mathcal{Z}_2^{LR}$ also has two undesirable consequences. First, Eq.~\eqref{exactZ2} implies $m_R=-m_L$, so that the trace of the mass matrix $\mathcal{M}_\psi$ in Eq.~\eqref{darkMassMatrix} vanishes and the two physical states remain exactly degenerate. The dark fermion therefore remains a pure Dirac state rather than acquiring the non-zero inelastic splitting required for inelastic DM. Second, spontaneous breaking of the exact $\mathcal{Z}_2^{LR}$ discrete symmetry by $\langle\phi\rangle=v_\phi\neq0$ produces two degenerate vacua, $\phi=\pm v_\phi$, and consequently leads to a cosmological domain-wall network that would eventually overclose the universe.

We therefore consider $\mathcal{Z}_2^{LR}$ Dark Parity to be an approximate symmetry which is explicitly broken by a small parameter $\epsilon \ll 1$. This small explicit breaking in the dark-sector Yukawa structure can be parameterized as
\begin{equation}
\label{epsRelation}
Y_R=-Y_L(1-\epsilon),
\qquad
|\epsilon|\ll1.
\end{equation}

The resulting small departure from the exact symmetry simultaneously allows a non-zero pseudo-Dirac mass splitting and a suppressed diagonal scalar coupling. The detailed consequences of this breaking for inelastic self-interactions and direct detection are discussed in Sec.~\ref{sec:DM}, while the associated domain-wall dynamics and the gravitational-wave signal from their annihilation are studied in Sec.~\ref{sec:DW}.

The construction thus establishes a common origin for the neutrino and DM sectors. In particular, the same scalar vacuum expectation value $v_\phi$ enters the effective Dirac neutrino mass in Eq.~\eqref{eq:mnuEff} and the Majorana mass terms of the dark fermion in Eq.~\eqref{darkMassMatrix}. The approximate $\mathcal{Z}_2^{LR}$ symmetry then organizes the dark-sector interactions such that the inelastic coupling can remain sizable while the elastic coupling is parametrically suppressed. This connection is the central structural feature underlying the phenomenology developed in the subsequent sections.

\section{Neutrino Phenomenology}
\label{sec:nupheno}

Having established the neutrino mass matrix in Sec.~\ref{subsec:neutrino_model}, we next investigate whether the modular $A_4$ construction can reproduce the observed neutrino oscillation data. Since the charged-lepton mass matrix is diagonal for the field assignments adopted in Table~\ref{tab:fiels}, the leptonic mixing matrix is directly identified with the unitary matrix that diagonalizes the light-neutrino mass matrix.
\subsection{Neutrino mass and mixing}
\label{subsec:nuscan}
To numerically diagonalize the neutrino mass matrix given in Eq.~\eqref{eq:mnuEff}, we construct the Hermitian matrix
\begin{equation}
\label{eq:nu_mass}
\mathcal{M}_\nu
=
m_\nu^{\rm eff}
\left(m_\nu^{\rm eff}\right)^\dagger,
\end{equation}
which is diagonalized according to
\begin{equation}
U_{\rm PMNS}^\dagger
\mathcal{M}_\nu
U_{\rm PMNS}
=
{\rm diag}
\left(m_1^2,m_2^2,m_3^2\right),
\end{equation}
where $m_i$ denote the three light-neutrino mass eigenvalues. Since the charged-lepton mass matrix is diagonal in our framework, the leptonic mixing matrix is directly identified with the Pontecorvo-Maki-Nakagawa-Sakata (PMNS) matrix. We adopt the standard parametrization of the PMNS matrix:
\begin{equation}
 U_{\rm PMNS} = \begin{pmatrix}
     c_{12}c_{13} & s_{12}c_{13} & s_{13}e^{-i\delta_{CP}}\\
     - s_{12}c_{23} - c_{12}s_{23}s_{13}e^{i\delta_{CP}} & c_{12}c_{23} - s_{12}s_{23}s_{13}e^{i\delta_{CP}} & s_{23}c_{13}\\
     s_{12}s_{23} - c_{12}c_{23}s_{13}e^{i\delta_{CP}} & - c_{12}s_{23} - s_{12}c_{23}s_{13}e^{i\delta_{CP}} & c_{23}c_{13}
 \end{pmatrix}  
\end{equation}
  where $c_{ij}\equiv\cos\theta_{ij}$ and $s_{ij}\equiv\sin\theta_{ij}$.
Since the light neutrinos are Dirac fermions in the present framework, there are no physical Majorana phases.

After diagonalizing $\mathcal{M}_\nu$, the neutrino oscillation parameters are extracted from the elements of $U_{\rm PMNS}$. The three leptonic mixing angles are obtained through
\begin{equation}
\sin^2\theta_{13}=|U_{13}|^2,\qquad
\sin^2\theta_{12}=\frac{|U_{12}|^2}{1-|U_{13}|^2},\qquad
\sin^2\theta_{23}=\frac{|U_{23}|^2}{1-|U_{13}|^2}.
\end{equation}
The Dirac CP-violating phase is obtained from the Jarlskog invariant,
\begin{equation}
J_{CP}=\mathrm{Im}\left(U_{e1}U_{\mu2}U_{e2}^{*}U_{\mu1}^{*}\right)
=s_{23}c_{23}s_{12}c_{12}s_{13}c_{13}^{2}\sin\delta_{CP}\label{eqjcp}\,.
\end{equation}

The mass-squared differences are calculated from the eigenvalues as
\begin{equation}
\Delta m_{21}^2=m_2^2-m_1^2,
\qquad
\Delta m_{31}^2=m_3^2-m_1^2,
\end{equation}
for the normal-ordering convention adopted in the numerical analysis.

\begin{table}[h]
    \centering
    \begin{tabular}{|c|c|c|}
    \hline
         Parameters&Input &Allowed by {\tt NuFit6.1}\cite{Esteban:2024eli,NuFIT6.1}  \\
         \hline
         $\alpha$&$[10^{-5},1]$&$[0.84,1.25]\times10^{-4}$\\
         $\beta$&$[10^{-5},1]$&$[1.26,2.06]\times10^{-4}$\\  
         $\gamma$&$[10^{-5},1]$&$[0.989,1.56]\times10^{-4}$\\
         $\alpha_\phi$&$[10^{-3},1]$&$[0.0259,0.0269]$\\
         $\beta_\phi$&$[10^{-3},1]$&$[0.195,0.227]$\\
         $x$&$[-0.5,+0.5]$&$[0.0419,0.06383]$\\
         $y$&$[0,2.5]$&$[0.746,0.769]$\\
         $v_{\phi}$&$[10,1000]$ TeV&$[30.8,42.7]$ TeV \\
         \hline
    \end{tabular}
    \caption{Input ranges used in the numerical scan and the corresponding allowed regions of the model parameters after imposing the neutrino-oscillation constraints.}
    \label{tab:para}
\end{table}

\begin{figure}[t]
    \centering
    \includegraphics[scale=0.65]{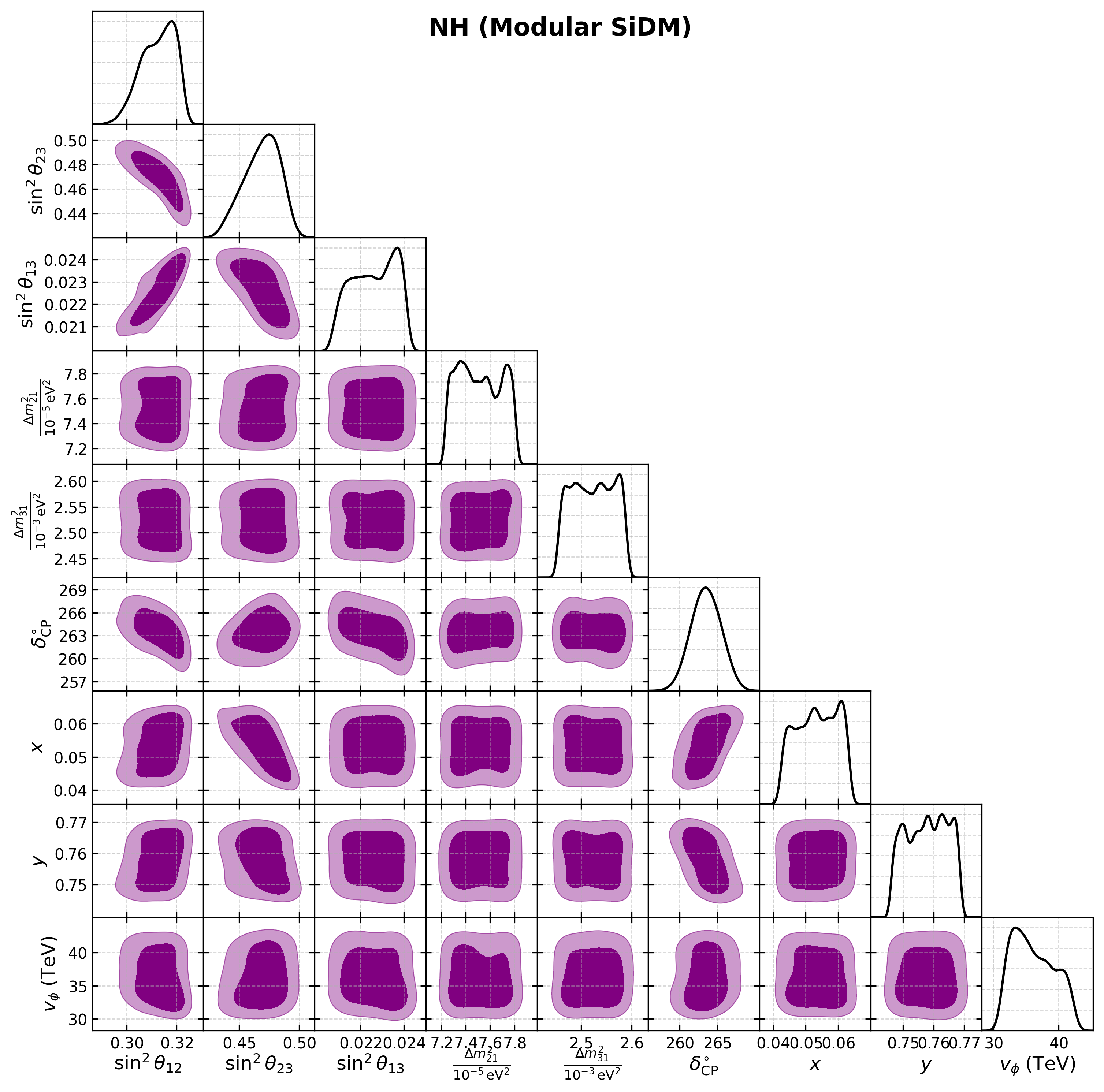}
    \caption{Triangle plot illustrating the allowed parameter space for the normal hierarchy (NH) in the modular SIDM framework. The diagonal panels show the one-dimensional distributions of the allowed parameter values, while the off-diagonal panels display the pairwise correlations among the neutrino oscillation observables and the model parameters. The shaded regions correspond to the 68\% and 95\% confidence regions obtained from the random numerical parameter scan.}
    \label{fig:trignu}
\end{figure}

The numerical scan is performed over the independent model parameters
\begin{equation}
\left\{
\alpha,\beta,\gamma,
\alpha_\phi,\beta_\phi,
x,y,v_\phi
\right\},
\end{equation}
where the complex modulus is written as
$
\tau=x+iy.
$
For simplicity, we fix the mass of the heavy vector-like fermions to
$
M_N=1~{\rm TeV}$. This fixed choice is sufficient for the purpose of demonstrating the viability of the neutrino-sector construction, since the heavy scale enters the effective Dirac mass through the combination shown in Eq.~\eqref{eq:mnuEff}. The input ranges used in the scan and the corresponding allowed regions obtained after imposing the neutrino-oscillation constraints are summarized in Table~\ref{tab:para}.

We impose the $3\sigma$ allowed ranges of the neutrino oscillation parameters from {\tt NuFit~6.1}~\cite{Esteban:2024eli,NuFIT6.1}, considering normal ordering. The choice of normal ordering is motivated by the present global-fit preference. The corresponding analysis for inverted ordering can be performed within the same framework however it is not required for the phenomenological analysis presented below.

The resulting allowed parameter space is shown in the triangle plot in Fig.~\ref{fig:trignu}. The diagonal panels display the marginalized distributions of the accepted scan points, while the off-diagonal panels show the corresponding pairwise correlations among the neutrino observables and the model parameters. The shaded regions correspond to the 68\% and 95\% confidence intervals obtained from the numerical scan. The scan demonstrates that the modular structure can reproduce the observed neutrino mixing angles and mass-squared differences while simultaneously restricting the modulus $\tau$ and the scalar vacuum expectation value $v_\phi$ to comparatively narrow regions.
In particular, the atmospheric mixing angle is preferentially located toward the lower-octant region, approximately between $41^\circ$ and $45^\circ$, while the Dirac CP phase is concentrated around
$\delta_{\rm CP}\simeq257^\circ-268^\circ$.  For the viable parameter space, we obtain $J_{\rm CP}\simeq [-0.035,-0.032]$, consistent with the allowed ranges of the neutrino mixing parameters.
The non-trivial correlations between the mixing observables and the real and imaginary components of the modulus, $x$ and $y$, reflect the underlying modular-flavor structure. In contrast, the correlations involving $v_\phi$ are comparatively weaker within the neutrino-only scan. Nevertheless, $v_\phi$ is particularly important for the full framework because it also controls the dark-sector mass terms through Eq.~\eqref{darkMassMatrix}. Consequently, the allowed neutrino range of $v_\phi$ provides a direct input for the DM analysis developed in Sec.~\ref{sec:DM}.

This connection through the VEV of the scalar plays an important role for the phenomenology of the model. The scalar vacuum expectation value is not introduced solely as an independent dark-sector scale, rather it simultaneously participates in the generation of the light Dirac neutrino masses and in the splitting of the dark pseudo-Dirac states. Thus, the neutrino-sector scan already constrains one of the central quantities entering the subsequent DM phenomenology.

\subsection{Thermal $N_{\rm eff}$}
\label{neff}
An additional consequence of the Dirac nature of the light neutrinos is the presence of three right-handed neutrinos $\nu_R$. Their masses naturally remain very small in the present construction and hence their contribution to the cosmological radiation density is controlled by the temperature at which they decouple from the thermal bath~\cite{Luo:2020sho, Borah:2025fkd, Mahapatra:2023oyh}. We therefore study the resulting contribution to the effective number of relativistic species, $\Delta N_{\rm eff}$.

The right-handed neutrinos do not possess the standard electroweak interactions and can remain out of equilibrium with the SM plasma at sufficiently high temperatures. In the present framework, however, they interact with the heavy messenger sector and the scalar $\phi$ through the neutrino-sector interaction in Eq.~\eqref{lagNu}. Owing to the $A_4$ structure, the three generations of $\nu_R$ need not have identical effective couplings. From the symmetric and anti-symmetric contractions in Eq.~\eqref{lagEx}, the effective Yukawa couplings relevant for the thermalization of the three right-handed neutrinos can be written as:
\begin{eqnarray}\label{eq:yuRH}
\nonumber    y_{\nu_R}^1&=&\alpha_\phi Y_{\phi2}+\alpha_{\phi}Y_{\phi3}+\beta_\phi Y_{\phi2}-\beta_\phi Y_{\phi3},\\
    y_{\nu_R}^2&=&\alpha_\phi Y_{\phi3}+\alpha_{\phi}Y_{\phi1}+\beta_\phi Y_{\phi3}-\beta_\phi Y_{\phi1},\\
\nonumber    y_{\nu_R}^3&=&\alpha_\phi Y_{\phi1}+\alpha_{\phi}Y_{\phi2}+\beta_\phi Y_{\phi1}-\beta_\phi Y_{\phi2}.
\end{eqnarray}
The different values of these effective couplings imply different decoupling temperatures for the three $\nu_R$ species.

For each right-handed neutrino, we determine the decoupling temperature $T_{\nu_R}^d$ by comparing its interaction rate with the Hubble expansion rate. We then evaluate its residual contribution to the radiation density using the instantaneous-decoupling approximation. Since the right-handed neutrinos decouple much earlier than the active neutrinos in the parameter region of interest, the entropy released by the subsequent annihilation of particles in the thermal bath dilutes their temperature relative to that of the active neutrinos. For a relativistic right-handed neutrino species, its contribution is therefore
\begin{equation}
\label{eq:Dneff}
\Delta N_{\rm eff}^{(i)}
=
\left(
\frac{g_{*s}(T_{\nu_L}^d)}
     {g_{*s}(T_{\nu_R,i}^d)}
\right)^{4/3},
\end{equation}
where $T_{\nu_L}^d\simeq1~{\rm MeV}$ denotes the active-neutrino decoupling temperature. The total contribution from the three right-handed neutrinos is then
\begin{equation}
\label{eq:Dnefftotal}
\Delta N_{\rm eff}
=
\sum_{i=1}^{3}
\Delta N_{\rm eff}^{(i)}.
\end{equation}
Equation~\eqref{eq:Dneff} makes explicit why the three generations need not contribute equally: a species that decouples earlier experiences a larger entropy dilution and consequently contributes less to $\Delta N_{\rm eff}$.

\begin{figure}[h]
    \centering
    \includegraphics[scale=0.45]{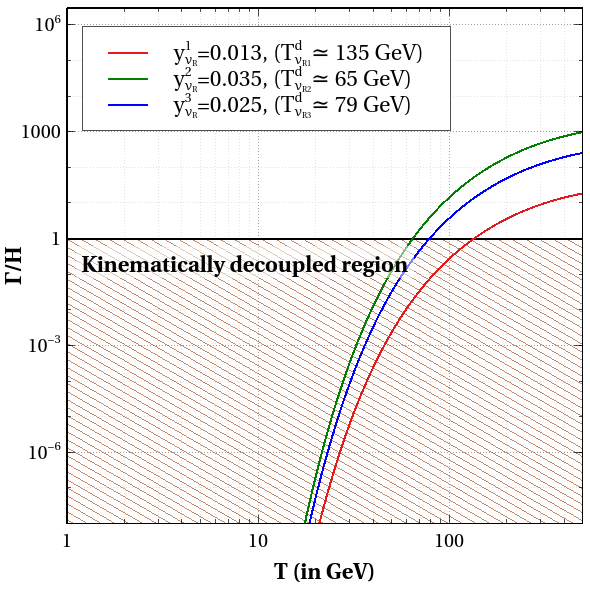}
    \caption{Decoupling profiles of the three right-handed neutrinos for a representative set of Yukawa couplings. The corresponding decoupling temperatures are indicated in the figure. }
    \label{fig:decP}
\end{figure}

The resulting decoupling profiles for the three right-handed neutrinos are shown in Fig.~\ref{fig:decP}. For illustration, one representative set of Yukawa couplings obtained from the allowed neutrino parameter space is used in the figure. The corresponding numerical ranges of the effective couplings, decoupling temperatures, and individual contributions to $\Delta N_{\rm eff}$ are summarized in Table~\ref{tab:neff}.

\begin{table}[h]
    \centering
    \begin{tabular}{|c|c|c|c|c|}
         \hline
         Couplings&Allowed Range& $T_{\nu_R}^d$ in GeV&$\Delta N_{\rm eff}$ &Total $\Delta N_{\rm eff}$ \\
         \hline
         $y_{\nu_R}^1$& $[0.0127, 0.0138]$&$[128,138]$&$[0.0470, 0.0472]$&\\
          $y_{\nu_R}^2$& $[0.0318, 0.0374]$&$[62.5, 68]$&$[0.0498, 0.0503]$&$[0.1455,0.147]$\\
           $y_{\nu_R}^3$& $[0.0236, 0.0301]$&$[71, 82]$&$[0.0487, 0.0495]$&\\
           \hline
    \end{tabular}
    \caption{Numerical estimation of individual contributions to $\Delta N_{\rm eff}$ for the allowed ranges of the Yukawa couplings associated with different species of $\nu_R$. Total contribution for three generations are shown in the last column.}
    \label{tab:neff}
\end{table}

The resulting total contribution, $\Delta N_{\rm eff}\simeq0.146,$ is a characteristic prediction of the parameter region considered here. This lies within the sensitivity of
upcoming experiments such as CMB-S4 \cite{Abazajian:2019eic} and CMB-HD \cite{CMB-HD:2022bsz}. Importantly, this contribution is not an additional independent ingredient of the model rather it follows from the same Dirac-neutrino construction that generates the light neutrino masses. The scalar $\phi$, which appears in the effective Dirac mass in Eq.~\eqref{eq:mnuEff}, also controls the interactions responsible for the thermal history of the right-handed neutrinos.

The $N_{\rm eff}$ analysis therefore provides an additional cosmological constraint on the neutrino-sector parameter space before considering the DM phenomenology. In particular, the allowed values of $\alpha_\phi$, $\beta_\phi$, and $v_\phi$ are simultaneously relevant for the generation of the light neutrino masses, the thermalization of the right-handed neutrinos, and the dark-sector phenomenology through the common scalar $\phi$. This shared dependence is one of the characteristic features of the framework.

\section{Dark Matter}\label{sec:DM}
The dark sector contains a Dirac fermion $\Psi$ and the real scalar mediator $\phi$. As discussed in Sec.~\ref {sec:model}, the breaking of the approximate $\mathcal{Z}_2^{LR}$ symmetry by $\langle\phi\rangle=v_\phi$ generates Majorana mass terms for the two chiral components of $\Psi$. The resulting fermion is therefore naturally described as a pseudo-Dirac state. The same approximate symmetry suppresses the diagonal scalar coupling while retaining an unsuppressed off-diagonal interaction, providing the basic mechanism for inelastic DM scattering and self-interactions.
\subsection{Pseudo-Dirac Dark Matter}
\label{subsec:pseudoDirac}

After $\phi$ acquires its vacuum expectation value, the mass terms of the dark fermion in the Weyl basis $(\Psi_L,\Psi_R^c)$ are  
given in Eq.\eqref{darkMassMatrix}. 
Transitioning to the physical mass basis requires diagonalizing the mass matrix $\mathcal{M}_{\psi}$
.Consequently, the transformation from the Weyl interaction basis to the physical Majorana mass eigenstates $ (\chi_1, \chi_2)^T$ is mediated by a unitary matrix $U$:
\begin{equation}
    \begin{pmatrix} \psi_L \\ \psi_R^c \end{pmatrix} = \begin{pmatrix} \cos\theta & \sin\theta \\ -\sin\theta & \cos\theta \end{pmatrix} \begin{pmatrix} \chi_1 \\ \chi_2 \end{pmatrix},
\end{equation}
where the orthogonal mixing angle $\theta$ is determined by the relation:
\begin{equation}
\tan 2\theta = \frac{2M_\psi}{m_R - m_L}\,.
\end{equation}
Since the off-diagonal bare mass $M_\psi$ is much larger than the VEV-induced Majorana masses, the ratio diverges, driving the system to near-maximal mixing ($\theta \approx \pi/4$).

To leading order in $m_{L,R}/M_\psi$, the two physical masses are:
\begin{equation}
m_{\chi_1} \simeq M_\psi - \frac{m_L + m_R}{2}, \quad m_{\chi_2} \simeq M_\psi + \frac{m_L + m_R}{2},
\end{equation}
and hence mass splitting is,
\begin{equation}
\delta
\equiv
m_{\chi_2}-m_{\chi_1}
\simeq
m_L+m_R.
\label{eq:DMsplitting}
\end{equation}
Expanding the scalar mediator field around its VEV ($\phi = (v_\phi + h_\phi)/\sqrt{2}$), the small mass splitting is therefore directly related to the departure from the exact $\mathcal{Z}_2^{LR}$ limit. Using the parametrization introduced in Sec.~\ref{sec:model},
\begin{equation}
Y_R=-Y_L(1-\epsilon),
\qquad
\epsilon\ll1,
\label{eq:YRepsilonDM}
\end{equation}
we obtain
\begin{equation}
\delta
\simeq\sqrt{2}
\epsilon Y_Lv_\phi.
\label{eq:deltaepsilon}
\end{equation}

Thus the interaction Lagrangian in the mass basis is given by:
\begin{equation}
    \mathcal{L}_{\rm int} \supset -\frac{1}{\sqrt{2}} \left(Y_L\overline{(\Psi_L)^C}h_\phi\Psi_L+Y_R\overline{(\Psi_R)^C}h_\phi\Psi_R\right) + \text{h.c.}
\end{equation}
Applying the unitary transformation $U$ to rewrite the chiral states in terms of $\chi_1$ and $\chi_2$
\begin{align}
    \mathcal{L}_{\rm int} \supset &-\frac{1}{\sqrt{2}} h_\phi \Big[ (Y_L \cos^2\theta + Y_R \sin^2\theta) \overline{\chi_1}  \chi_1 + (Y_L \sin^2\theta + Y_R \cos^2\theta) \overline{\chi_2}  \chi_2 \nonumber\\&+ 2\sin\theta\cos\theta (Y_L - Y_R) \overline{\chi_1}  \chi_2 \Big] + \text{h.c.}
\end{align}
Taking the maximal mixing limit ($\sin\theta \approx \cos\theta \approx 1/\sqrt{2}$), we obtain:
\begin{equation}
    \mathcal{L}_{\rm int} \supset -\frac{1}{2\sqrt{2}} h_\phi \left[ (Y_L + Y_R)(\bar{\chi}_1 \chi_1 + \bar{\chi}_2 \chi_2) + 2(Y_L - Y_R)\bar{\chi}_1 \chi_2 \right]\,.
\end{equation}
Substituting Eq.~\eqref{eq:YRepsilonDM}, in the above equation, finally the interaction term can be written as:
\begin{equation}
\mathcal{L}_{\rm int}
\supset
-\frac{Y_L}{2\sqrt{2}}h_\phi
\left[
\epsilon
\left(
\overline{\chi_1}\chi_1+\overline{\chi_2}\chi_2
\right)
+
2(2-\epsilon)\overline{\chi_1}\chi_2
\right]
+\mathrm{h.c.}
\label{eq:DMinteractionepsilon}
\end{equation}
This expression makes the central feature of the construction transparent. The diagonal Yukawa interaction is suppressed by the small symmetry-breaking parameter ($\propto \epsilon Y_L$),
whereas the off-diagonal interaction remains of order $Y_L$. Consequently, elastic scattering of $\chi_{1,2}$ from ordinary matter can be strongly suppressed without suppressing the dark-sector self-interaction. At the same time, the small but nonzero value of $\epsilon$ generates the mass splitting required for inelastic scattering.

For convenience, we define the effective off-diagonal coupling as
\begin{equation}
y_{\rm eff}
\equiv
\frac{Y_L}{\sqrt{2}}(2-\epsilon)
\simeq
\sqrt{2}\,Y_L,
\label{eq:yeff}
\end{equation}
where the last relation holds for $\epsilon\ll1$. We further introduce
\begin{equation}
\alpha
\equiv
\frac{y_{\rm eff}^2}{4\pi}.
\label{eq:alphaDM}
\end{equation}
In terms of these quantities, the symmetry-breaking parameter can be related to the physical splitting as
\begin{eqnarray}
\epsilon=\frac{\delta}{\sqrt{4\pi\alpha}v_\phi}
\end{eqnarray}
This relation is useful in connecting the direct-detection phenomenology to the underlying symmetry-breaking parameter. The resulting dark sector therefore contains two nearly degenerate Majorana states, with $\chi_1$ taken to be the stable DM state and $\chi_2$ its slightly heavier partner. For the parameter region of interest, both states can remain cosmologically populated, while their dominant low-energy interaction with the mediator is inelastic. The phenomenology is consequently governed by the interplay between the mediator mass $m_\phi$, the DM mass $m_{\chi_1}$, the coupling $\alpha$, and the mass splitting $\delta$.

\subsection{Inelastic Self-Interacting Dark Matter}
\label{subsec:iSIDM}
To resolve the small-scale structure anomalies of the Cold Dark Matter paradigm (such as the core-cusp problem), the DM must exhibit a velocity-dependent self-interaction cross-section~\cite{Schutz:2014nka, Blennow:2016gde, Zhang:2016dck, Dutta:2021wbn}. Because the mediator mass, $m_\phi$ is small, the scattering enters the non-perturbative regime, requiring the evaluation of the multi-state Schrödinger equation. The presence of the light scalar mediator and the off-diagonal $\chi_1$--$\chi_2$ coupling leads to inelastic self-interactions in the dark sector. In the nonrelativistic limit, the two-state system can be described by a coupled-channel potential, with the mass splitting $\delta$ determining the threshold for transitions between the ground and excited states. The potential for the two pseudo-Dirac fermion DM with a light scalar mediator is: \cite{Schutz:2014nka, Blennow:2016gde, Zhang:2016dck, Arkani-Hamed:2008hhe}
\begin{equation}
	V(r)=\left(
	\begin{array}{cc}
		0 & -\frac{\alpha}{r} e^{-m_\phi r}\\
		-\frac{\alpha}{r} e^{-m_\phi r} &2\delta  \\
	\end{array}
	\right).\,
\end{equation}
The two-body Schrödinger equation for relative motion is
\begin{equation}
	\frac{1}{m_\chi}\nabla^2 \Psi(\vec{r})=\big(V(r) - m_\chi v^2)\Psi(\vec{r}),
\end{equation}
where $m_{\chi}$ is the mass of the DM ignoring the tiny mass splitting $\delta$, $v$ is the individual velocity of either of the DM particles in the center of mass frame and $\Psi(\vec{r})$ is the wave function. Defining dimensionless parameters, $\epsilon_v=\frac{v}{ \alpha}$, $\epsilon_\delta=\sqrt{\frac{2\delta}{m_\chi \alpha^2}}$, $\epsilon_{\phi}=\frac{m_{\phi}}{m_\chi \alpha}$ and writing $r\Psi(\vec{r})=\psi(r)$, the s-wave Schrödinger equation is given by
\begin{equation}
	\psi''(r)=\left(
	\begin{array}{cc}
		-\epsilon^2_v & -\frac{e^{\epsilon_\phi r}}{r}\\
		-\frac{e^{\epsilon_\phi r}}{r} &~~~\epsilon^2_\delta -\epsilon^2_v \\
	\end{array}
	\right)\psi(r).
\end{equation}

The phenomenologically relevant quantity for self-interacting DM is the cross section per unit DM mass (${\sigma}/{m_{\chi_1}}$),
which is constrained by observations of galaxy clusters, merging systems, and dwarf galaxies. We therefore determine the regions of $(m_{\chi_1},m_\phi,\alpha,\delta)$ parameter space capable of producing the required velocity-dependent self-interaction cross-section while simultaneously satisfying the relic-density and direct and indirect detection constraints discussed in the following discussions. The relevant cross sections are given in appendix \ref{appen:isidm}. For a more general analysis, one may refer to \cite{Schutz:2014nka}.

\begin{figure}[h]
\centering
\includegraphics[scale=0.43]{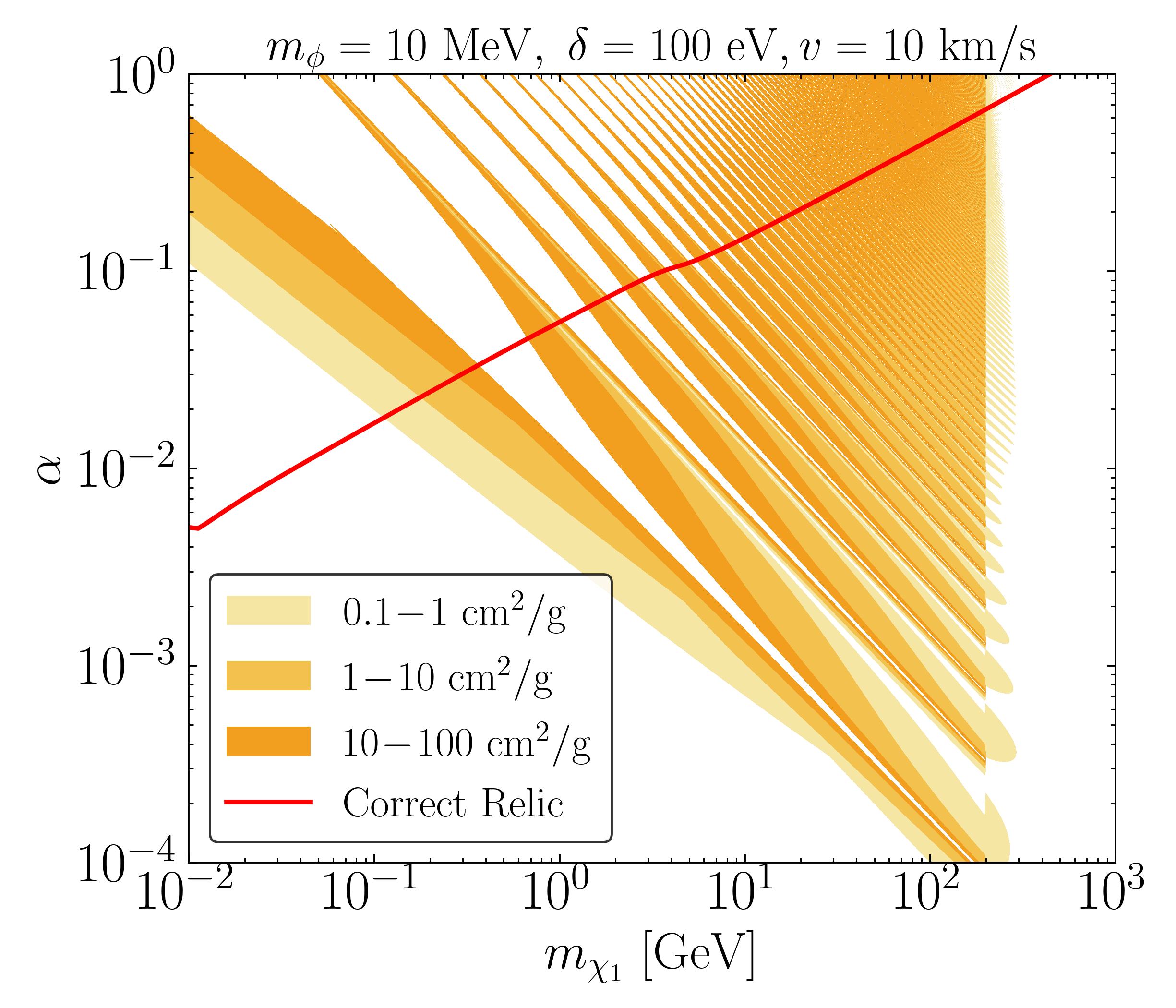}
\includegraphics[scale=0.43]{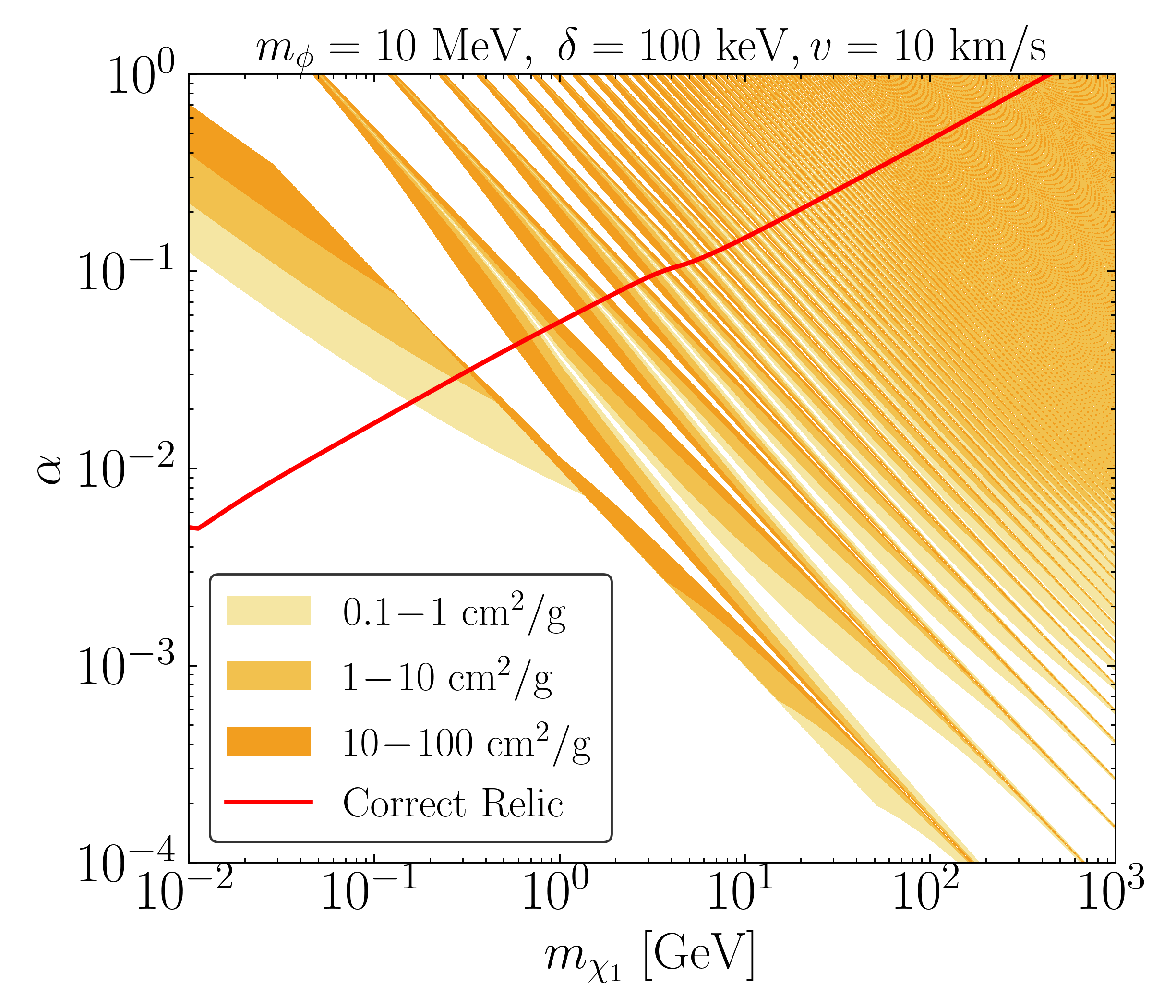}\
\includegraphics[scale=0.43]{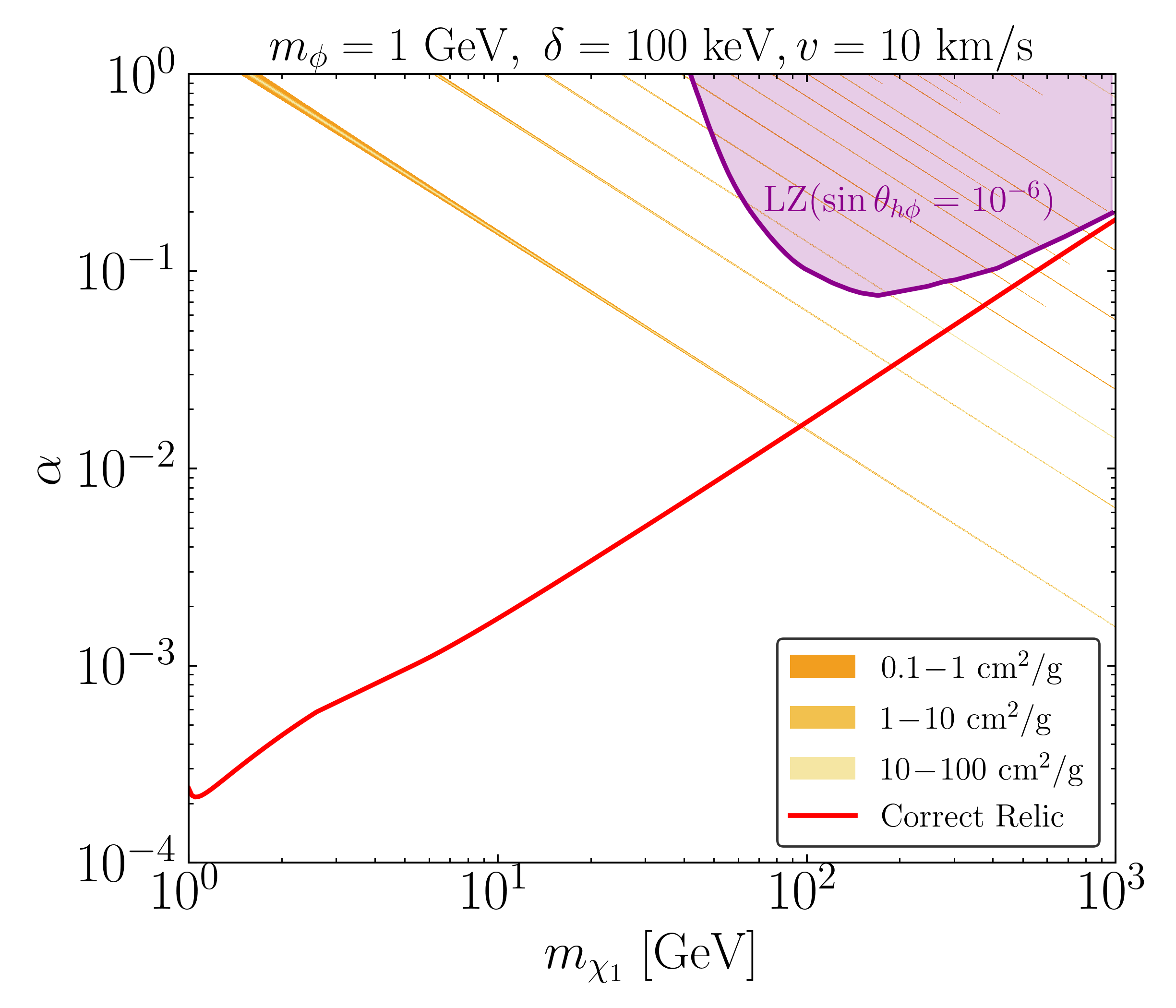}
\includegraphics[scale=0.43]{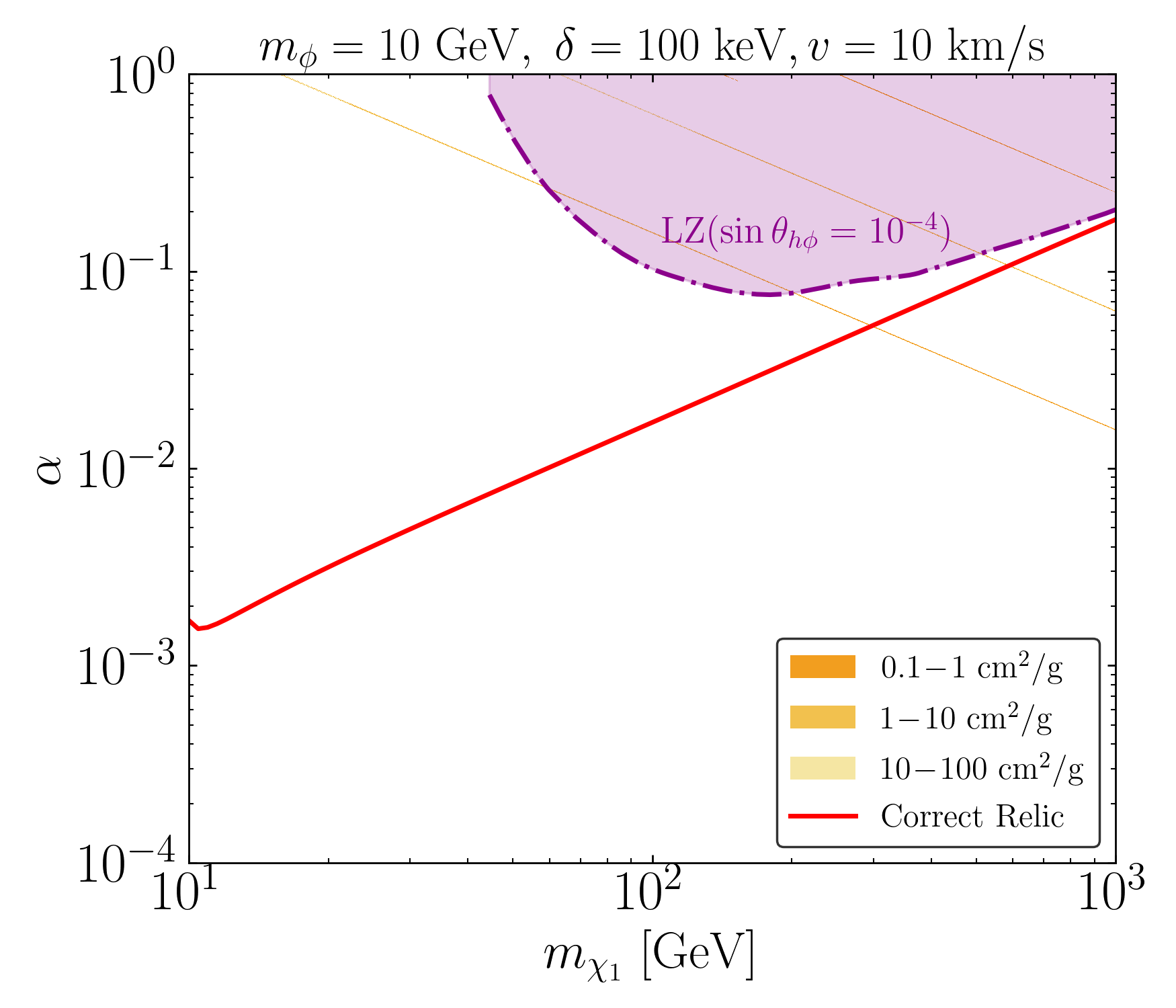}
\caption{Allowed parameter space in the $\alpha$--$m_{\chi_1}$ plane for the required DM self-interaction cross section. The red curves indicate parameter combinations yielding the observed DM relic abundance. The purple shaded regions represent the parameter space already ruled out by the LZ(2025) constraints.}
\label{fig:sidm_param}
\end{figure}

Using these self-interaction cross sections and using the required $\sigma/m$ from astrophysical observations at different scales, we constrain the parameter space of the model. The self-interaction properties of the DM are controlled primarily by the effective coupling $\alpha$ and the mediator-to-DM mass ratio, while the mass splitting $\delta$ determines the kinematics of the coupled $\chi_1$--$\chi_2$ system. To illustrate the viable parameter space, Fig.~\ref{fig:sidm_param}
shows the regions in the $\alpha$--$m_{\chi_1}$ plane required to achieve significant self-interactions in dwarf galaxies, evaluated at a characteristic velocity of $v = 10\text{ km/s}$. The shaded yellow bands depict three representative cross-section yield $0.1$--$1~{\rm cm^2/g}$, $1$--$10~{\rm cm^2/g}$, and $10$--$100~{\rm cm^2/g}$.  The different panels correspond to different choices of the mediator mass and the mass splitting.

As depicted in the upper panel of Fig.~\ref{fig:sidm_param}, for a light mediator, $m_\phi=10~{\rm MeV}$, the self-interaction cross section
can be sizable over a broad range of DM masses and couplings. This
can be understood from the long-range nature of the scalar-mediated Yukawa
potential, for which decreasing $m_\phi$ increases the interaction range and
thereby enhances the non-perturbative scattering effects. As a result, the
required self-interaction cross sections can be obtained with relatively small
values of $\alpha$ over a wide range of $m_{\chi_1}$. The corresponding
parameter space also exhibits a pronounced band structure characteristic of
the resonant regime of Yukawa-mediated scattering. The effect of the mass splitting is particularly visible by comparing the two
upper panels. Increasing the splitting from $\delta=100~{\rm eV}$ to
$\delta=100~{\rm keV}$ substantially modifies the location and width of the
self-interaction bands. The splitting changes the coupled-channel dynamics of
the two Majorana states and consequently shifts the positions of the
non-perturbative resonances. Thus, even for the same mediator mass, the
self-interaction cross section is not determined solely by $\alpha$ and
$m_{\chi_1}$, but is also sensitive to $\delta$.

In the bottom panels of Fig.~\ref{fig:sidm_param}, increasing the mediator mass to $m_\phi = 1\text{ GeV}$ and $10\text{ GeV}$ pushes the parameter space into the perturbative Born regime ($\epsilon_\phi = m_\phi / (m_{\chi_1} \alpha) \gg 1$). Here, the quantum resonances are completely washed out. For heavier mediators, the interaction becomes shorter ranged and the
self-interaction rate is correspondingly reduced and hence require progressively
larger DM masses and/or larger values of $\alpha$ to obtain the same
range of self-interaction cross sections. In particular, the broad
self-interaction regions present for $m_\phi=10~{\rm MeV}$ are substantially
reduced as the mediator mass is increased.

Overall, Fig.~\ref{fig:sidm_param} demonstrates that the model can accommodate the phenomenologically interesting range $\sigma/m_{\chi_1} \sim 0.1 - 100\text{ cm}^2/\text{g}$ for suitable combinations of $(m_{\chi_1}, m_\phi, \alpha, \delta)$. This parameter space should be scrutinized further when identifying the parameter region that simultaneously satisfies the relic-density requirement (indicated by the solid red curves) and the direct-detection constraints (indicated by the purple shaded regions) discussed below.

\subsection{Dark Matter Relic Density}
\label{subsec:relic}
The relic abundance of the dark sector is determined by the coupled thermal evolution of the two nearly degenerate Majorana states, $\chi_1$ and $\chi_2$. In the parameter region of interest, the mass splitting $\delta=m_{\chi_2}-m_{\chi_1}$ is much smaller than the DM mass, while the off-diagonal coupling to the light mediator $h_\phi$ remains unsuppressed. At temperatures well above the mass splitting $T\gg\delta$, transitions between the two states are efficient and the two components remain in chemical equilibrium. 

The various dark-sector processes relevant for determining the relic abundance include DM annihilation into mediator pairs, co-annihilation, conversion between the two dark states, and scattering processes involving the mediator. The relative importance of these processes depends on the DM mass, mediator mass, coupling $\alpha$, and the mass splitting $\delta$.

We define the comoving number densities by
$Y_{\chi_i}\equiv{n_{\chi_i}}/{s}$
where $s$ is the entropy density, and introduce the dimensionless inverse temperature $
x\equiv{m_{\chi_1}}/{T}$.
Thus the DM yield, $Y_{\chi_i}$ is obtained by solving the following Boltzmann equations:
\begin{eqnarray}
    \frac{dY_{\chi_1}}{dx}&=&-\frac{sx}{\mathcal{H}(m_{\chi_1})}\Bigg[\langle\sigma{v}\rangle_{\overline{\chi_1}\chi_1\rightarrow h_\phi h_\phi}\left( Y_{\chi_1}^2-(Y^{\rm eq}_{\chi_1})^2 \right)+\langle\sigma{v}\rangle_{\overline{\chi_1}\chi_1\rightarrow \overline{\chi_2}\chi_2}\left( Y_{\chi_1}^2-Y_{\chi_2}^2\frac{(Y^{\rm eq}_{\chi_1})^2}{(Y^{\rm eq}_{\chi_2})^2} \right)\nonumber\\
    &+& \langle\sigma{v}\rangle_{\overline{\chi_1}\chi_2\rightarrow h_\phi h_\phi}\left( Y_{\chi_1}Y_{\chi_2}-Y^{\rm eq}_{\chi_1}Y^{\rm eq}_{\chi_2} \right)- \langle\sigma{v}\rangle_{{\chi_2}h_\phi\rightarrow {\chi_1} h_\phi}\left( Y_{\chi_2}Y_{h_\phi}^{\rm eq}-Y_{\chi_1}Y_{h_\phi}^{\rm eq}\frac{Y^{\rm eq}_{\chi_2}}{Y^{\rm eq}_{\chi_1}} \right)\nonumber\\
    &-& \frac{\Gamma_{\chi_2\rightarrow\chi_1}}{s} \frac{K_1(m_{\chi_2}/T)}{K_2(m_{\chi_2}/T)} \left( Y_{\chi_2}-Y_{\chi_1}\frac{Y^{\rm eq}_{\chi_2}}{Y^{\rm eq}_{\chi_1}} \right) \Bigg],\label{eq:Boltzmann1}\\
    \frac{dY_{\chi_2}}{dx}&=&-\frac{sx}{\mathcal{H}(m_{\chi_1})}\Bigg[\langle\sigma{v}\rangle_{\overline{\chi_2}\chi_2\rightarrow h_\phi h_\phi}\left( Y_{\chi_2}^2-(Y^{\rm eq}_{\chi_2})^2 \right)-\langle\sigma{v}\rangle_{\overline{\chi_1}\chi_1\rightarrow \overline{\chi_2}\chi_2}\left( Y_{\chi_1}^2-Y_{\chi_2}^2\frac{(Y^{\rm eq}_{\chi_1})^2}{(Y^{\rm eq}_{\chi_2})^2} \right)\nonumber\\
    &-& \langle\sigma{v}\rangle_{\overline{\chi_1}\chi_2\rightarrow h_\phi h_\phi}\left( Y_{\chi_1}Y_{\chi_2}-Y^{\rm eq}_{\chi_1}Y^{\rm eq}_{\chi_2} \right)+ \langle\sigma{v}\rangle_{{\chi_2}h_\phi\rightarrow \chi_1 h_\phi}\left( Y_{\chi_2}Y_{h_\phi}^{\rm eq}-Y_{\chi_1}Y_{h_\phi}^{\rm eq}\frac{Y^{\rm eq}_{\chi_2}}{Y^{\rm eq}_{\chi_1}} \right)\nonumber\\
    &+& \frac{\Gamma_{\chi_2\rightarrow\chi_1}}{s} \frac{K_1(m_{\chi_2}/T)}{K_2(m_{\chi_2}/T)} \left( Y_{\chi_2}-Y_{\chi_1}\frac{Y^{\rm eq}_{\chi_2}}{Y^{\rm eq}_{\chi_1}} \right) \Bigg],\label{eq:Boltzmann2}
\end{eqnarray}
where $s=2\pi^2/45g_{*s}m_{\chi_1}^3x^{-3}$, $\mathcal{H}(m_{\chi_1})=1.66\sqrt{g_*}m_{\chi_1}^2/m_{\rm pl}$ are the entropy density and Hubble expansion rate respectively with $m_{\rm pl}=1.22\times10^{19}$ GeV is the Planck mass. The detailed thermal averages are evaluated using the corresponding annihilation, conversion, and scattering cross sections. The conversion and scattering terms in Eqs.~\eqref{eq:Boltzmann1} and \eqref{eq:Boltzmann2} have opposite signs in the two equations, as required by number transfer between the two states. They therefore do not change the total dark-sector number density directly, but they determine how the total abundance is distributed between $\chi_1$ and $\chi_2$. In contrast, (co-)annihilation processes such as $\chi_i\chi_j\rightarrow h_\phi h_\phi$ reduce the total dark-sector abundance and determine the freeze-out density.

For sufficiently small $\delta/T$, the two states remain in approximate chemical equilibrium during freeze-out, and their equilibrium abundance ratio is
\begin{equation}
\frac{Y_{\chi_2}^{\rm eq}}
{Y_{\chi_1}^{\rm eq}}
=
\left(1+\frac{\delta}{m_{\chi_1}}\right)^{3/2}
e^{-\delta/T}.
\label{eq:equilibriumratio}
\end{equation}
Since $\delta\ll m_{\chi_1}$, the dominant temperature dependence of the ratio arises from the Boltzmann factor. During the epoch relevant for thermal freeze-out, the two states can consequently have comparable abundances when the conversion processes remain efficient.

As established by the interaction Lagrangian in Eq.~\eqref{eq:DMinteractionepsilon}, the diagonal scalar couplings are strictly suppressed by the explicit symmetry-breaking parameter $\epsilon$. Consequently, the  co-annihilation cross sections $\langle \sigma v \rangle_{\overline{\chi_i}\chi_j \to h_\phi h_\phi}$ are proportional to $\epsilon^2$ and provide a subdominant contribution to the overall DM depletion. The thermal freeze-out is instead overwhelmingly dominated by the unsuppressed off-diagonal couplings appearing in annihilation process $\chi_i \chi_i \to h_\phi h_\phi$, which scales with the unsuppressed coupling $Y_L$. Concurrently, the rapid conversion and scattering processes ($\chi_1 \chi_1 \leftrightarrow \chi_2 \chi_2$ and $\chi_2 h_\phi \leftrightarrow \chi_1 h_\phi$) act efficiently to maintain chemical equilibrium, balancing the relative abundances of the two dark states during the freeze-out epoch without changing the total number of dark-sector particles.   

Although the Boltzmann equations explicitly incorporate the decay of the heavier state via the $\Gamma_{\chi_2 \to \chi_1 \gamma \gamma}$ and $\Gamma_{\chi_2 \to \chi_1 \nu \nu}$ term, these decays have virtually no impact on the number density evolution in the early universe. For the parameter space of primary interest, where the mass splitting is $\delta \lesssim \mathcal{O}(100\text{ keV})$, both the loop-induced diphoton decay ($\chi_2 \to \chi_1 \gamma \gamma$) and the tree-level decay into active neutrinos ($\chi_2 \to \chi_1 \nu \bar{\nu}$) are profoundly kinematically suppressed. The latter is further suppressed by the heavy Dirac Inverse Seesaw messenger scale $M_N$. We discuss these lifetimes in detail within the indirect detection framework in Sec.~\ref{subsec:dmdecay}. Because the $\chi_2$ lifetime vastly exceeds the age of the universe, the excited state effectively behaves as a stable species during all relevant cosmological epochs.

By numerically integrating these coupled Boltzmann equations, we derive the exact combinations of $m_{\chi_1}$, $m_\phi$, and $\alpha$ that yield the correct total relic abundance.
\begin{equation}
    \Omega_{\chi_{i}}h^2\simeq 0.118\left( \frac{Y^{\infty}_{\chi_i}}{4.2\times10^{-10}} \right)\left( \frac{m_{\chi_i}}{1~{\rm GeV}} \right)
\end{equation}

These solutions form the solid red relic density contours overlaid in Fig.~\ref{fig:sidm_param}. Because the $\chi_2$ state remains cosmologically stable and the initial chemical equilibrium condition in Eq.~\eqref{eq:equilibriumratio} enforces $Y_{\chi_1} \approx Y_{\chi_2}$ for $T \gg \delta$, the present-day total DM relic density is simply the sum of both components:
\begin{equation}
\Omega_{\rm DM}h^2 = \Omega_{\chi_1}h^2 + \Omega_{\chi_2}h^2.
\end{equation}
This confirms that the pseudo-Dirac states persist as a multi-component dark sector today, with each component constituting approximately half of the total DM energy density in the galactic halo.

\subsection{Dark Matter Detection Prospects}
\subsubsection{CMB and Indirect Detection Constrtaints}\label{subsec:dmdecay}
Strong DM annihilation into SM particles is stringently constrained by measurements of the Cosmic Microwave Background (CMB) due to entropy injection during the recombination epoch~\cite{Madhavacheril:2013cna,Slatyer:2015jla,Planck:2018vyg, Elor:2015bho, Profumo:2017obk}, as well as by present-day indirect search experiments targeting DM annihilation in galactic halos~\cite{Fermi-LAT:2015att,HESS:2018cbt,Profumo:2017obk}. In our framework, the dominant annihilation channel for the fermion DM is into a pair of scalar mediators ($\chi_i \chi_i \to h_\phi h_\phi$). Because the DM consists of Majorana fermions and the mediator is a scalar, this annihilation process is strictly $p$-wave suppressed, scaling with the velocity squared ($v^2$). In the late universe and during recombination, where DM velocities are exceedingly small, this cross section effectively vanishes~\cite{Choquette:2016xsw, An:2016kie}. This inherent kinematic suppression naturally evades both CMB and contemporary indirect detection bounds, offering a robust theoretical motivation for scalar mediator scenarios.

\begin{figure}[h]
\centering
\includegraphics[scale=0.445]{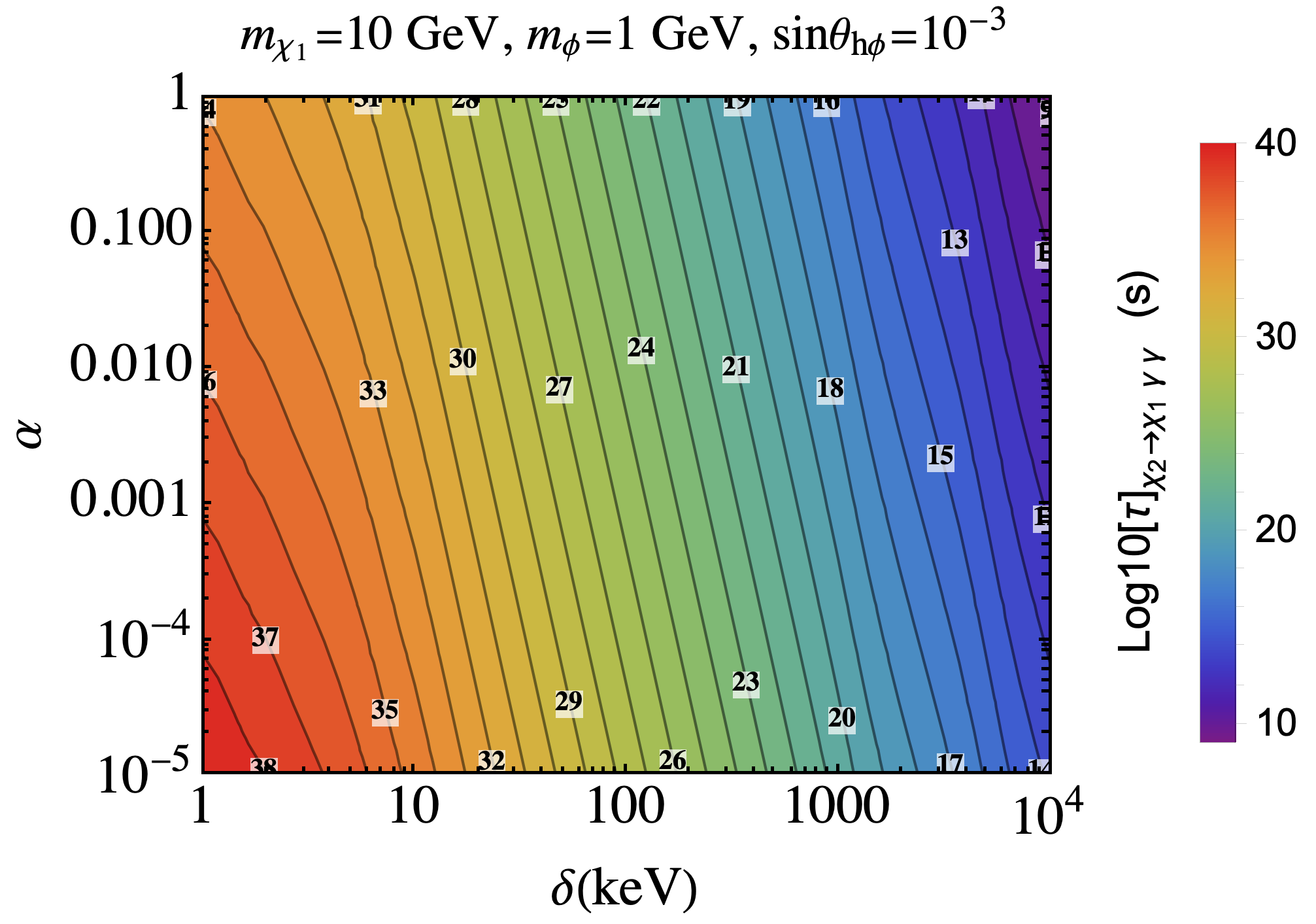}
\includegraphics[scale=0.445]{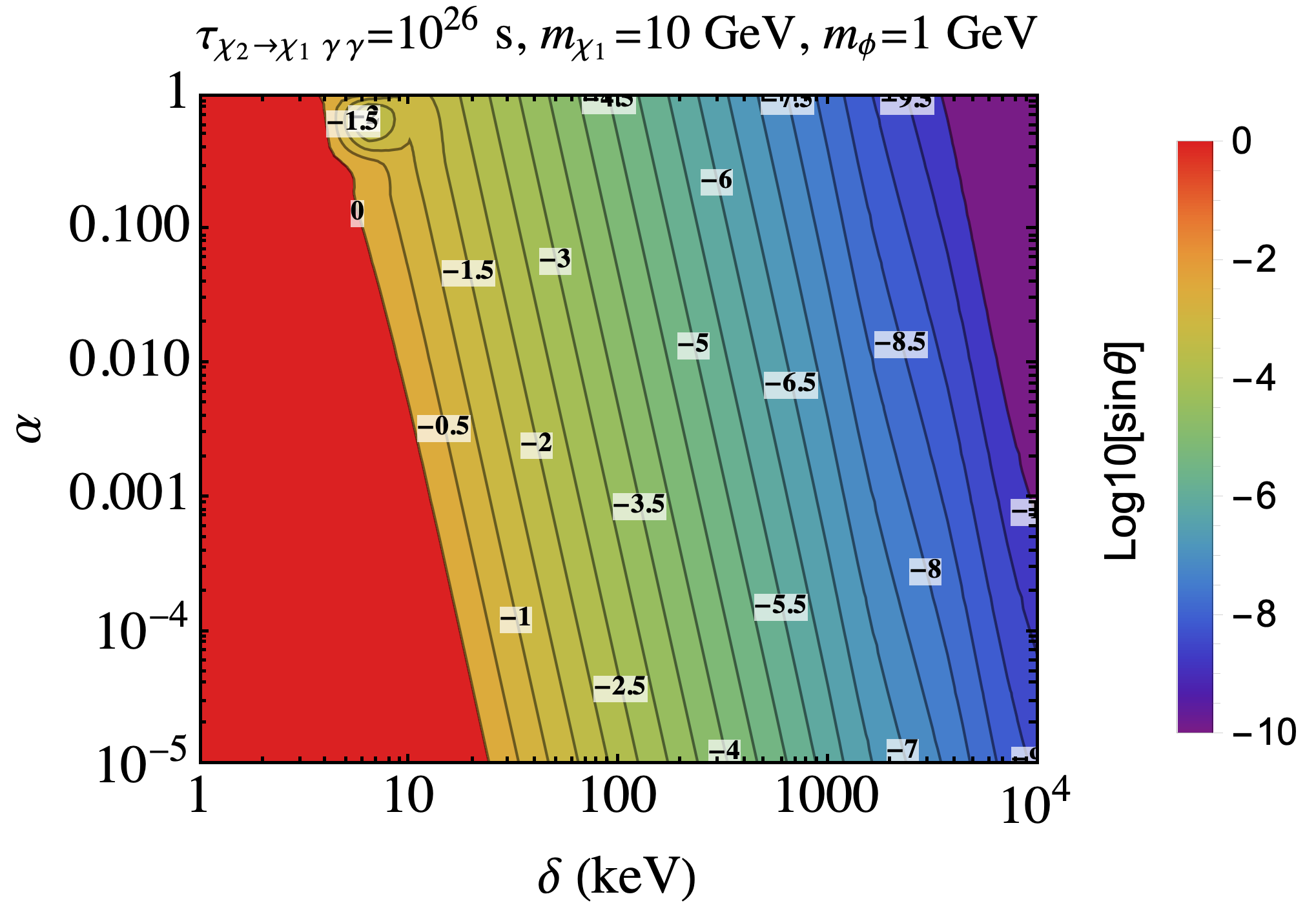}
\caption{The $\chi_2 \to \chi_1 \gamma \gamma$ decay lifetime in the $\alpha - \delta$ plane for $m_{\chi_1} = 10$ GeV and $m_\phi = 1$ GeV. The left panel show the logarithmic lifetime (in seconds) for scalar mixing angle $\sin\theta_{h\phi} = 10^{-3}$. The right panel displays the required mixing angle $\sin\theta_{h\phi}$ to maintain a fixed lifetime of $\tau = 10^{26}$ s.}
\label{fig:lifetime}
\end{figure}

Due to the small pseudo-Dirac mass splitting ($\delta$) and identical thermal interactions, $\chi_1$ and $\chi_2$ freeze out with approximately equal number densities, both surviving as present-day DM components. However, the heavier $\chi_2$ state is kinematically permitted to decay into the lighter $\chi_1$ state. Within our modular Dirac inverse seesaw architecture, the available channels are $\chi_2 \to \chi_1 \nu \bar{\nu}$ and $\chi_2 \to \chi_1 \gamma \gamma$. The effective vertex for $\chi_2 \to \chi_1 \nu \bar{\nu}$ can be written as
\begin{eqnarray}
 \mathcal{Y}_{\rm eff}=y_{\rm eff}\frac{1}{m_\phi^2}Y_\phi \frac{1}{M_N}Y_H\frac{v_h}{\sqrt{2}}\,,   
\end{eqnarray}
with which the decay width is estimated to be \cite{Fuss:2025xwe}
\begin{eqnarray}
\Gamma_{\chi_2\rightarrow\chi_1\nu\nu}\simeq \frac{1}{320\pi^3}\mathcal{Y}_{\rm eff}^2\delta^5. 
\end{eqnarray}
This gives a lifetime of
\begin{eqnarray}
\tau_{\chi_2\rightarrow\chi_1\nu\nu}\simeq 2.13\times10^{21}{~\rm s} \left[ \left( \frac{y_{\rm eff}}{10^{-2}} \right)^2 \left( \frac{1~{\rm GeV}}{m_\phi} \right)^4 \left( \frac{Y_{\phi}}{10^{-4}} \right)^2 \left( \frac{1~{\rm TeV}}{M_N} \right)^2 \left( \frac{Y_{H}}{10^{-4}} \right)^2 \left( \frac{\delta}{100{~\rm keV}} \right)^5  \right]^{-1}.
\end{eqnarray}
With these choices of parameters, $\chi_2$ can be sufficiently long-lived, allowing the constraints from diffuse neutrino observations to be safely evaded \cite{Fuss:2025xwe}. Consequently, the loop-induced diphoton decay $\chi_2 \to \chi_1 \gamma \gamma$, facilitated by the scalar mixing angle $\theta_{h\phi}$ with the SM Higgs presents a more severe constraint. This decay width is given by~\cite{Krnjaic:2025zjl}:
\begin{eqnarray}
\Gamma_{\chi_2\rightarrow\chi_1\gamma\gamma}= \frac{\alpha^2}{1024\pi^3m_{\chi_2}^3}\int_0^{\delta m^2}dq^2 q^4 \sqrt{\lambda[m_{\chi_2}^2,m_{\chi_1}^2,q^2]}|C_{\gamma\gamma}|^2\frac{(m_{\chi_2}-m_{\chi_1})^2-q^2}{(q^2-m_\phi^2)^2},   
\end{eqnarray}
with,
\begin{eqnarray}
    C_{\gamma\gamma}=\frac{\alpha_{\rm em}\sin\theta_{h\phi}}{2\pi v_h}\left[\sum_f N_c^f Q_f^2 \mathcal{A}_{1/2}(\tau_f)+ \mathcal{A}_W (\tau_W ) \right],
\end{eqnarray}
where, $\alpha_{\rm em}=1/137.0359$, $\tau_i=q^2/4m_i^2$ and,
\begin{eqnarray}
 \nonumber   \mathcal{A}_{1/2}&=&2\tau^{-2}\left(\tau+(\tau-1)f(\tau)\right),\\
    \mathcal{A}_W&=&-\tau^{-2}\left(2\tau^2+3\tau+3(2\tau-1)f(\tau)\right).
\end{eqnarray}
In the above relations, the function $f(\tau)$ is defined as
\begin{eqnarray}
f(\tau)=\left\{ \begin{array}{lc}\arcsin^2 \sqrt{\tau}; &   \tau \leq 1 \\
\\ -\frac{1}{4}\left[\log \frac{1+\sqrt{1-\tau^{-1}}}{1-\sqrt{1-\tau^{-1}}}-i\pi\right]^2; &   \tau > 1 
\end{array}
\right.
\end{eqnarray}

In Fig.~\ref{fig:lifetime}, we evaluate the constraints on the $\chi_2$ lifetime in the $\alpha - \delta$ plane. To ensure $\chi_2$ survives as a dominant DM component to the present epoch, its lifetime must significantly exceed the age of the universe ($\sim 10^{17}$ s). Furthermore, to evade stringent X-ray and gamma-ray bounds from galactic halo observations, lifetimes typically must exceed $\sim 10^{26}$ s~\cite{Slatyer:2016qyl,Essig:2013goa,He:2020sat}. The left panel demonstrates that the lifetime is highly sensitive to the mass splitting $\delta$. As $\delta$ decreases, the available phase space for the decay shrinks drastically, substantially increasing the lifetime. For the parameter space of primary interest corresponding to our inelastic direct detection analysis ($\delta \lesssim 100$ keV), the lifetime comfortably exceeds $10^{26}$ seconds, even for relatively large dark fine-structure couplings ($\alpha \sim 0.1$) and mixing angles ($\sin\theta_{h\phi} = 10^{-3}$). The right panel confirms that for splittings below a few hundred keV, the maximum scalar mixing angle required to satisfy the $10^{26}$ s bound resides well within the allowed region of standard Higgs portal constraints, allowing $\chi_2$ as a viable, stable co-dark matter candidate.

\subsubsection{Direct Detection}
\label{subsec:dmdd}
\noindent
Direct-detection experiments provide one of the most stringent tests of
DM models with appreciable couplings to the SM.
In the present framework, the scalar mediator $h_\phi$ communicates
between the dark sector and the visible sector through its mixing with
the SM Higgs boson. In the absence of the approximate
$\mathcal{Z}_2^{LR}$ symmetry, the scalar interaction would generically
induce an unsuppressed elastic scattering process
$\chi_1 N\rightarrow\chi_1 N$, which is strongly constrained by
xenon-based experiments. The approximate dark-sector left-right
symmetry introduced in Sec.~\ref{sec:model} suppresses this diagonal
coupling, while allowing a sizable off-diagonal interaction
$\chi_1\leftrightarrow\chi_2$. Consequently, the leading nuclear
scattering process relevant for direct detection is the inelastic
transition $\chi_1+N\rightarrow\chi_2+N$. 

The importance of the mass splitting can be understood directly from
kinematics. Since the incident DM particle must provide the
energy required to excite $\chi_1$ into $\chi_2$, the minimum velocity
required to produce a nuclear recoil of energy $E_R$ is
\begin{equation}
    v_{\rm min}(E_R)
    =
    \frac{1}{\sqrt{2m_NE_R}}
    \left(
    \frac{m_NE_R}{\mu_{\chi N}}+\delta
    \right),
    \label{eq:vmin_inelastic}
\end{equation}
where, $\mu_{\chi N}=\frac{m_{\chi_1}m_N}{m_{\chi_1}+m_N}$ is the reduced mass of the DM--nucleus system. For $\delta=0$, Eq.~\eqref{eq:vmin_inelastic} reduces to the familiar expression for elastic DM scattering. For nonzero $\delta$,
however, the additional term proportional to $\delta$ shifts the
required scattering velocity towards the high-velocity tail of the
Galactic DM distribution. Once $v_{\rm min}$ becomes
comparable to the maximum speed of DM particles in the
laboratory frame, the scattering rate is strongly suppressed. Thus,
a relatively large nuclear cross section can remain compatible with
direct-detection limits when the inelastic splitting is sufficiently
large.
This mechanism is particularly relevant for the present model because
the same pseudo-Dirac structure that suppresses elastic scattering also
allows sizable self-interactions through the light mediator
$h_\phi$. Hence, the direct-detection constraint does not require the
dark-sector coupling to be small; instead, it can be relaxed
kinematically through the mass splitting. This provides the essential
connection between the self-interacting and inelastic nature of the
DM considered here.

The differential rate per unit detector mass is given by \cite{Tucker-Smith:2001myb}
\begin{eqnarray}
    \frac{d\mathcal{R}}{dE_R}=N_T\frac{\rho_{\chi_1}}{m_{\chi_1}}\int_{v_{\rm min}}^{v_{\rm max}}dv ~vf(v)\frac{d\sigma}{dE_R},
\end{eqnarray}
where, $N_T$ is the number of target nuclei per unit detector mass,
$\rho_{\chi_1}$ denotes the local energy density associated with the
ground-state component, and $f(v)$ is the DM speed
distribution in the detector frame. Since the two pseudo-Dirac states
are nearly degenerate and constitute comparable fractions of the
present-day relic abundance, we take
\begin{equation}
    \rho_{\chi_1}\simeq \rho_{\chi_2}
    \simeq \frac{\rho_{\rm DM}}{2},
    \qquad
    \rho_{\rm DM}=0.4~{\rm GeV\,cm^{-3}},
    \label{eq:rho_DM_components}
\end{equation}
unless otherwise stated. The speed distribution $f(v)$ is given by:
\begin{eqnarray}
    f(v)=\frac{v}{\sqrt{\pi}v_ev_0}e^{-\frac{v_e^2+v^2}{v_0^2}}\left( e^{\frac{2vv_e}{v_0^2}}- e^{-\frac{2vv_e}{v_0^2}} \right),
\end{eqnarray}
with $v_0=220$ km/s, and the time-averaged Earth's speed relative to the galactic rest frame is $v_e=v_\odot=v_0+12 {~\rm km/s}$. For the scalar-mediated interaction considered here, the spin-independent differential cross-section can be expressed as:
\begin{eqnarray}
  \frac{d\sigma}{dE_R}=\frac{m_N}{2v^2}\frac{\sigma_n}{\mu_n^2}\frac{(f_pZ+f_n(A-Z))^2}{f_n^2}F^2(E_R),
\end{eqnarray}
with the Helm form factor:
\begin{eqnarray}
    F^2(E_R)=\left( \frac{3j_1(qr_0)}{qr_0} \right)^2e^{-s^2q^2},
\end{eqnarray}
where, $q=\sqrt{2m_NE_R}$, $s=1$ fm, $r_0=\sqrt{r^2-5s^2}$, $r=1.2A^{1/3}$ fm. Here $\sigma_n$ is the relic-neutron
cross-section given by:
\begin{eqnarray}
    \sigma_n=\frac{\mu_n^2}{\pi A^2}\left(Zf_p-(A-Z)f_n\right)^2.
\end{eqnarray}
Here $\mu_n=m_{\chi_1}m_n/(m_{\chi_1}+m_n)$, and the effective nucleon couplings are:
\begin{eqnarray}
f_{p,n}=\sum_{q=u,d,s}f^{p,n}_{T_q}\alpha_q\frac{m_{p,n}}{m_q}+\frac{2}{27}f^{p,n}_{T_G}\sum_{q=c,t,b}\alpha_q\frac{m_{p,n}}{m_q},
\end{eqnarray}
with the quark-level effective coupling determined by the scalar mixing:
\begin{eqnarray}
    \alpha_q=y_{\rm eff}\frac{m_q}{v_h}\left(\frac{1}{m_\phi^2}-\frac{1}{m_h^2}\right)\sin\theta_{h\phi}\cos\theta_{h\phi}.
\end{eqnarray}
Integrating with $v_{\rm max}=\infty$, the differential rate becomes: 
\begin{eqnarray}
 \frac{d\mathcal{R}}{dE_R}=\frac{N_Tm_N\rho_\chi}{4v_0m_{\chi_1}}\frac{\sigma_n}{\mu_n^2}\frac{(f_pZ+f_n(A-Z))^2}{f_n^2}F^2(E_R) \left( \frac{{\rm erf}[x_{\rm min}+\eta]-{\rm erf}[x_{\rm min}-\eta]}{\eta} \right),  
\end{eqnarray}
where, $x_{\rm min}=v_{\rm min}/v_0$, $\eta=v_e/v_0$.
Restricting the integral to the galactic escape velocity $v_{\rm max}=v_{\rm esc}+v_e$ (using $v_{\rm esc}=540~{\rm km/s}$), we obtain:
\begin{eqnarray}
 \frac{d\mathcal{R}}{dE_R}&=&\frac{N_Tm_N\rho_\chi}{4v_0m_{\chi_1}}\frac{\sigma_n}{\mu_n^2}\frac{(f_pZ+f_n(A-Z))^2}{f_n^2}F^2(E_R) \nonumber\\&& \times\left( \frac{{\rm erf}[x_{\rm min}+\eta]-{\rm erf}[x_{\rm min}-\eta]-{\rm erf}[x_{\rm max}+\eta]+{\rm erf}[x_{\rm max}-\eta]}{\eta} \right), \label{eq:diff_rate} 
\end{eqnarray}
where $x_{\rm max}=v_{\rm max}/v_0$. The total expected rate is computed by integrating over recoil energies:
\begin{eqnarray}
    \mathcal{R}=\int_{E_R}\epsilon(E_R) \frac{d\mathcal{R}}{dE_R}dE_R,
\end{eqnarray}
where $\epsilon(E_R)$ is the detector efficiency. The total number of events is $N_{\rm event}\simeq \mathcal{R}\times\mathcal{E}_{\rm LZ}$, where the exposure $\mathcal{E}_{\rm LZ}$ for LZ is $4.2\pm0.1$ tonne-years \cite{LZ:2024zvo}. 

The absolute kinematic cutoff for inelastic scattering dictates a maximum mass splitting:
\begin{eqnarray}
\delta_{\rm max}=\frac{\mu_{\chi N} v_{\rm max}^2}{2}\label{eq:del_max}
\end{eqnarray}
For $\delta>\delta_{\rm max}$, the inelastic up-scattering process is
kinematically forbidden. Equation~\eqref{eq:del_max} therefore
provides a useful upper boundary on the mass splitting accessible to
a given target and DM mass.

We use the stringent constraints from the LUX-ZEPLIN experiment
to test the viable parameter space of the inelastic DM
scenario. The previous LZ search based on a $4.2$ tonne-year exposure
provides the standard spin-independent limits at low recoil energies~\cite{LZ:2024zvo},
while the most recent LZ analysis employs a $2.84$ tonne-year
exposure and extends the nuclear-recoil search window to approximately
$270$~keV~\cite{LZ:2026axp}. 

\begin{figure}[h]
\centering
\includegraphics[scale=0.5]{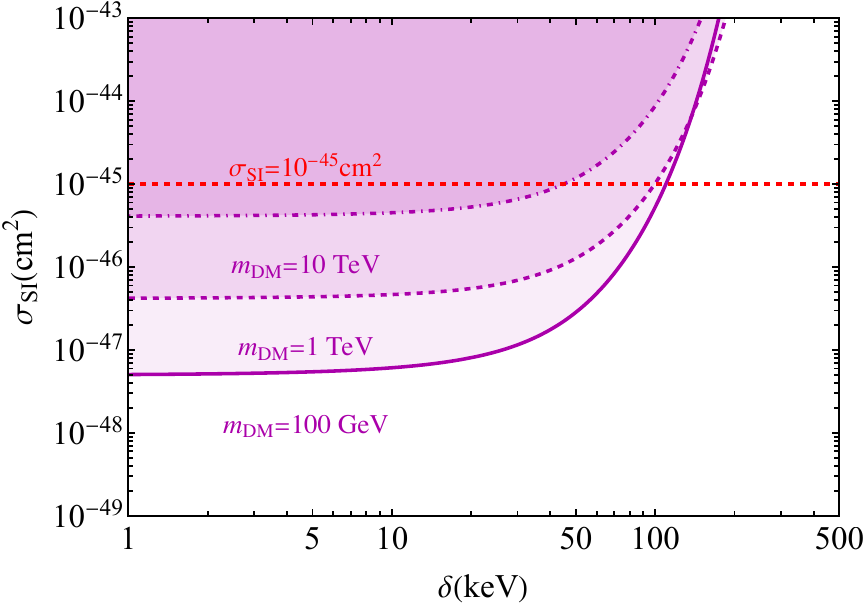}
\includegraphics[scale=0.5]{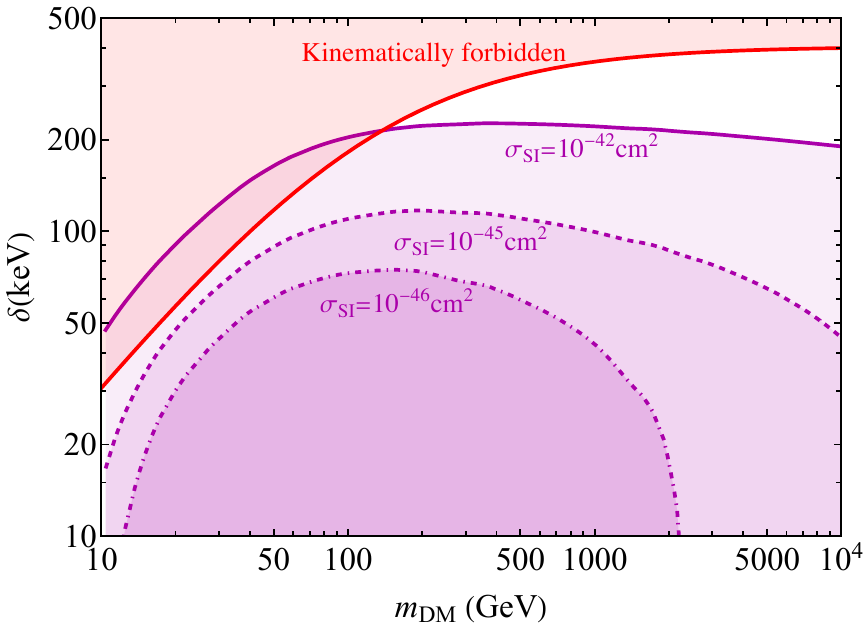}
\includegraphics[scale=0.5]{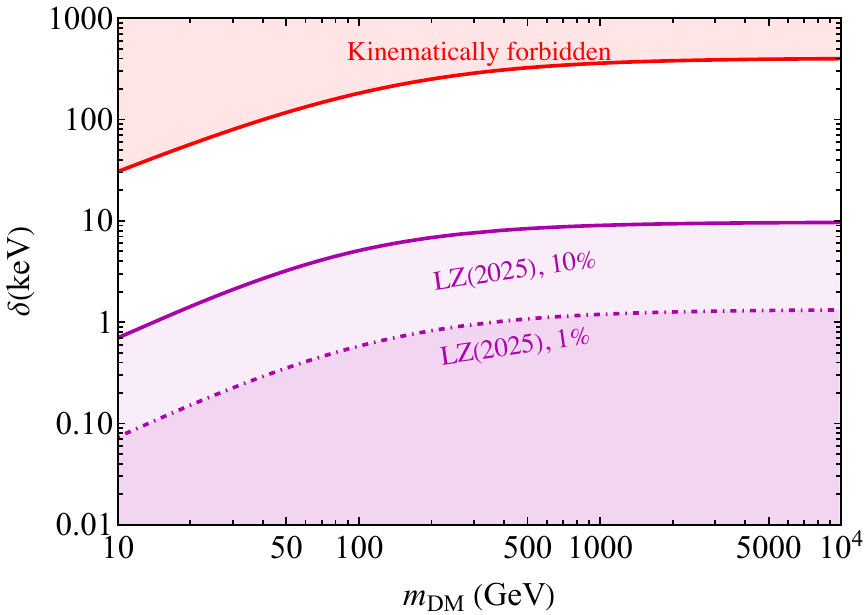}
\caption{Impact of the inelastic mass splitting on the spin-independent
direct-detection rate. 
\textit{Top left:} spin-independent cross section as a function of
$\delta$ for $m_{\chi_1}=100~{\rm GeV}$, $1~{\rm TeV}$, and
$10~{\rm TeV}$. The shaded regions correspond to parameter choices
for which the predicted event rate exceeds the adopted LZ(2025) constraint.
The horizontal dashed line denotes the reference value
$\sigma_{\rm SI}=10^{-45}\,{\rm cm^2}$.
\textit{Top right:} Exclusion limits in the $(m_{\chi_1},\delta)$ plane
for fixed reference cross sections
$10^{-42}$, $10^{-45}$, and $10^{-46}\,{\rm cm^2}$.
\textit{Bottom:} corresponding exclusion limits obtained using the
mass-dependent LZ(2025) constraint with small upward relaxations of the
experimental constraint. The results demonstrate the kinematic
suppression of nuclear scattering as the inelastic splitting is
increased.}
\label{fig:inel_SI_xsec}
\end{figure}

To illustrate the impact of the inelastic mass splitting on direct
detection, we first consider the predicted spin-independent
cross section required to obtain a fixed event rate. The top-left
panel of Fig.~\ref{fig:inel_SI_xsec} shows the corresponding
$\sigma_{\rm SI}$ as a function of $\delta$ for
100 GeV (solid magenta), 1 TeV (dashed magenta), and 10 TeV (dashed-dotted magenta).
For each DM mass, the shaded region corresponds to parameter
choices for which the predicted number of events exceeds the adopted
LZ(2025) upper limit. The increase of the required cross section with
$\delta$ follows directly from Eq.~\eqref{eq:vmin_inelastic}: a larger
splitting increases the minimum velocity required for
$\chi_1 N\rightarrow\chi_2N$, thereby reducing the fraction of halo
particles capable of producing the recoil. Consequently, a larger
microscopic scattering cross section $\sigma_{\rm SI}$ is required to compensate for
this effect. For illustration, the horizontal line in the top-left panel denotes a reference cross-section
$\sigma_{\rm SI}=10^{-45}\,{\rm cm^2}$. The intersection of this
reference cross section with the corresponding exclusion curves
defines a characteristic splitting $\delta_c$ above which the
reference cross section is no longer excluded. For instance, a cross-section of $\sim 10^{-45}{\rm ~cm^2}$ for a DM mass of 100 GeV is already ruled out by the LZ. However, in the inelastic DM framework, this cross-section is allowed with a $\delta>\delta_c$. This demonstrates the central advantage of the inelastic
scenario: a nucleon cross section that would be strongly constrained
in the elastic limit can remain viable once the splitting pushes the
scattering into the high-velocity tail of the halo distribution. 

In the top-right panel of Fig.~\ref{fig:inel_SI_xsec}, we instead fix
the reference cross sections to $10^{-42}{\rm~ cm^2}$ (solid magenta), $10^{-45}{\rm~ cm^2}$ (dashed magenta) and $10^{-46}{\rm~ cm^2}$ (dashed-dotted magenta) and determine the
corresponding boundary in the $(m_{\chi_1},\delta)$ plane. The regions below each curve give a large number of events that are ruled out by LZ(2025). The
excluded region moves towards larger $\delta$ as the cross section is
increased. This behavior follows because a larger cross section
produces more scattering events at fixed $\delta$, hence consequently, a
larger mass splitting is required to obtain the same suppression of
the event rate.

The bottom panel shows the corresponding result when the experimental limit is used, which is dependent on the DM mass rather than fixing a
single reference cross section. We consider small upward relaxations of the adopted LZ(2025) limit (1\% and 10\%. Relaxing the LZ(2025) limit by $x\%$ we mean, we consider the cross-section limits as $\sigma_{\rm SI}^{x\%}=\sigma_{\rm SI}^{\rm LZ}+\sigma_{\rm SI}^{\rm LZ}\times \frac{x}{100}$) and determine the resulting boundary in the
$(m_{\chi_1},\delta)$ plane. The solid magenta and dashed dotted magenta regions represent the 10\% and 1\% limits, respectively. This representation is useful for
identifying the region in which the inelastic suppression becomes
important while retaining the full DM mass dependence of
the direct-detection sensitivity.

\begin{figure}[h]
\centering
\includegraphics[scale=0.55]{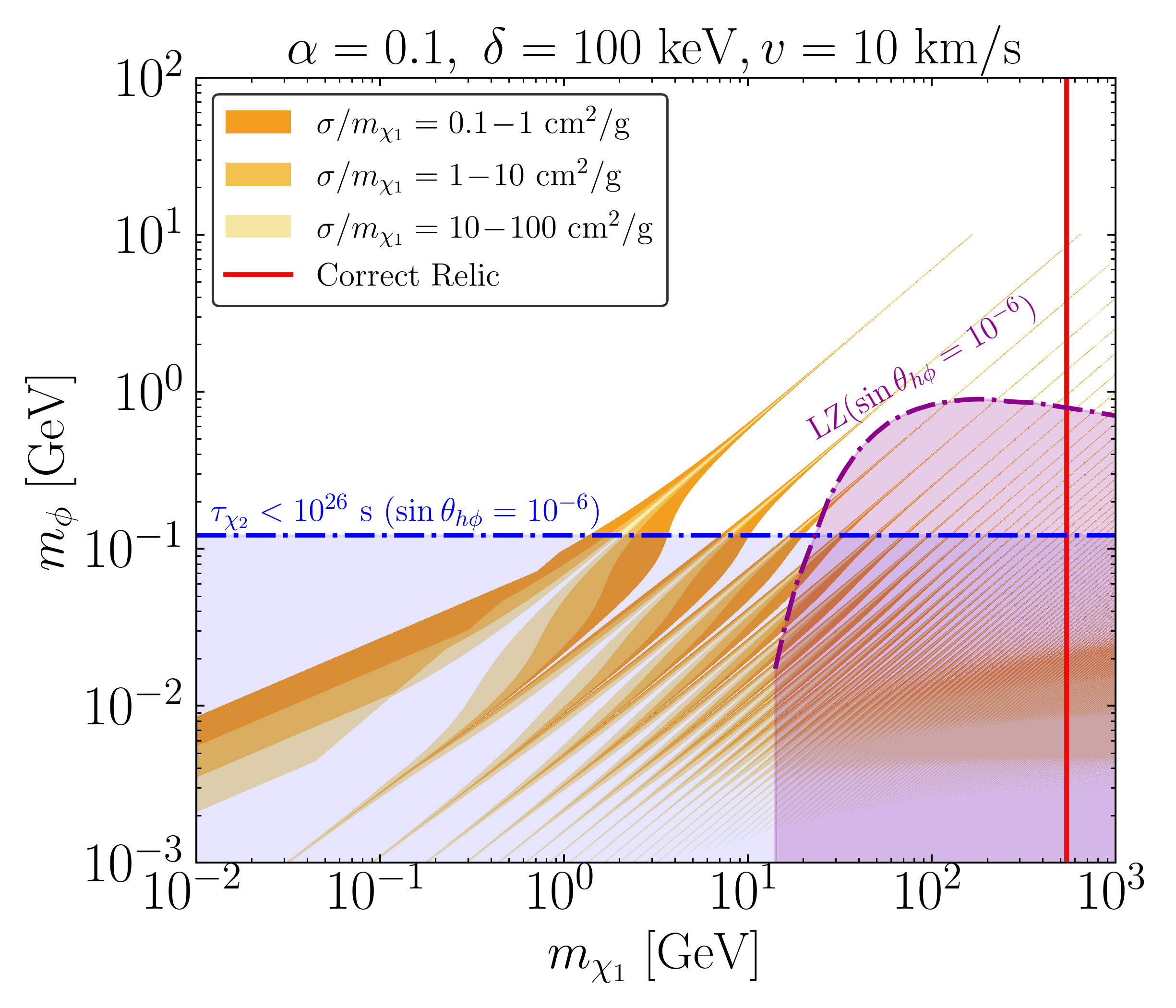}
\caption{Summary of viable parameter space in the $m_\phi - m_{\chi_1}$ plane for $\alpha = 0.1$, with fixed $\delta = 100$ keV and $v = 10$ km/s. Overlaid are the SIDM cross-section regions (yellow bands), the correct relic density contour (solid red), the indirect lifetime constraint ($\tau_{\chi_2} < 10^{26}$ s, blue shaded), and the LZ (2025) direct detection exclusion \cite{LZ:2024zvo} zone (purple shaded).}
\label{fig:sidm_param2}
\end{figure}

In Fig.~\ref{fig:sidm_param2}, we synthesize all prior phenomenological constraints into a comprehensive summary of the model's viable parameter space in the $m_\phi - m_{\chi_1}$ plane for fixed couplings $\alpha = 0.1$, assuming a mass splitting of $\delta = 100$ keV. The overlapping regions demonstrate where the model simultaneously achieves the desired SIDM cross sections for dwarf halos (yellow bands), satisfies the Planck thermal relic abundance (red line), ensures the $\chi_2$ lifetime exceeds the $10^{26}$ s indirect detection bound (above the blue dash-dotted line), and successfully evades the inelastic LZ(2025) \cite{LZ:2024zvo} exclusion limit (outside the purple shaded region for $\sin\theta_{h\phi} = 10^{-6}$). 

Finally, building upon this robust phenomenological foundation, we will utilize this exact inelastic DM formalism in the subsequent section to explain the recently reported 248 keV nuclear recoil event observed by LZ, identifying the specific, viable parameter configurations within our model that can naturally account for this signal.

\section{The LZ230616 Nuclear Recoil Event}\label{sec:lzevent}

The inelastic nature of the DM candidate in our framework makes
the model particularly interesting in light of the recent high-energy
nuclear-recoil event reported by the LUX-ZEPLIN (LZ) experiment \cite{LZ:2026axp}. Using
an exposure of $2.84~{\rm tonne\!-\!yr}$, the latest LZ analysis extends
the nuclear-recoil search window up to approximately $270$~keV and
reports a single event consistent with a nuclear recoil of
$E_R=248\pm23_{\rm stat}\pm23_{\rm sys}$~keV~\cite{LZ:2026axp}. The event occurs in a
region with a low expected background. A profile-likelihood analysis
performed by the LZ collaboration finds a maximum local significance
of $3.4\sigma$ and a global significance of $2.6\sigma$. 

In our model, the ground-state DM particle $\chi_1$ can
undergo the endothermic transition
$\chi_1 + N \longrightarrow \chi_2 + N$. 
As discussed in Sec.~\ref{subsec:dmdd}, the mass splitting modifies the
minimum velocity required to produce a recoil of energy $E_R$ according
to Eq.~\eqref{eq:vmin_inelastic}. A
nonzero $\delta$ therefore suppresses low-energy recoils and shifts the
signal towards the high-velocity tail of the Galactic DM
distribution. This feature is particularly relevant for a recoil as
large as $248$~keV and provides the basic motivation for investigating
the LZ230616 event within our pseudo-Dirac DM framework.

To determine the region of parameter space that can accommodate the
observed recoil, we perform an extended maximum likelihood analysis
in the model parameter space. Our analysis is intended to determine
the DM parameters that yield a recoil spectrum compatible
with the observed event and is therefore distinct from the full LZ
profile-likelihood analysis, which incorporates the detailed
background model and experimental nuisance parameters.

We define the parameter vector
\begin{equation}
    \Theta =
    \left\{
        m_{\chi_1},\,
        \delta,\,
        \alpha,\,
        m_\phi,\,
        \sin\theta_{h\phi}
    \right\}.
    \label{eq:LZ_parameters}
\end{equation}
For a given point in this parameter space, the predicted number of
events per unit recoil energy is
\begin{equation}
    \frac{dN(\Theta)}{dE_R}
    =
    \mathcal{E}_{\rm LZ}\,
    \epsilon(E_R)\,
    \frac{d\mathcal R(\Theta)}{dE_R},
    \label{eq:dNdER_LZ}
\end{equation}
where $d\mathcal R/dE_R$ is the differential recoil rate derived in
Sec.~\ref{subsec:dmdd}, $\epsilon(E_R)$ is the detector efficiency,
and $\mathcal{E}_{\rm LZ}=2.84~{\rm tonne\!-\!yr}$
is the exposure used in the high-energy recoil LZ analysis.
The expected total number of signal events in the analysis window is
then,
\begin{equation}
    N_{\rm tot}(\Theta)
    =
    \int_{E_R^{\rm min}}^{E_R^{\rm max}}
    dE_R\,
    \frac{dN(\Theta)}{dE_R},
    \label{eq:Ntot_LZ}
\end{equation}
where,
\begin{equation}
    E_R^{\rm min}=5.4~{\rm keV},
    \qquad
    E_R^{\rm max}=270~{\rm keV}.
\end{equation}
For the single event of interest, the extended unbinned likelihood can
be written as \cite{Barlow:1990vc}
\begin{equation}
    \mathcal L(\Theta)
    =
    e^{-N_{\rm tot}(\Theta)}
    \prod_{i=1}^{n_0}
    \left.
    \frac{dN(\Theta)}{dE_R}
    \right|_{E_R=E_i},
    \label{eq:LZ_extended_likelihood}
\end{equation}
where, $n_0=1$ and $E_i = 248 \pm 23 ~(\rm stat) \pm23 ~(sys)$ keV.

We construct the negative log-likelihood as:
\begin{equation}
    -\ln\mathcal L(\Theta)
    =
    N_{\rm tot}(\Theta)
    -
    \left.
    \ln\left[
    \frac{dN(\Theta)}{dE_R}
    \right]\right|_{E_R=248~{\rm keV}},
    \label{eq:minuslogL}
\end{equation}
and minimize it to extract the best-fit (BF) model parameters, $\hat{\Theta} \equiv (\hat{m}_{\chi_1}, \hat{\delta}, \hat{\alpha}, \hat{m}_\phi, \widehat{\sin\theta}_{h\phi})$, yielding the maximum likelihood:
\begin{equation}
    -\ln\mathcal L_{\rm max}
    =
    -\ln\mathcal L(\widehat{\Theta}).
    \label{eq:Lmax}
\end{equation}
We define the test statistic
\begin{equation}
    {\rm TS}(\Theta)
    =
    2
    \left[
        -\ln\mathcal L(\Theta)
        +\ln\mathcal L_{\rm max}
    \right].
    \label{eq:TS_LZ}
\end{equation}
The confidence regions shown below are obtained by fixing the remaining three parameters and varying the two parameters displayed in each plane. For two parameters of interest, we use ${\rm TS}=2.30, 6.18, 11.83$ for the $1\sigma$, $2\sigma$, and $3\sigma$ regions respectively.

\begin{figure}[h]
\centering
\includegraphics[width=0.485\linewidth]{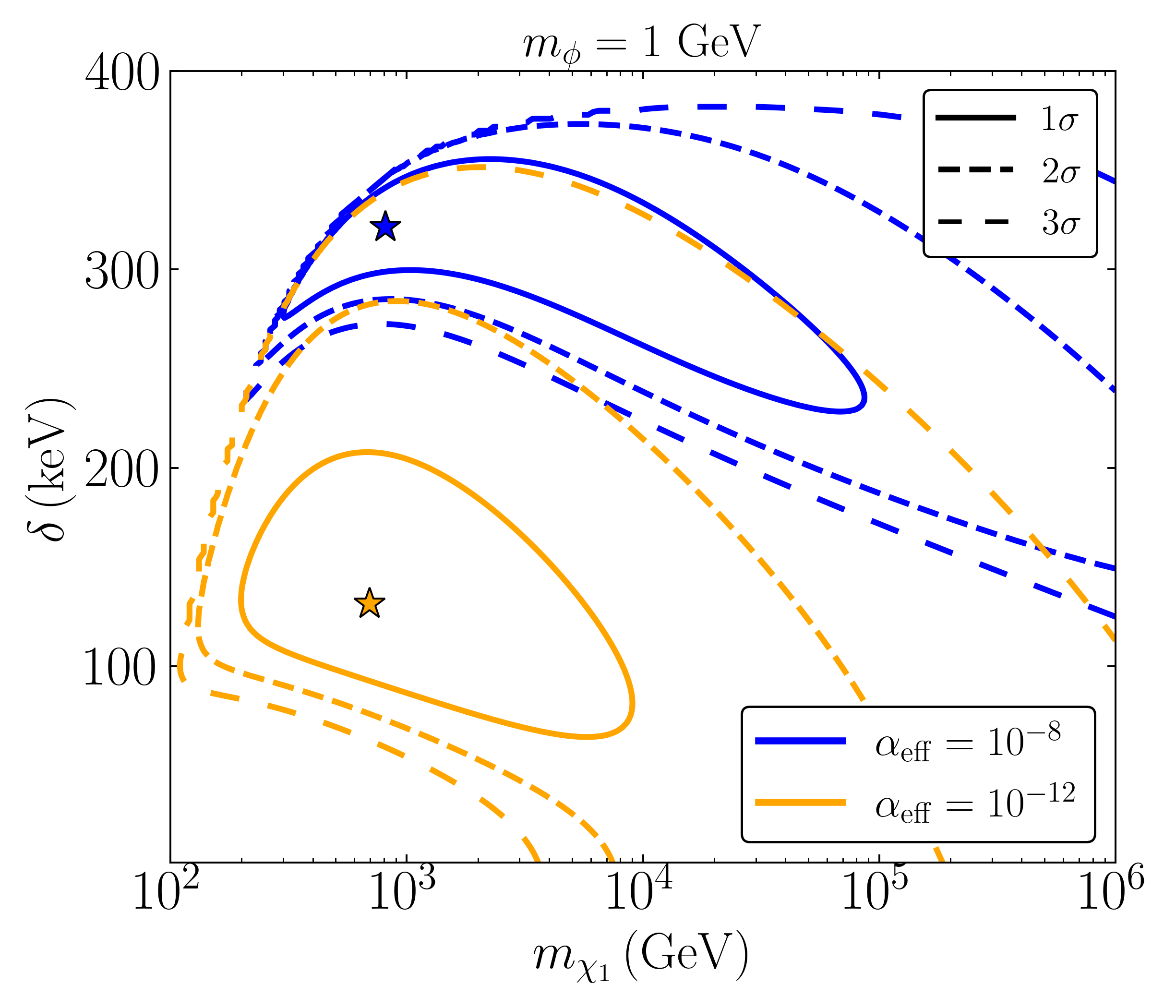}
\includegraphics[width=0.485\linewidth]{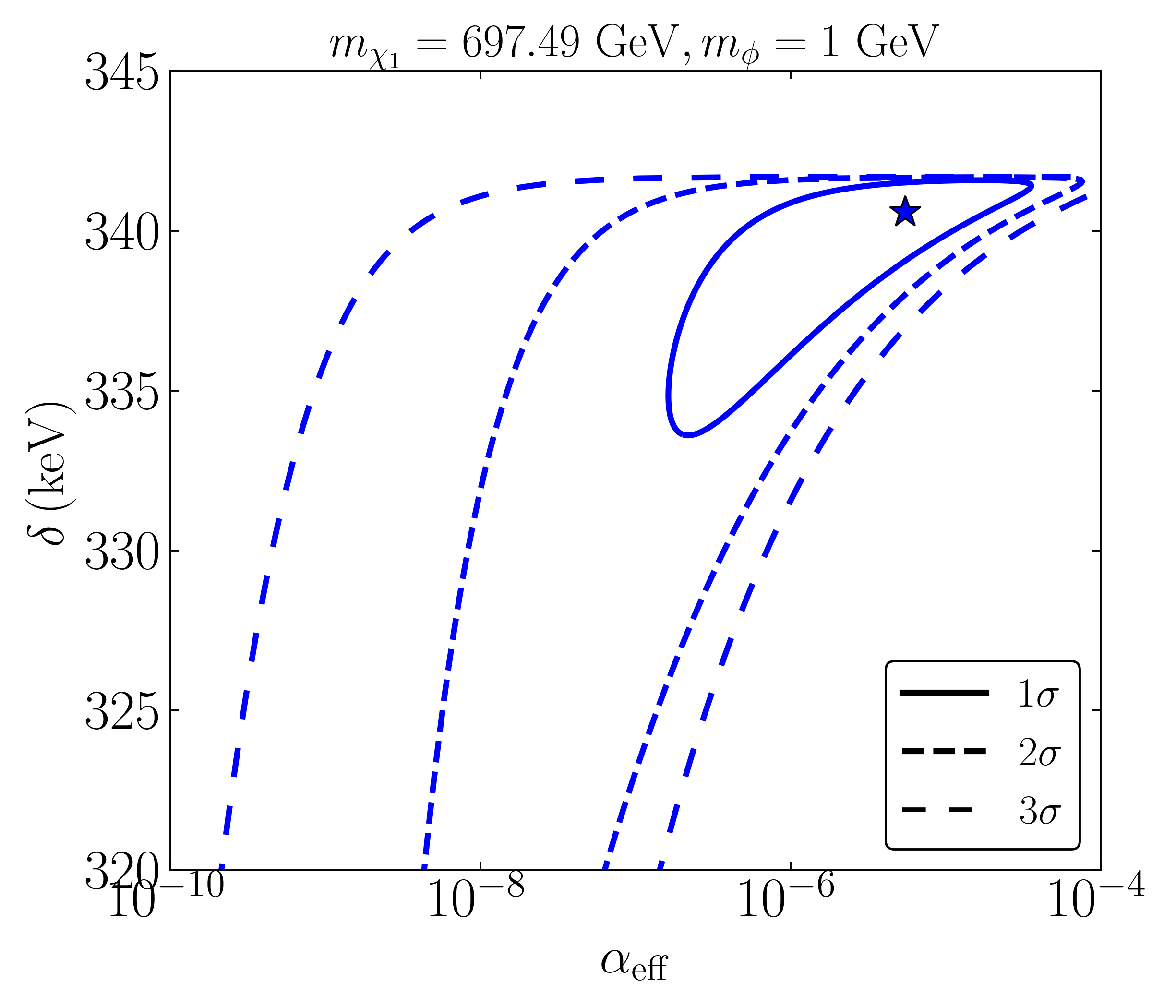}
\caption{[\textit{Left}:] $1\sigma$, $2\sigma$ and $3\sigma$ contours accommodating the LZ230616 event  in the $\delta-m_{\chi_1}$ plane for two choices of effective coupling $\alpha_{\rm eff}=10^{-8},10^{-12}$. [\textit{Right}:] $1\sigma$, $2\sigma$ and $3\sigma$ contours accommodating the LZ230616 event in the $\delta-\alpha_{\rm eff}$ plane for DM mass of 697.49 GeV. In both cases, we keep $m_\phi=1$ GeV.}
\label{fig:LZ_excess0}
\end{figure}

An important simplification occurs in our model. The spin-independent
inelastic scattering rate depends on the dark-sector coupling and the
scalar mixing through the combination $\mathcal R
    \propto
    \alpha\,\sin^2(2\theta_{h\phi})$.
We therefore define the effective coupling $
    \alpha_{\rm eff}
    \equiv
    \alpha\,\sin^2(2\theta_{h\phi}),
$
which completely determines the direct-detection rate for fixed
$m_{\chi_1}$, $\delta$, and $m_\phi$. Consequently, the event-level
likelihood can first be studied in the four-dimensional parameter
space
\begin{equation}
    \left\{
        m_{\chi_1},\,
        \delta,\,
        \alpha_{\rm eff},\,
        m_\phi
    \right\}.
\label{eq:reduced_LZ_parameter_space}
\end{equation}
The separate dependence on $\alpha$ and $\sin\theta_{h\phi}$ is then
recovered by imposing the relic-density and the excited-state lifetime
requirements discussed in Sec.~\ref{sec:DM}.

In the \textit{left} panel of Fig. \ref{fig:LZ_excess0}, the resulting confidence regions are shown in the
$(m_{\chi_1},\delta)$ plane for $m_\phi=1$ GeV and two representative values of the effective coupling, $\alpha_{\rm eff}=\{10^{-8},10^{-12}\}$. The solid, dashed, long-dashed contours correspond to the $1\sigma$, $2\sigma$ and $3\sigma$ regions and star points represent the best-fit points for the two effective couplings, $\alpha_{\rm eff}$. The $1\sigma$ region that could explain the observed LZ event for the $\alpha_{\rm eff}=10^{-8}$ is shown by the blue solid line. The best fit ($\textcolor{blue}{\star}$) value corresponds to \{$m^{\rm BF}_{\chi_1}=811.03~{\rm GeV}, \delta^{\rm BF}= 321.46~{\rm keV}$\}. The corresponding $1\sigma$  ranges of the parameters are $\{\delta^{1\sigma}=[229.57,353.88]{~\rm keV},m^{1\sigma}_{\chi_1}=[303.68, 86034.64]{~\rm GeV}\}$. The very broad upper range in $m_{\chi_1}$ reflects the fact that once
$m_{\chi_1}\gg m_N$, the reduced mass approaches the nuclear mass and the kinematic dependence on the DM mass becomes weak. The characteristic shape of the contours can be understood from the
kinematics and normalization of the inelastic scattering rate. The largest mass splitting for which a recoil of energy $E_R$ remains kinematically accessible follows from $v_{\rm min}(E_R)\leq v_{\rm max}$ and is given by,
\begin{eqnarray}
\delta_{\rm max}(E_R)=v_{\rm max}\sqrt{2m_NE_R}-E_R\left(1+\frac{m_N}{m_{\chi_1}}\right).
\end{eqnarray}

It is observed that the $\delta$ values initially increase with $m_{\chi_1}$ and then decrease. The increasing behavior is dictated by the kinematics, while the decreasing behavior is due to the suppression of the event rate. The maximum kinematically allowed value of $\delta$ increases with $m_{\chi_1}$. However, for larger DM masses ($m_{\chi_1}\gg m_N$), $\delta_{\rm max}(E_R)$, becomes nearly independent of $m_{\chi_1}$, causing the kinematic ceiling to saturate and remain approximately fixed. At the same time, the number of events decreases with increasing DM mass as $N_{\rm event}\propto 1/m_{\chi_1}$. Therefore, to obtain the observed number of events, $\delta$ needs to decrease. This behavior is more clearly visible in the $2\sigma$ and $3\sigma$ contours. For $\alpha_{\rm eff}=10^{-12}$, the contours are shown in orange. The BF values in this case are $\{m_{\chi_1}^{\rm BF}=697.49~{\rm GeV}, \delta^{\rm BF}=131.44~{\rm keV}\}$ with the $1\sigma$ ranges of the parameters $\{\delta^{1\sigma}=[65.16,207.52]{~\rm keV},m^{1\sigma}_{\chi_1}=[200.22,8907.35]{~\rm GeV}\}$. It is interesting to note that the best-fit value of $\delta$ is significantly reduced to $\sim131$ keV for $\alpha_{\rm eff}=10^{-12}$, compared to $\sim321$ keV for $\alpha_{\rm eff}=10^{-8}$. This is a direct consequence of the smaller scattering strength. A smaller $\alpha_{\rm eff}$ suppresses the
signal rate, and the resulting loss of events is compensated by reducing $\delta$. Since $v_{\rm min}$ decreases as $\delta$ decreases,
a larger fraction of the high-velocity halo can participate in the scattering. Thus, the same observed recoil can be accommodated with
a smaller mass splitting when the effective coupling is reduced.

In the \textit{right} panel of Fig. \ref{fig:LZ_excess0}, we show the likelihood for the LZ 248 keV excess in the $\delta-\alpha_{\rm eff}$ plane.
Here we have fixed the DM mass and mediator mass to be $\{m_{\chi_1}=697.49~{\rm GeV},m_\phi=1~{\rm GeV}\}$. 
For the fixed DM mass, once
$
    v_{\rm min}(248~{\rm keV})>v_{\rm max},
$
the $248$~keV recoil cannot be produced and the predicted signal rate
vanishes and the likelihood is undefined. This occurs at approximately $\delta\simeq342~{\rm keV}.$

The confidence contours exhibit a clear anticorrelation between
$\delta$ and $\alpha_{\rm eff}$. Increasing $\alpha_{\rm eff}$
enhances the scattering rate, whereas increasing $\delta$ suppresses
the available phase space. These two effects therefore compensate one
another in the region preferred by the event.
As $\alpha_{\rm eff}$ decreases, the number of signal events decreases. This is compensated by reducing the mass splitting, which enhances the scattering rate. The $1\sigma$ allowed ranges of the parameters are $\{\delta^{1\sigma}=[333.63,341.54]{~\rm keV},\alpha^{1\sigma}_{\rm eff}=[1.64\times10^{-7}, 3.54\times10^{-5}]\}$ while at  $2\sigma$ we obtain $\{\delta^{2\sigma}=[290.69,341.59]{~\rm keV},\alpha^{2\sigma}_{\rm eff}=[2.29\times10^{-9}, 7.48\times10^{-5}]\}$.

\begin{figure}[h]
\centering
\includegraphics[width=0.7\linewidth]{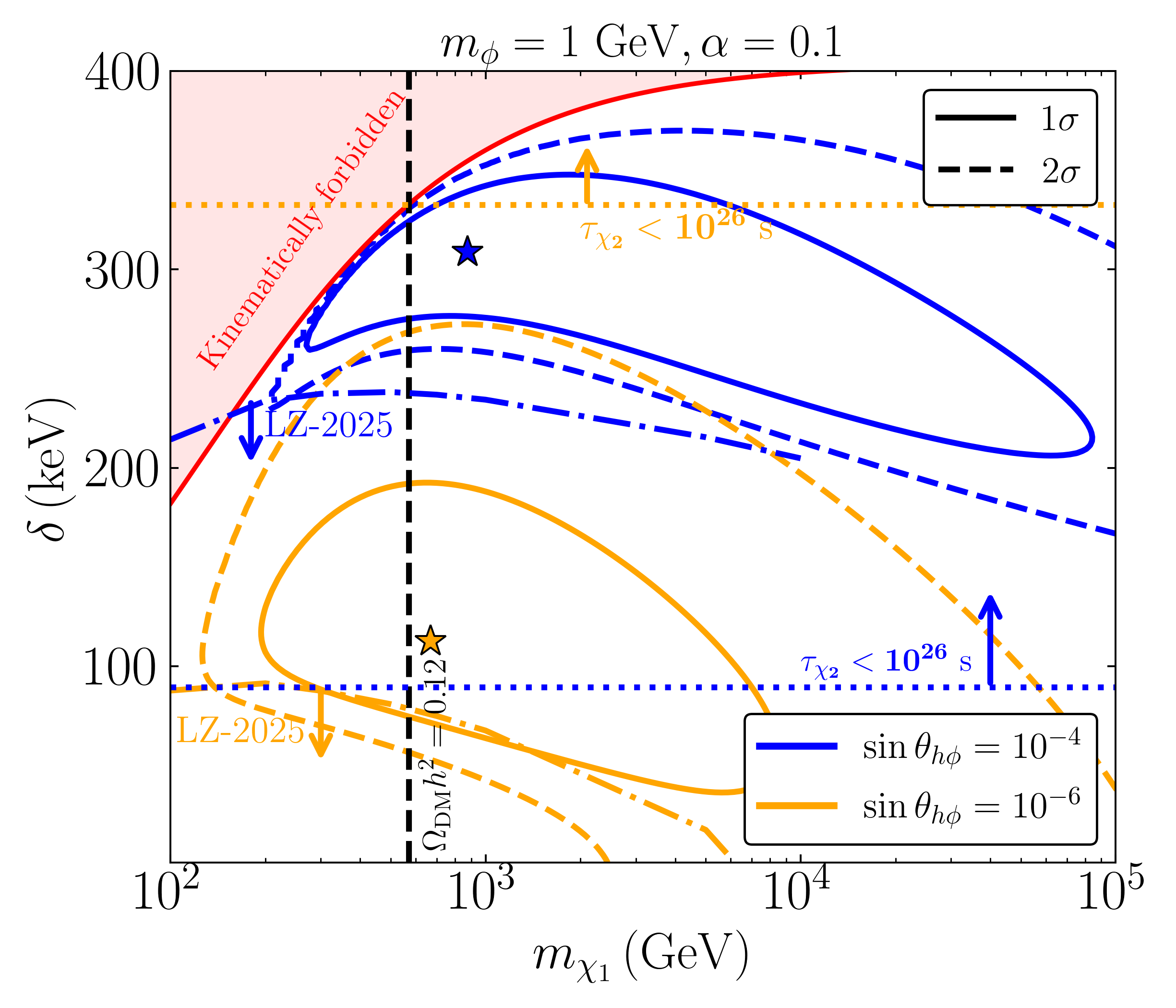}
\caption{$1\sigma$ and $2\sigma$ contours that could explain the observed event at LZ in the $\delta-m_{\chi_1}$ plane. Here, we have chosen two values of $\sin\theta_{h\phi}=\{10^{-4},10^{-6}\}$. The other two parameters are fixed at ${m_\phi=1{\rm~GeV},\alpha=0.1}$. The LZ (2025) constraint and $\chi_2$ decay constraint are also shown. See the main text for more details.}
\label{fig:LZ_excess}
\end{figure}
We now move to Fig. \ref{fig:LZ_excess}, where the parameter space, which can explain the observed 248 keV recoil event at LZ, is shown in the plane of $\delta$ vs $m_{\chi_1}$. We show the $1\sigma$ and $2\sigma$ contours for two sets of mixing angles $\sin\theta_{h\phi}=10^{-4},10^{-6}$ shown with blue and orange color, respectively. Here we fix $m_\phi=1$ GeV, $\alpha=0.1$.
The red shaded region corresponds to kinematically inaccessible region, $\delta>\frac{1}{2}\mu_{\chi N} v_{\rm max}^2$ for which the endothermic transition $\chi_1 N\rightarrow \chi_2 N$ is forbidden, even for the fastest moving particles with velocity $v_{\rm max}$ in the adopted halo model. Therefore, no nuclear recoil of any energy is produced. The contours for larger
$\sin\theta_{h\phi}$ occur at larger values of $\delta$, reflecting
the enhanced scattering rate associated with the larger Higgs--scalar
mixing. Conversely, reducing the mixing suppresses the direct-detection rate and therefore requires a smaller mass splitting to
retain an event rate compatible with LZ230616. The blue and orange stars represents the best fit values for $\sin\theta_{h\phi}=10^{-4}~{\rm and}~10^{-6}$ respectively. The best fit value of the mass splitting and DM mass for $\sin\theta_{h\phi}=10^{-4}$ is \{$m_{\chi_1}^{\rm BF}=874.98~{\rm GeV},\delta^{\rm BF}=308.67{~\rm keV}$\} with the $1\sigma$ range $\{m_{\chi_1}^{1\sigma}=[276.82,82143.44]{~\rm GeV},\delta^{1\sigma}=[207.52,345.86]{~\rm keV}]\}$.
The same for $\sin\theta_{h\phi}=10^{-6}$ is \{$m_{\chi_1}^{\rm BF}=670.14~{\rm GeV},\delta^{\rm BF}=112.72{~\rm keV}$\} with the allowed $1\sigma$ range $\{m_{\chi_1}^{1\sigma}=[200.22,8119.85]{~\rm GeV},\delta^{1\sigma}=[37.09,191.48]{~\rm keV}]\}$. 
The direct-detection constraint from the previous LZ(2025) analysis is shown
by the corresponding dashed-dotted curves in Fig.~\ref{fig:LZ_excess}. For $\sin\theta_{h\phi}=10^{-4}$, the low-$\delta$ portion of the
event-compatible region predicts too many nuclear recoils and is
therefore excluded by the LZ(2025) direct-detection bound. In addition, the
excited state $\chi_2$ is cosmologically long-lived in the parameter
region of interest. 
Hence, the $\chi_2$ lifetime has to be at least comparable to the age of the Universe. We take a conservative lower limit on the $\chi_2$ lifetime of $10^{26}$ s~\cite{Slatyer:2016qyl,Essig:2013goa,He:2020sat}. The blue dotted line represents $\tau_{\chi_2}=10^{26}$ s. The region above this line corresponds to $\tau_{\chi_2}<10^{26}$ s, and hence is ruled out. 
For $\sin\theta_{h\phi}=10^{-4}$, the combination of the direct-detection and lifetime constraints removes the parameter region that would otherwise accommodate the LZ event. Thus, this benchmark does not provide a simultaneous solution to the LZ event and the cosmological lifetime requirement. In contrast, for
$\sin\theta_{h\phi}=10^{-6}$, the direct-detection constraint removes
only a small portion of the $2\sigma$ region, while a substantial
region remains compatible with both the LZ event and the excited-state
lifetime constraint. For coupling $\alpha=0.1$, correct DM relic can be obtained for $m_{\chi_1}=572$ GeV, which is shown by a vertical black dashed line. The correct relic density line lies within the $1\sigma$ region preferred by the LZ230616 event for this benchmark.

\begin{figure}[h]
\centering
\includegraphics[width=0.48\linewidth]{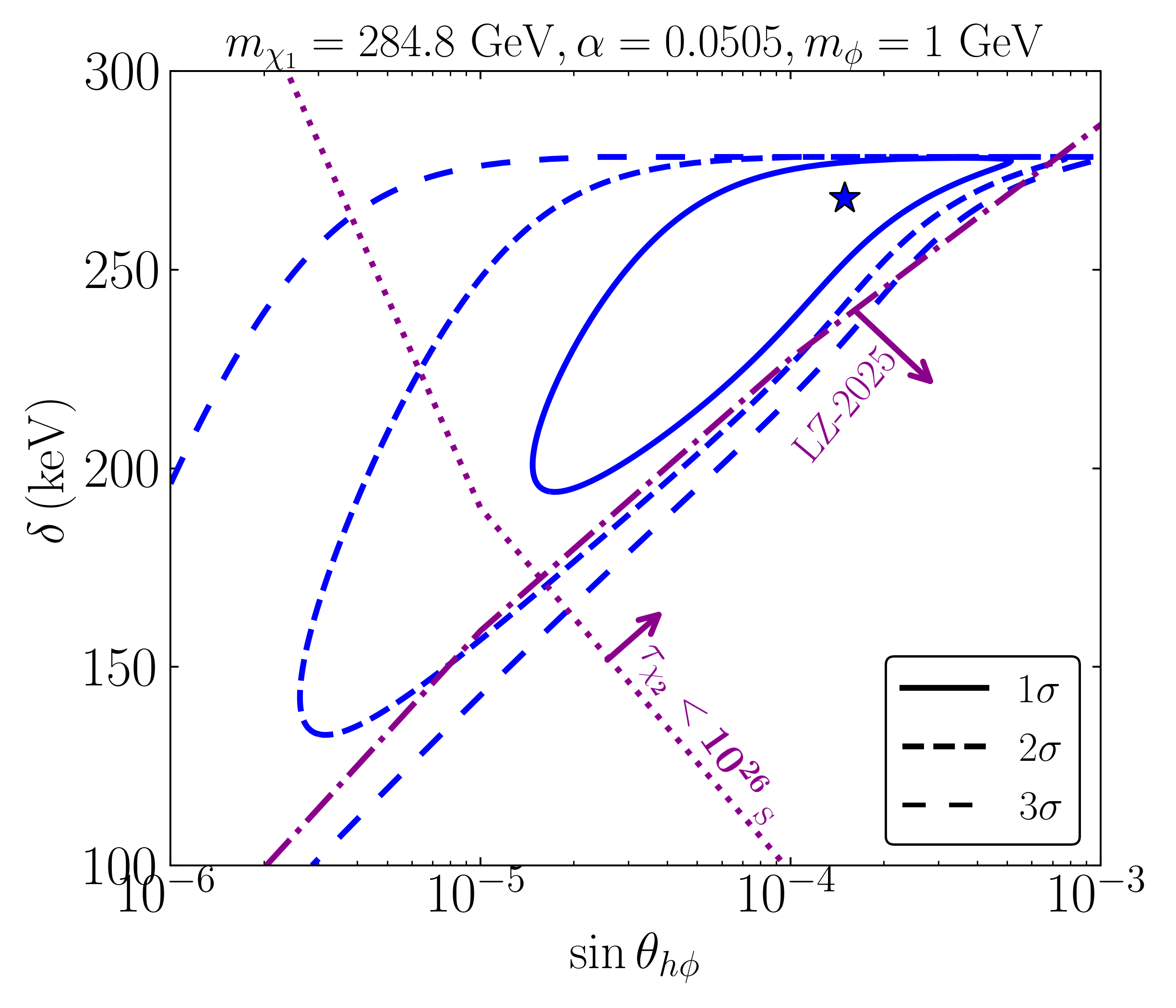}
\includegraphics[width=0.48\linewidth]{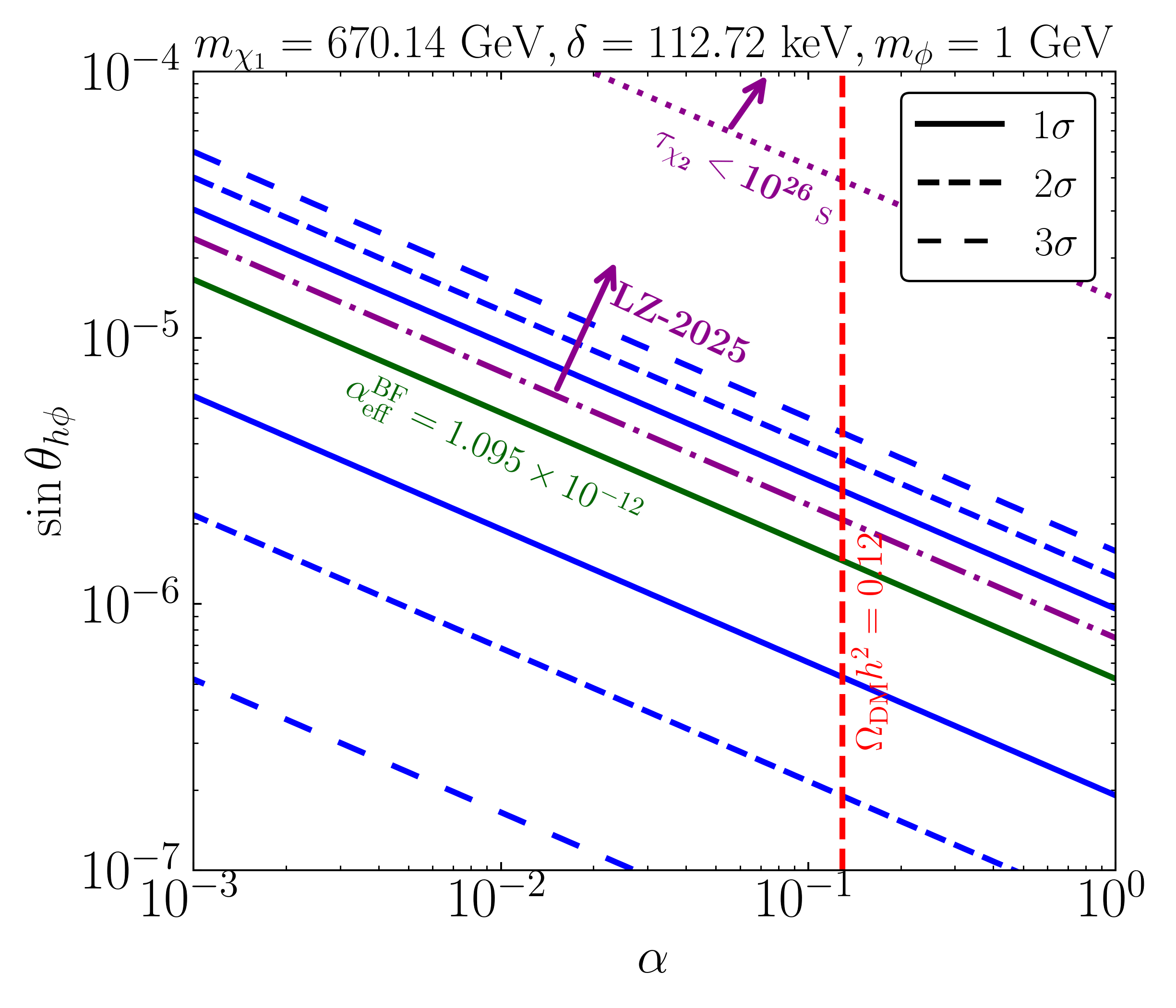}
\caption{[\textit{Left}:] $1\sigma$, $2\sigma$  and $3\sigma$ contours that could explain the observed event at LZ in the $\delta-\sin\theta_{h\phi}$ plane. Here we have fixed three parameters as $\{m_{\chi_1}=284.8~{\rm GeV},m_\phi=1{\rm~GeV},\alpha=0.0505\}$. [\textit{Right}:] $1\sigma$, $2\sigma$  and $3\sigma$ contours in the plane of $\sin\theta_{h\phi}-\alpha$ plane. Here we have fixed three parameters as $\{m_{\chi_1}=670.14~{\rm GeV},\delta=112.72~{\rm keV},m_\phi=1{\rm~GeV}\}$.}
\label{fig:LZ_excess1}
\end{figure}

We further explore the interplay between the scalar
mixing, the dark-sector coupling and the mass-splitting to explain the LZ event. In the \textit{left} panel of Fig. \ref{fig:LZ_excess1}, we fix the parameters at $\{m_{\chi_1}=284.8{\rm~ GeV}, \alpha=0.0505, m_\phi=1{~\rm GeV}\}$ and search for the region of parameter space that can accommodate LZ230616 event in the $\delta-\sin\theta_{h\phi}$ plane. The solid blue, dashed blue and long-dashed blue contours correspond to the $1\sigma$, $2\sigma$, and $3\sigma$ confidence regions of the parameter space. The best fit point is $\{\delta^{\rm BF}=268.01{~\rm keV},\sin\theta_{h\phi}^{\rm BF}=1.49\times10^{-4}\}$. The LZ (2025) exclusion is shown with a dashed-dotted blue line, and the blue dotted line represents the $\chi_2$ lifetime constraint. The lifetime constraint excludes the $1\sigma$ and a significant part of $2\sigma$ regions of the parameter space. Only a part of the  $2\sigma$ and $3\sigma$ region with $\sin\theta_{h\phi}\lesssim10^{-5}$ remains consistent with the LZ event, direct detection, and the lifetime constraint. This BP also gives the correct DM relic abundance.

We now fix the DM mass at $m_{\chi_1}=670.14$ GeV, $\delta=112.72$ keV, and $m_\phi=1$ GeV, and show the regions of the parameter space that are consistent with the LZ event in the $\sin\theta_{h\phi}-\alpha$ plane in the \textit{right} panel of Fig. \ref{fig:LZ_excess1}. Note that, since $\alpha_{\rm eff}=\alpha \sin^22\theta$, the fixed values of these parameters give $\alpha_{\rm eff}^{\rm BF}=1.095\times10^{-12}$. This BF value can be realized with multiple combinations of $\alpha$ and $\sin\theta_{h\phi}$. Thus, in this case, $\alpha^{\rm BF}$ and $\sin\theta_{h\phi}^{\rm BF}$ lie along a line as shown by the green color rather than at a single point, unlike the earlier cases. The value of $\alpha$ consistent with the correct DM relic abundance is shown by the red dashed line. The lifetime constraint is shown by the magenta dotted line, while the LZ (2025) excluded region is shown by the magenta dashed-dotted line. The LZ (2025) constraint excludes part of
the upper region of the $2\sigma$ and $3\sigma$ bands. The best-fit as well as $1\sigma$ region remain consistent with the LZ event, DD, lifetime, and DM relic abundance constraints. In Table \ref{tab:BF}, we list the best-fit (BF) parameters for all the benchmark points discussed above.

\begin{table}[h]
    \centering
    \resizebox{\textwidth}{!}{
    \begin{tabular}{|c|c|c|c|c|}
         \hline
         Figure&Fixed parameters& Best Fit parameters\\
         \hline
         Fig. \ref{fig:LZ_excess0}(left) ($\textcolor{blue}{\star}$) & $m_{\phi}=1$ GeV, $\alpha_{\rm eff}=10^{-8}$ & $m_{\chi_1}=811.03$ GeV, $\delta=321.46$ keV\\\hline
        Fig. \ref{fig:LZ_excess0}(left) ($\textcolor{orange}{\star}$) & $m_{\phi}=1$ GeV, $\alpha_{\rm eff}=10^{-12}$ & $m_{\chi_1}=697.49$ GeV, $\delta=131.44$ keV\\\hline
        Fig. \ref{fig:LZ_excess0}(right) ($\textcolor{blue}{\star}$) & $m_{\phi}=1$ GeV, $m_{\chi_1}=697.49$ GeV & $\alpha_{\rm eff}=5.503\times10^{-6}$, $\delta=340.582$ keV\\\hline
        Fig. \ref{fig:LZ_excess} ($\textcolor{blue}{\star}$) & $m_{\phi}=1$ GeV, $\alpha=0.1$, $\sin\theta_{h\phi}=10^{-4}$ & $m_{\chi_1}=874.98$ GeV, $\delta=308.67$ keV\\\hline
        Fig. \ref{fig:LZ_excess} ($\textcolor{orange}{\star}$) & $m_{\phi}=1$ GeV, $\alpha=0.1$, $\sin\theta_{h\phi}=10^{-6}$ & $m_{\chi_1}=670.14$ GeV, $\delta=112.72$ keV\\\hline
        Fig. \ref{fig:LZ_excess1} (left) ($\textcolor{blue}{\star}$) & $m_{\phi}=1$ GeV, $m_{\chi_1}=284.8$ GeV, $\alpha=0.0505$  & $\sin\theta_{h\phi}=1.49\times10^{-4}$, $\delta=268.01$ keV\\\hline
        Fig. \ref{fig:LZ_excess1} (right) & $m_{\phi}=1$ GeV, $m_{\chi_1}=670.14$ GeV, $\delta=112.72$ keV  & $\alpha\sin^22\theta=1.095\times10^{-12}$\\\hline
    \end{tabular}}
    \caption{Best fit values of the parameters for all the BPs discussed related to LZ230616 event.}
    \label{tab:BF}
\end{table}

\begin{figure}[h]
\centering
\includegraphics[width=0.7\linewidth]{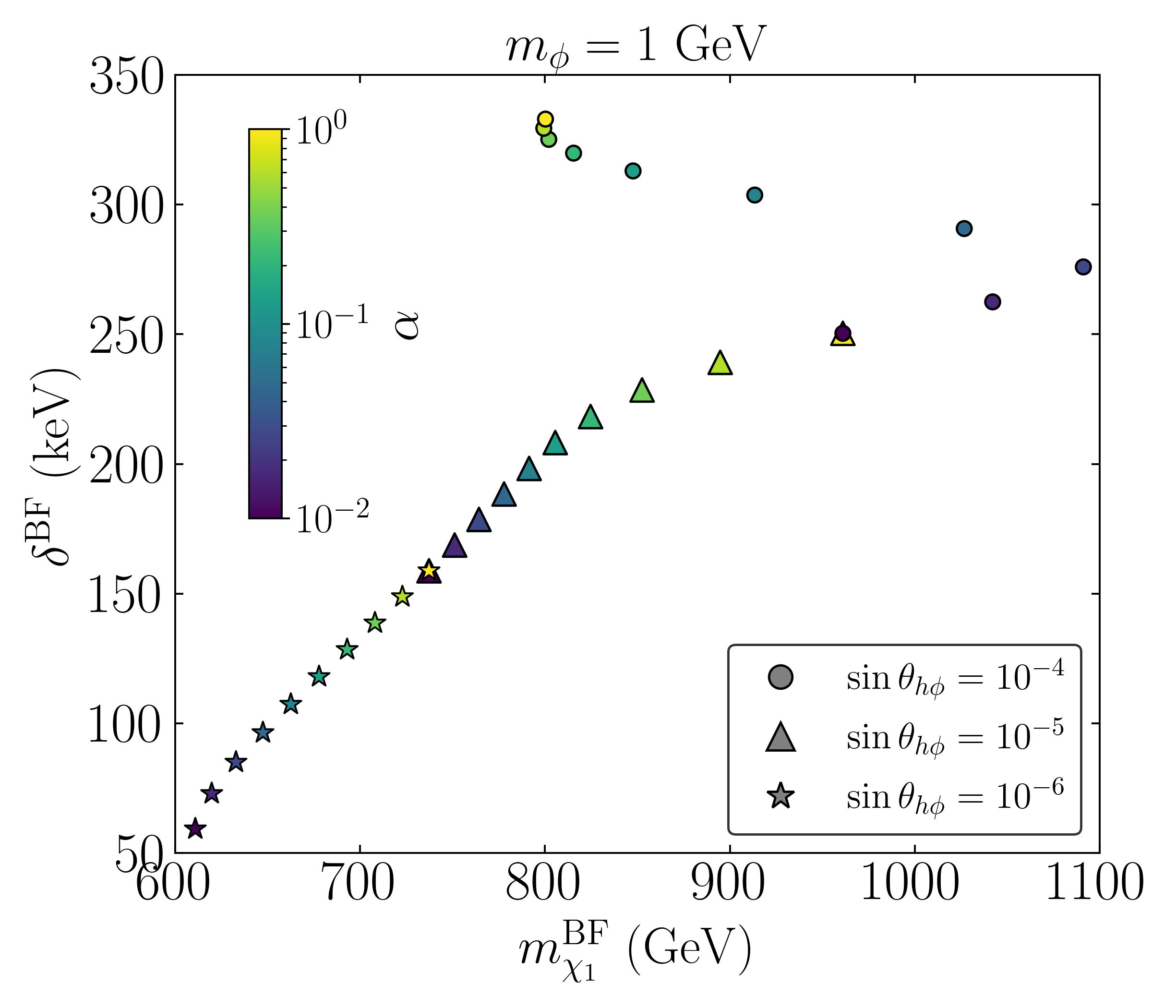}
\caption{Best-fit values of the mass splitting $\delta$ and DM mass
$m_{\chi_1}$ that accommodate the $248$~keV LZ230616 recoil for
$m_\phi=1~{\rm GeV}$. The three marker types correspond to
$\sin\theta_{h\phi}=10^{-4}$, $10^{-5}$, and $10^{-6}$.
The color scale represents the corresponding dark-sector coupling
$\alpha$, varied over $10^{-2}-1$.}
\label{fig:lzexcess_3}
\end{figure}

The correlations discussed above can be summarized by scanning over
the dark-sector coupling and extracting the best-fit values of
$m_{\chi_1}$ and $\delta$. Fig.~\ref{fig:lzexcess_3} shows the
result for fixed
$m_\phi=1~{\rm GeV}$
and three representative values of the scalar mixing, $\sin\theta_{h\phi}
    =
    \{10^{-4},\,
    10^{-5},\,
    10^{-6}\}.$
The corresponding best-fit points are represented by different
markers, while the color scale indicates the value of the dark-sector coupling $\alpha$, which is varied over $\alpha\in[10^{-2},1]$. The best-fit values exhibit a characteristic correlation between $m_{\chi_1}$ and $\delta$. At relatively small DM masses, the
preferred splitting increases with $m_{\chi_1}$.This behavior is
primarily kinematic, {\it i.e.} the reduced mass $\mu_{\chi N}$ increases with
$m_{\chi_1}$, allowing a larger mass splitting while maintaining
kinematic access to a $248$~keV recoil. For larger DM masses, however, $\mu_{\chi N}\rightarrow m_N$ and the kinematic upper limit on $\delta$ saturates. Further increase in $m_{\chi_1}$ suppresses the event rate. Therefore, beyond this point, the best fit value of $m_{\chi_1}$ starts decreasing, while $\delta$ continues to increase and approaches its kinematic upper limit, resulting in the characteristic turnover in the best fit values of $m_{\chi_1}$. Since an increase in $\delta$ also suppresses the event rate, this reduction is compensated by an increase in $\alpha$, as is clearly visible from the color variation in the plot.

Taken together, the results demonstrate that the $248$~keV recoil can
be accommodated by a pseudo-Dirac DM state with a mass
splitting in the $\mathcal O(100~{\rm keV})$ range and a DM
mass extending from the few-hundred-GeV scale to the multi-TeV scale,
depending on the scalar mixing and dark-sector coupling. More
importantly, the LZ-preferred region cannot be considered in
isolation. The same parameters are simultaneously constrained by the
relic abundance, the self-interaction requirement, direct detection,
and the cosmological lifetime of the excited state.

\section{Dark Parity Domain Walls and Gravitational Waves}\label{sec:DW}

The spontaneous breaking of any discrete symmetry leads to the formation of stable domain walls (DWs). DWs are two-dimensional topological defects that can quickly dominate the energy density of the Universe and thereby disrupt the standard cosmological evolution. In our scenario, the dark parity, $\mathcal{Z}_2^{LR}$ symmetric potential involving $\phi$ is given as,
\begin{eqnarray}
    V_{\mathcal{Z}_2^{LR}}(\phi)=-\frac{\mu_\phi^2}{2}\phi^2+\frac{\lambda_\phi}{4}\phi^4.
\end{eqnarray}
The potential has two degenerate minima at $\pm v_\phi$. During the symmetry breaking, the field can choose either of these two minima independently in different spatial regions. As a result, different regions of the Universe can settle into different vacua, with some regions having $\phi=+v_\phi$ and others having $\phi=-v_\phi$. The boundaries separating these regions form DWs. The equation of motion for the DW is
\cite{Vilenkin:2000jqa,Gelmini:1988sf,Larsson:1996sp,Saikawa:2017hiv,Nakayama:2016gxi,Paul:2024iie,Ma:2025bjf,Borah:2026kfo},
\begin{eqnarray}
    \frac{d^2\phi}{dx^2}-\frac{d V_{\mathcal{Z}_2^{LR}}}{d \phi}=0
\end{eqnarray}
with the boundary condition
$\lim_{x\rightarrow\pm\infty}\phi(x)=\pm v_\phi.$

Solving the equation of motion, we get
\begin{eqnarray}
\phi(x)=v_\phi\tanh(\alpha x),
\end{eqnarray}
where, $\alpha\simeq\sqrt{\frac{\lambda_\phi}{2}}v_\phi$.
The DW is extended along the $x=0$ plane, and the two vacua are realized at $x\rightarrow\pm\infty$. The width of the DW is estimated as $\delta^\prime\sim\left(\frac{\sqrt{\lambda_\phi}v_\phi}{\sqrt{2}}\right)^{-1}$.
The surface energy density, also referred to as the tension of the DWs, is calculated to be,
\begin{eqnarray}
\sigma_{\rm DW}=\frac{4}{3}\sqrt{\frac{\lambda_\phi}{2}}v_\phi^3\simeq\frac{2}{3}m_{\phi}v_{\phi}^2,
\end{eqnarray}
where, $m_{\phi}=\sqrt{2\lambda_\phi}v_\phi$.\\ 

To prevent these topological defects from overclosing the Universe, they must be made unstable by introducing explicit $\mathcal{Z}_2^{LR}$-breaking terms. In a cohesive theoretical framework, this explicit breaking is not arbitrary rather it must be communicated across all sectors. The explicit breaking term lifts the degeneracy between the two vacua, creating a pressure difference across the domain walls. This pressure makes them unstable and leads to their annihilation. At the level of the scalar potential, the corresponding lifting of the
vacuum degeneracy can be parametrized by the leading odd operator,
\begin{eqnarray}
V_{\slashed{\mathcal{Z}_2}^{LR}}(\phi)=\frac{\mu_1}{2\sqrt{2}}\phi^3,
\end{eqnarray}
where $\mu_1$ is a mass dimension one coupling\footnote{This bias term can be generated at one loop with connection to Dirac neutrino mass \cite{Borah:2026kfo}.}. Then the total potential becomes
\begin{eqnarray}
    \mathcal{V}(\phi)=V_{\mathcal{Z}_2^{LR}}(\phi)+V_{\slashed{\mathcal{Z}_2}^{LR}}(\phi).
\end{eqnarray}
This lifts the degeneracy of the minima by,
\begin{eqnarray}
V_{\rm bias}\equiv |\mathcal{V}(-v_\phi)-\mathcal{V}(v_\phi)|=\frac{\mu_1v_\phi^3}{\sqrt{2}}.
\end{eqnarray}
The domain walls must annihilate before they come to dominate the energy density of the Universe and also before BBN. The annihilation of the domain walls releases their energy, generating a stochastic background of gravitational waves (GWs) which can be potentially detected at future GW experiments. The peak amplitude of the GW spectrum at present, $t_0$, is given by \cite{Saikawa:2017hiv}
\begin{eqnarray}
\Omega^{\rm peak}_{\rm GW}h^2(t_0)&=&7.18824\times10^{-18}\mathcal{A}^2\tilde{\epsilon}_{GW} \bigg(\frac{\sigma_{\rm DW}}{1{\rm TeV^3}}\bigg)^2\bigg(\frac{g_{*s}(T_{\rm ann})}{10}\bigg)^{-\frac{4}{3}}\bigg(\frac{T_{\rm ann}}{10^{-2}\rm GeV}\bigg)^{-4},\label{eq:peakamp}
\end{eqnarray}
where $\mathcal{A}=0.8 \pm 0.1$ is the area parameter, $\tilde{\epsilon}_{GW}\simeq0.7\pm0.4$\cite{Hiramatsu:2013qaa} is the efficiency parameter, $T_{\rm ann}$ is the temperature at which the DWs annihilate, and $g_{*s}(T_{\rm ann})$ is the relativistic entropy d.o.f at the epoch of DWs annihilation. Assuming the DWs disappear at temperature $T_{\rm ann}$, the peak frequency of the GW spectrum at present is estimated as 
\begin{eqnarray}
	f_{\rm peak}(t_0)&=&1.78648\times10^{-10}{\rm Hz}\bigg(\frac{g_{*s}(T_{\rm ann})}{10}  \bigg)^{-\frac{1}{3}}\bigg(\frac{g_{*}(T_{\rm ann})}{10}\bigg)^{\frac{1}{2}}\bigg(\frac{T_{\rm ann}}{10^{-2}\rm GeV}\bigg).\label{eq:peakfre}
\end{eqnarray}

\begin{figure}[h]
\centering
\includegraphics[width=0.6\linewidth]{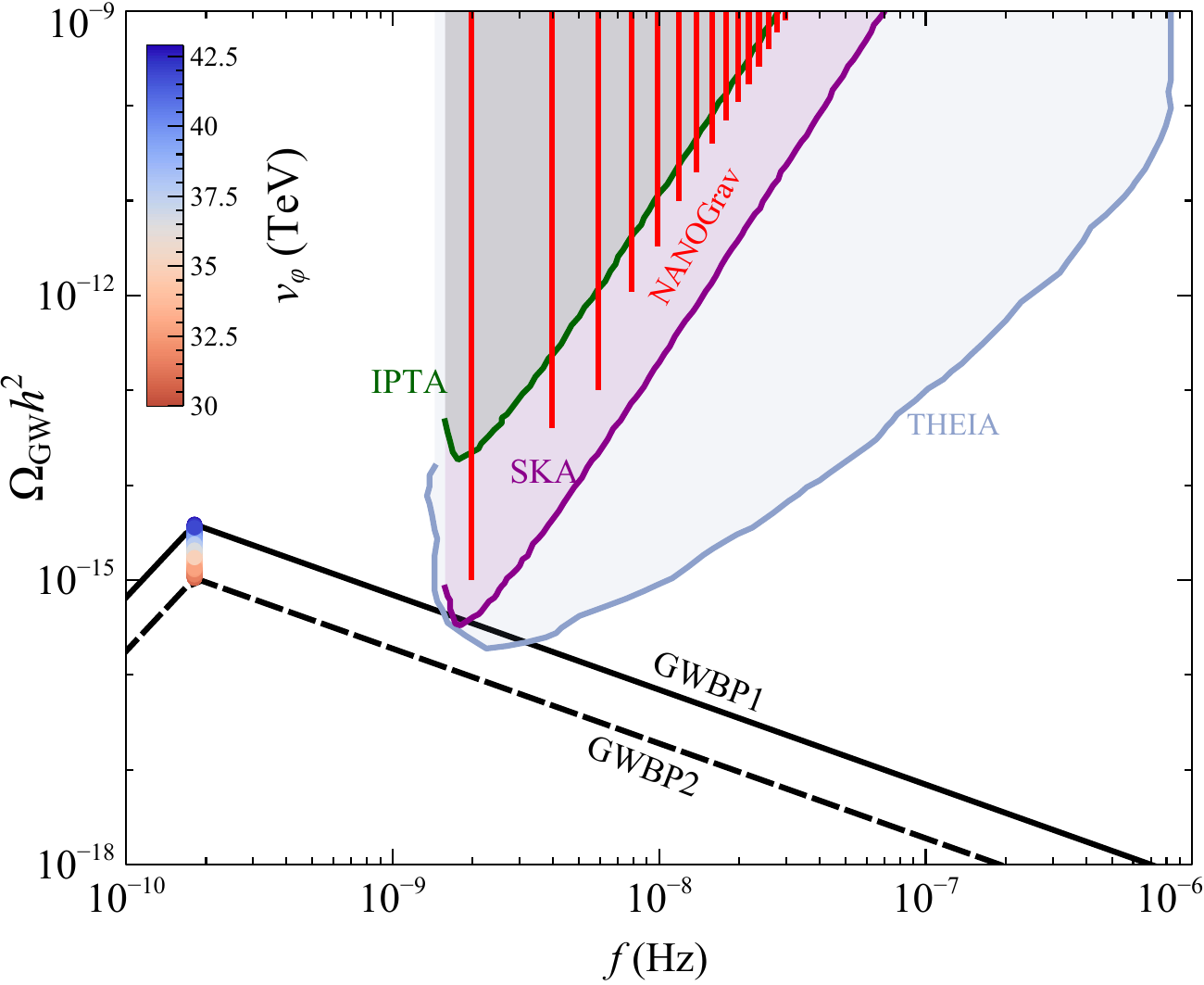}
\caption{Gravitational wave spectrum for two benchmark point consistent with neutrino mass, and DM phenomenology. Experimental sensitivities  are shown with colored-shaded contours. The colored points represents the peak amplitude and peak frequencies of the data points shown in Fig. \ref{fig:trignu}. The color code depicts the values of VEV, $v_\phi$ in TeV.}
\label{fig:gwbp}
\end{figure}

The amplitude of the GW for any frequency at present varies as
\begin{equation}
\Omega_{\rm GW}h^2(t_0,f) =\Omega^{\rm peak}_{\rm GW}h^2(t_0)\left\{
	\begin{array}{l}
	\frac{f_{\rm peak}}{f}~~~~~~,~f>f_{\rm peak}\\
	\bigg(\frac{f}{f_{\rm peak}}\bigg)^3,~f<f_{\rm peak}.\\
	\end{array}
	\right.
\end{equation}

In our phenomenological analysis, within the adopted scan and benchmark assumptions, the neutrino-sector constraints restrict the VEV to lie between 30.8 TeV and 42.7 TeV. Setting $m_\phi=30$ GeV and $T_{\rm ann}=10$ MeV, we evaluate the GW spectrum for this viable parameter space (previously derived in Fig.~\ref{fig:trignu}) and present the results in Fig.~\ref{fig:gwbp}. To bracket the observable range, we highlight two benchmark points representing the largest and smallest allowed VEVs $v_\phi=42.719$ TeV (GWBP1) and $v_\phi=30.832$ TeV (GWBP2) respectively. The GW amplitude as a function of frequency is shown for these two BPs with black solid (GWBP1) and black dashed (GWBP2) lines. For GWBP1, the surface energy density reaches $\sigma_{\rm DW}=36.5~{\rm TeV}^3$, pushing the resulting GW spectrum (black solid line) well into the sensitivity reach of upcoming interferometers like THEIA and pulsar timing arrays like SKA. Conversely, for GWBP2, the reduced tension ($\sigma_{\rm DW}=19~{\rm TeV}^3$) suppresses the spectrum (black dashed line) below near-future detection thresholds. Consequently, while terrestrial direct detection experiments probe the localized DM interactions today, future gravitational wave observatories hold the immense potential to independently verify the high-energy symmetry breaking scale that governs the entire unified framework.

The GW signal should therefore be viewed not as an isolated prediction of
the scalar potential, but as an additional detection channel for the
underlying dark-parity-breaking dynamics. Neutrino oscillation data
constrain $v_\phi$ through the Dirac inverse-seesaw relation, the dark
matter phenomenology constrains the combination of $v_\phi$, $\alpha$,
$\delta$, and $m_\phi$, while the domain-wall dynamics converts the same
symmetry-breaking sector into a stochastic GW signal. A future observation
of such a background would consequently provide information that is
complementary to direct detection, indirect detection, relic-density
constraints, and neutrino measurements. 

\section{Conclusion}\label{sec:conclusion}
We have presented a unified framework that simultaneously addresses the origin of Dirac neutrino masses, the self-interacting nature of DM to solve the small-scale issues of CDM, and the recently reported 248 keV nuclear recoil event at LUX-ZEPLIN while respecting all relevant phenomenological, experimental and cosmological constraints. By employing a non-holomorphic modular $A_4$ flavor symmetry, the model intrinsically connects the neutrino and dark sectors through a common singlet scalar mediator $\phi$. A structural feature of our construction is that the scalar VEV $v_\phi$ acts as the common bridge between the neutrino and dark-sector dynamics establishing non-trivial connections between neutrino predictions, DM
phenomenology, and early-Universe cosmology.

In the neutrino sector, the Dirac inverse-seesaw mechanism generates the
observed neutrino masses without requiring extremely small fundamental
Dirac Yukawa couplings. The modular $A_4$ structure restricts the flavor
parameters and leads to characteristic predictions for the neutrino
oscillation observables. In particular, our numerical analysis favors the
lower octant of the atmospheric mixing angle
($\theta_{23} \in [41^\circ, 45^\circ]$), together with a relatively narrow
range of the Dirac CP phase
($\delta_{\rm CP} \in [257^\circ, 268^\circ]$) and constrains
the scalar VEV $v_\phi$. Moreover, because the observed neutrino masses tie the scalar Yukawa couplings directly to $v_\phi$, the decoupling temperature of the right-handed neutrinos is inherently fixed, yielding a characteristic cosmological radiation contribution of $\Delta N_{\rm eff} \simeq 0.146$ lies in a potentially testable range for future precision
measurements of the $\Delta N_{\rm eff}$.

The dark sector is governed by an approximate $\mathcal{Z}_2^{LR}$ parity
under which the two chiral components of the dark fermion are interchanged.
In the exact symmetry limit, the dark fermion forms a degenerate
pseudo-Dirac pair and the diagonal scalar coupling is suppressed. A small explicit breaking of the $\mathcal{Z}_2^{LR}$ dark parity by $\epsilon \ll 1$, in conjunction with its spontaneous breaking by $v_\phi$, induces a tiny Majorana mass splitting that renders the fermion DM pseudo-Dirac. This splitting elegantly kinematically suppresses elastic nuclear scattering while preserving unsuppressed off-diagonal couplings. We find that this structure can naturally  accommodate
velocity-dependent self-interactions in the phenomenologically relevant
range ($\sigma/m \sim 0.1 - 100~{\rm cm}^2/{\rm g}$) required to resolve small-scale structure anomalies, while the inelastic threshold suppresses nuclear scattering at
low recoil energies. The observed relic abundance can be reproduced after
including all annihilation, co-annihilation, conversion, and the decay processes. The heavier state can furthermore remain
cosmologically long-lived, providing an additional indirect probe of the
pseudo-Dirac dark sector.

A particularly interesting consequence of the inelastic realization is
its high-recoil direct-detection signature.
We performed a dedicated phenomenological likelihood analysis
of the LZ230616 event. We find that the model contains viable regions in
which an inelastic DM signal can accommodate the observed recoil
energy of $248$~keV while remaining compatible with the relic-density
requirement, direct-detection constraints, indirect-detection bounds, and
the lifetime of the excited state. Representative confidence
regions favor DM masses in the several-hundred-GeV to TeV range
and splittings of $\mathcal{O}(100)$~keV. For example, a representative point within the $1\sigma$ confidence region occurs around $m_{\chi_1} \simeq 572~{\rm GeV}$ and $\delta \simeq 113~{\rm keV}$ with scalar mixing ($\sin\theta_{h\phi} \simeq 10^{-6}$) that successfully accounts for the anomalous high-energy recoil while strictly satisfying the correct relic abundance, standard direct detection limits, indirect detection constraints and cosmological lifetime constraints for the excited state ($\tau_{\chi_2} > 10^{26}~{\rm s}$). This demonstrates the existence of a viable inelastic regime in which the
kinematic threshold naturally shifts the scattering probability towards
the observed high recoil energy. If future
high-energy direct-detection data establish an excess with a recoil
spectrum characteristic of endothermic scattering, the parameter region
identified here would provide a concrete realization in which such a
signal is correlated with self-interacting DM and the underlying
dark-parity-breaking mechanism.

Finally, the explicit violation of dark parity lifts the vacuum degeneracy of the scalar potential, destabilizing the cosmological domain wall network formed during the spontaneous breaking of $\mathcal{Z}_2^{LR}$. The subsequent annihilation of these topological defects produces a stochastic gravitational wave background. Because the domain wall tension is determined entirely by $v_\phi$ which is already constrained by the neutrino sector, the predicted gravitational wave spectrum provides a definitive, independent observational signature within the sensitivity reach of upcoming observatories like THEIA and SKA. Ultimately, this framework establishes a highly predictive, multi-messenger phenomenological landscape where precision neutrino measurements, ongoing direct detection searches, and future gravitational wave astronomy can collectively probe the scenario.

\section*{Acknowledgments}
The work of B.K is supported in part by the Polish National Science Center (NCN) under grant 2020/37/B/ST2/02371, the Research Excellence Initiative of the University of Silesia in Katowice and the Swiss National Science Foundation (SNSF) under grant MAPS IZ11Z0\_230193.
S.M. acknowledges support from the IIT Goa Startup Grant [2025/SG/SM/057].
P.K.P. acknowledges the
Ministry of Education, Government of India, for providing financial support for his research via the Prime Minister’s Research Fellowship (PMRF) scheme. Claude Opus 5.0 was used to validate some of the calculations and numerical implementations presented in this work.


\appendix
\section*{Appendix}

\section{Non-Holomorphic Modular Flavor Symmetry Framework}\label{nonholo}

In this section, we review the general formalism of non-holomorphic modular flavor symmetry introduced by Qu and Ding \cite{Ding:2023htn}, where Yukawa couplings are generalized from standard holomorphic modular forms to polyharmonic Maa{\ss} forms of level $N$.

Under the modular group $\Gamma = \text{SL}(2, \mathbb{Z})$, the complex modulus $\tau$ transforms via M{\"o}bius transformations reads:
\begin{eqnarray}
    \tau \to \gamma \tau = \frac{a\tau + b}{c\tau + d}, \quad \gamma = \begin{pmatrix} a & b \\ c & d \end{pmatrix} \in \text{SL}(2, \mathbb{Z}).
\end{eqnarray}
Considering any fields $\psi_i$, which are assigned a modular weight $k_{\psi}$ and transform under the finite modular group $\Gamma'_N$ (or $\Gamma_N$) as:
\begin{eqnarray}
    \psi_i \to (c\tau + d)^{-k_{\psi}} \rho_{\psi}(\gamma)_{ij} \psi_j,
\end{eqnarray}
where $\rho_{\psi}(\gamma)$ denotes the unitary representation matrix of $\Gamma'_N$ (or $\Gamma_N$). Unlike standard supersymmetric (SUSY) modular flavor theories, the Yukawa couplings $Y_{\mathbf{r}}^{(k, m)}(\tau, \bar{\tau})$ are non-holomorphic polyharmonic Maa{\ss} forms of level $N$, modular weight $k$, and depth $m$, transforming as a multiplet in the irreducible representation $\mathbf{r}$ of $\Gamma'_N$ as :
\begin{eqnarray}
    Y_{\mathbf{r}}^{(k, m)}(\gamma \tau, \gamma \bar{\tau}) = (c\tau + d)^k \rho_{\mathbf{r}}(\gamma) Y_{\mathbf{r}}^{(k, m)}(\tau, \bar{\tau}).
\end{eqnarray}
As mention earlier, polyharmonic Maa{\ss} forms satisfy Laplace-type differential equations rather than strict holomorphicity conditions ($\bar{\partial}_{\bar{\tau}} Y = 0$), the weight $k$ can be positive, zero, or negative.

In a non-supersymmetric framework, the effective Yukawa Lagrangian for a generic fermion mass term $\bar{\psi}_{L} Y(\tau, \bar{\tau}) \psi_{R} H$ is required to be invariant under both gauge and modular transformations. Modular invariance imposes the weight-matching condition:
\begin{eqnarray}
    k_{Y} + k_{\psi_L} + k_{\psi_R} + k_{H} = 0,
\end{eqnarray}
where $k_Y$, $k_{\psi_L}$, $k_{\psi_R}$, and $k_H$ represent the modular weights of the respective coupling and fields mentioned in the suffix. Furthermore, the tensor product of representations must contain the trivial singlet $\mathbf{1}$:
\begin{eqnarray}
\left( \rho_Y \otimes \rho_{\psi_L}^* \otimes \rho_{\psi_R} \otimes \rho_H \right) \supset \mathbf{1}.    
\end{eqnarray}
Since, $k_Y$ can take negative integer values, the weight constraints on matter fields are significantly relaxed compared to standard holomorphic models. This feature allows novel operators without expanding the field content. 

The non-holomorphic modular framework consistently accommodates generalized CP (gCP) symmetry. The gCP transformation acts on the modulus $\tau$ and matter fields $\psi$ as \cite{Novichkov:2019sqv}:
\begin{eqnarray}
    \tau \xrightarrow{\text{CP}} -\bar{\tau}, \quad \psi(x) \xrightarrow{\text{CP}} X_{\mathbf{r}} \bar{\psi}(t, -\mathbf{x}),
\end{eqnarray}
where $X_{\mathbf{r}}$ is a unitary gCP transformation matrix satisfying $X_{\mathbf{r}} \rho_{\mathbf{r}}^*(\gamma) X_{\mathbf{r}}^{-1} = \rho_{\mathbf{r}}(\gamma_{\text{CP}} \gamma \gamma_{\text{CP}}^{-1})$. For polyharmonic Maa{\ss} forms, the gCP condition restricts the expansion coefficients $c_{\pm}(n)$ in the Maa{\ss} Fourier modes to be real parameters (up to an overall phase convention). This substantially reduces the number of free parameters, allowing the model to be highly predictive in the flavor-mixing observables and CP-violating phases.
As with holomorphic modular forms \cite{Feruglio:2017spp}, a CP transformation maps the polyharmonic Maa{\ss} form multiplet $Y_{\mathbf{r}}^{(k_Y)}(\tau)$ to its conjugate state
\begin{eqnarray}
    Y_{\mathbf{r}}^{(k_Y)}(-\tau^*) = X_{\mathbf{r}} Y_{\mathbf{r}}^{(k_Y)*}(\tau),
\end{eqnarray}
Hence, when expressed in the $S$- and $T$-symmetric basis with real Clebsch-Gordan coefficients for $\Gamma'_N$ (or $\Gamma_N$), generalized CP (gCP) symmetry restricts all coupling constants to be real. Consequently, the vacuum expectation value (VEV) of the modulus $\tau$ serves as the sole source of both modular and gCP symmetry breaking.

\subsection{Fourier Expansion of Polyharmonic Maa{\ss} Forms}
From the modular transformation condition for $\gamma = T^N$, a level-$N$ polyharmonic Maa{\ss} form $Y(\tau)$ is periodic under $\tau \to \tau + N$:
\begin{eqnarray}
    Y(\tau + N) = Y(\tau).
\end{eqnarray}
Consequently, $Y(\tau)$ admits a generalized Fourier expansion \cite{Liu:2019khw,Lu:2019vgm}:

\begin{eqnarray}
    Y(\tau) = \sum_{n \in \frac{1}{N}\mathbb{Z}} a_n(y) q^n, \quad q \equiv e^{2\pi i \tau}, \quad y = \operatorname{Im}(\tau).
\end{eqnarray}
Imposing the harmonic Laplacian condition forces the Fourier coefficients $a_n(y)$ to satisfy the differential equation:
\begin{eqnarray}
    \frac{d^2 a_n(y)}{dy^2} = \left( 4\pi n - \frac{k}{y} \right) \frac{d a_n(y)}{dy}.
\end{eqnarray}
Solving this differential equation yields the modes $a_n(y)$:
\begin{eqnarray*}
&&\text{For }~ n\ne0:\quad    a_n(y) = c^+(n) + c^-(n) \, \Gamma(1 - k, -4\pi n y)\\
&&\text{ For}~n=0:\quad    a_0(y) = \begin{cases} c^+(0) + c^-(0) y^{1-k}, & k \neq 1 \\ c^+(0) + c^-(0) \ln y, & k = 1 \end{cases}
\end{eqnarray*}
Here, $\Gamma(s, z)$ is the incomplete gamma function. For integer values $s = 1, 2, 3, 4, 5$, its explicit forms simplify to:
\begin{eqnarray}
    \begin{aligned} \Gamma(1, z) &= e^{-z} \\ \Gamma(2, z) &= (z + 1) e^{-z} \\ \Gamma(3, z) &= (z^2 + 2z + 2) e^{-z} \\ \Gamma(4, z) &= (z^3 + 3z^2 + 6z + 6) e^{-z} \\ \Gamma(5, z) &= (z^4 + 4z^3 + 12z^2 + 24z + 24) e^{-z} \end{aligned}
\end{eqnarray}
Combining these solutions yields the general Fourier expansion \cite{Bringmann:2017}:
\begin{eqnarray}
    Y(\tau) = \sum_{n \in \frac{1}{N}\mathbb{Z}} c^+(n) q^n + c^-(0) y^{1-k} + \sum_{n \in \frac{1}{N}\mathbb{Z} \setminus \{0\}} c^-(n) \, \Gamma(1 - k, -4\pi n y) q^n.
\end{eqnarray}
Finally, imposing the moderate growth condition at the cusp ($Y(\tau) = \mathcal{O}(y^\alpha)$ as $y \to \infty$) restricts the mode indices such that $c^+(n) = 0$ for $n < 0$ and $c^-(-n) = 0$ for $n < 0$ (i.e., $c^-(n) = 0$ for $n > 0$). Thus, the final Fourier expansion for a level-$N$ polyharmonic Maa{\ss} form of weight $k$ is given by:
\begin{eqnarray}\label{yupoly1}
    Y(\tau) = \sum_{\substack{n \in \frac{1}{N}\mathbb{Z} \\ n \ge 0}} c^+(n) q^n + c^-(0) y^{1-k} + \sum_{\substack{n \in \frac{1}{N}\mathbb{Z} \\ n < 0}} c^-(n) \, \Gamma(1 - k, -4\pi n y) q^n,
\end{eqnarray}
where $y^{1-k}$ is replaced by $\ln y$ when $k = 1$.

The important takeaway from this polyharmonic Maa{\ss} form given by Eq. \eqref{yupoly1} can be realize as follows:
\begin{itemize}
    \item The non-holomorphic terms are proportional to $c^-(0)$ and $c^-(n)$ vanish for traditional holomorphic modular forms.
    \item Polyharmonic Maa{\ss} forms generalize standard modular forms, spanning a strictly larger space of modular-invariant functions.
    \item Unlike holomorphic modular forms, whose products preserve modularity ($M_k \otimes M_{k'} \to M_{k+k'}$), the product of two polyharmonic Maa{\ss} forms generally does not yield a valid polyharmonic Maa{\ss} form of weight $k+k'$. This occurs because non-zero non-holomorphic terms (particularly for negative weights $k, k' < 0$) break the harmonic Laplacian condition.
\end{itemize}

\subsection{Polyharmonic Maa{\ss} form multiplets of level $N=3$}
\subsubsection{Generators and Representations under $A_4$}\label{appendA4}

The inhomogeneous finite modular group $\Gamma_3$ is isomorphic to the alternating group $A_4$ (the symmetry group of a regular tetrahedron). It is generated by two elements, $S$ and $T$, satisfying the presentation:
\begin{equation}
A_4 = \{ S, T ;\|; S^2 = T^3 = (ST)^3 = 1 \}.
\end{equation}

The group $A_4$ possesses four irreducible representations: three singlets ($\mathbf{1}, \mathbf{1}', \mathbf{1}''$) and one triplet ($\mathbf{3}$). In these representations, the generators $S$ and $T$ are specified by:
\begin{align}
\mathbf{1}:& \quad S = 1, \quad T = 1, \nonumber \\
\mathbf{1}':& \quad S = 1, \quad T = \omega, \nonumber \\
\mathbf{1}'':& \quad S = 1, \quad T = \omega^2, \nonumber \\
\mathbf{3}:& \quad S = \frac{1}{3}\begin{pmatrix} -1 & 2 & 2 \\ 2 & -1 & 2 \\ 2 & 2 & -1 \end{pmatrix}, \quad T = \begin{pmatrix} 1 & 0 & 0 \\ 0 & \omega & 0 \\ 0 & 0 & \omega^2 \end{pmatrix},
\end{align}
where $\omega = e^{2\pi i/3}$.

The multiplication rules for these representations with $\mathbf{3}_S$ and $\mathbf{3}_A$ denoting the symmetric and antisymmetric components, respectively are:
\begin{equation}
\mathbf{1} \otimes \mathbf{1} = \mathbf{1}, \quad \mathbf{1}' \otimes \mathbf{1}' = \mathbf{1}'', \quad \mathbf{1}' \otimes \mathbf{1}'' = \mathbf{1}, \quad \mathbf{1}'' \otimes \mathbf{1}'' = \mathbf{1}', \quad \mathbf{3} \otimes \mathbf{3} = \mathbf{1} \oplus \mathbf{1}' \oplus \mathbf{1}'' \oplus \mathbf{3}_S \oplus \mathbf{3}_A,
\end{equation}

Now, for two arbitrary triplets as $ A = (a_1, a_2, a_3)^T$ and $ B =(b_1, b_2, 
	b_3)^T$ respectively, their direct product can be decomposed into the direct sum mentioned above. The product rule  for this  two triplets  in the $S$ diagonal basis can be written as  \cite{He:2006dk, Borah:2024gql, Ishimori:2010au}
	\begin{eqnarray*}
	(A\times B)_{\bf{1}} &\backsim& a_1b_1+a_2 b_2+a_3b_3,\\
	(A\times B)_{\bf{1'}} &\backsim& a_1 b_1 + \omega^2 a_2 b_2 + \omega a_3 b_3,\\
	(A\times B)_{\bf{1''}} &\backsim& a_1 b_1 + \omega a_2 b_2 + \omega^2 a_3 b_3,\\
	(A\times B)_{\bf{3}_{1}} &\backsim& (a_2b_3+a_3b_2,a_3b_1+a_1b_3, a_1b_2+a_2b_1),\label{eq:3s}\\
	(A\times B)_{\bf{3}_{2}} &\backsim& (a_2b_3-a_3b_2, a_3b_1-a_1b_3, a_1b_2-a_2b_1)\label{eq:3a},
	\end{eqnarray*}

\subsubsection{ Holomorphic Modular Forms of Level 3}

All level-$3$ modular forms of weight $k$ can be built from degree-$k$ homogeneous polynomials in the modular functions $\vartheta(\tau)$ and $\varepsilon(\tau)$:
\begin{equation}
\vartheta(\tau) = 3\sqrt{2} \frac{\eta^3(3\tau)}{\eta(\tau)}, \qquad \varepsilon(\tau) = - \frac{3\eta^3(3\tau) + \eta^3(\tau/3)}{\eta(\tau)}.
\end{equation}

\begin{itemize}
    \item[$k_I=2$:] The three independent weight-2 modular forms form an $A_4$ triplet $Y^{(2)}_{\mathbf{3}} = (Y_1, Y_2, Y_3)^T$:
\begin{equation}
Y^{(2)}_{\mathbf{3}}(\tau) = \begin{pmatrix} \varepsilon^2(\tau) \ \sqrt{2}\vartheta(\tau)\varepsilon(\tau) \ -\vartheta^2(\tau) \end{pmatrix} \equiv \begin{pmatrix} Y_1(\tau) \ Y_2(\tau) \ Y_3(\tau) \end{pmatrix},
\end{equation}
with the $q$-expansions:
\begin{align}
Y_1(\tau) &= 1 + 12q + 36q^2 + 12q^3 + 84q^4 + 72q^5 + \dots, \nonumber \\
Y_2(\tau) &= -6q^{1/3}\left(1 + 7q + 8q^2 + 18q^3 + 14q^4 + \dots\right), \nonumber \\
Y_3(\tau) &= -18q^{2/3}\left(1 + 2q + 5q^2 + 4q^3 + 8q^4 + \dots\right).
\end{align}
\item[$k_I=4$:] Five independent forms are formed by contracting $Y^{(2)}_{\mathbf{3}}$ with itself:
\begin{align}
Y^{(4)}_{\mathbf{1}} &= Y_1^2 + 2Y_2 Y_3, \qquad Y^{(4)}_{\mathbf{1}'} = Y_3^2 + 2Y_1 Y_2, \nonumber \\
Y^{(4)}_{\mathbf{3}} &= \begin{pmatrix} Y_1^2 - Y_2 Y_3 \ Y_3^2 - Y_1 Y_2 \ Y_2^2 - Y_1 Y_3 \end{pmatrix}.
\end{align}
\item[$k_I=6$:] Seven independent forms decompose into a singlet $\mathbf{1}$ and two triplets $\mathbf{3}_I, \mathbf{3}_{II}$ under $A_4$:
\begin{align}
Y^{(6)}_{\mathbf{1}} &= Y_1^3 + Y_2^3 + Y_3^3 - 3Y_1 Y_2 Y_3, \nonumber \\
Y^{(6)}_{\mathbf{3} I} &= \begin{pmatrix} Y_1^3 + 2Y_1 Y_2 Y_3 , Y_1^2 Y_2 + 2Y_2^2 Y_3 , Y_1^2 Y_3 + 2Y_3^2 Y_2 \end{pmatrix},\\
Y^{(6)}_{\mathbf{3} II} &= \begin{pmatrix} Y_3^3 + 2Y_1 Y_2 Y_3 , Y_3^2 Y_1 + 2Y_1^2 Y_2 , Y_3^2 Y_2 + 2Y_2^2 Y_1 \end{pmatrix}.\nonumber
\end{align}
\end{itemize} 
\subsubsection{ Polyharmonic Maa{\ss} Forms of Level 3}\label{appenModular}

For integer weights $k > 2$, level-$3$ polyharmonic Maa{\ss} forms coincide with holomorphic modular forms. However, for non-positive or low weights ($k \le 2$), additional non-holomorphic structures appear:

\begin{itemize}
    \item[$k_I=2$:] The forms include the modified Eisenstein series $\widehat{E}_2(\tau)$ assigned to the trivial singlet $Y^{(2)}_{\mathbf{1}} \equiv \widehat{E}_2(\tau)$ alongside the triplet $Y^{(2)}_{\mathbf{3}}$.
    \item[$k_I=0$:] Besides the trivial singlet $Y^{(0)}_{\mathbf{1}} = 1$, the existence of $Y^{(2)}_{\mathbf{3}}$ implies a non-trivial triplet Maa{\ss} form $Y^{(0)}_{\mathbf{3}} = (Y_{\mathbf{3},1}^{(0)}, Y_{\mathbf{3},2}^{(0)}, Y_{\mathbf{3},3}^{(0)})^T$, with $q$-expansions:
\begin{align}
Y_{\mathbf{3},1}^{(0)} &= y - \frac{9\log 3}{4\pi} - \frac{1}{\pi}\sum_{n \neq 0} c_n^{(1)} e^{-4\pi n y} q^n, \nonumber \\
Y_{\mathbf{3},2}^{(0)} &= \frac{27 q^{1/3}e^{\pi y /3}}{\pi} \left( \frac{e^{-3\pi y}}{4 q} + \frac{e^{-7\pi y}}{5 q^2} + \dots \right) + \frac{9q^{1/3}}{2\pi}\left( 1 + \frac{7 q}{4} + \dots \right), \nonumber \\
Y_{\mathbf{3},3}^{(0)} &= \frac{9 q^{2/3}e^{2\pi y /3}}{2\pi}\left( \frac{e^{-2\pi y}}{q} + \frac{7 e^{-6\pi y}}{4 q^2} + \dots \right) + \frac{27 q^{2/3}}{\pi}\left( \frac{1}{4} + \frac{q}{5} + \dots \right).
\end{align}
\item[$k_I=-2$:]  Derived by lifting the weight-4 forms. Since $Y^{(4)}_{\mathbf{1}'}$ is a cusp form in $\Gamma(3)$, it cannot be lifted. The liftable forms yield the singlet $Y_{\mathbf{1}}^{(-2)}$ and triplet $Y_{\mathbf{3}}^{(-2)}$:
\begin{align}
Y_{\mathbf{1}}^{(-2)}(\tau) &= \frac{y^3}{3} - \frac{\pi\zeta(3)}{12\zeta(4)} - \frac{15\Gamma(3,4\pi y)}{4\pi^3 q} - \frac{135\Gamma(3,8\pi y)}{32\pi^3 q^2} - \dots - \frac{15 q}{2\pi^3} - \frac{135 q^2}{16\pi^3} - \dots, \nonumber \\
Y_{\mathbf{3},1}^{(-2)}(\tau) &= \frac{y^3}{3} + \frac{\pi\zeta(3)}{40\zeta(4)} + \frac{21\Gamma(3,4\pi y)}{16\pi^3 q} + \frac{189\Gamma(3,8\pi y)}{128\pi^3 q^2} + \dots + \frac{21 q}{8\pi^3} + \frac{189 q^2}{64\pi^3} + \dots, \nonumber \\
Y_{\mathbf{3},2}^{(-2)}(\tau) &= -\frac{729 q^{1/3}}{16\pi^3}\left( \frac{\Gamma(3,8\pi y/3)}{16 q} + \dots \right) - \frac{81q^{1/3}}{16\pi^3}\left( 1 + \frac{73 q}{64} + \dots \right), \nonumber \\
Y_{\mathbf{3},3}^{(-2)}(\tau) &= -\frac{81 q^{2/3}}{32\pi^3} \left( \frac{\Gamma(3,4\pi y/3)}{q} + \dots \right) - \frac{729 q^{2/3}}{8\pi^3}\left( \frac{1}{16} + \frac{7 q}{125} + \dots \right).
\end{align}
\item[$k_I=-4$:] Derived by lifting the weight-6 forms. Because $Y^{(6)}_{\mathbf{3}II}$ is a cusp form, only the singlet $Y^{(6)}_{\mathbf{1}}$ and the linear combination $Y^{(6)}_{\mathbf{3}} = Y^{(6)}_{\mathbf{3}I} - \frac{5}{13}Y^{(6)}_{\mathbf{3}II}$ can be lifted to weight $-4$ Maa{\ss} forms:
\begin{align}
Y_{\mathbf{1}}^{(-4)} &= \frac{y^5}{5} + \frac{\pi\zeta(5)}{80\zeta(6)} + \frac{63\Gamma(5,4\pi y)}{128\pi^5 q} + \dots + \frac{189 q}{16\pi^5} + \frac{6237 q^2}{512 \pi^5} + \dots, \nonumber \\
Y_{\mathbf{3},1}^{(-4)} &= \frac{y^5}{5} - \frac{3\pi\zeta(5)}{728\zeta(6)} - \frac{549}{3328 \pi^5} \left( \frac{\Gamma(5, 4 \pi y)}{q} + \dots \right) - \frac{1647}{416 \pi^5} \left( q + \frac{33 q^2}{32} + \dots \right), \nonumber \\
Y_{\mathbf{3},2}^{(-4)} &= \frac{72171 q^{1/3}}{212992 \pi^5} \left( \frac{\Gamma(5, 8 \pi y/3)}{q} + \dots \right) + \frac{6561 q^{1/3}}{832 \pi^5} \left( 1 + \frac{1057 q}{1024} + \dots \right), \nonumber \\
Y_{\mathbf{3},3}^{(-4)} &= \frac{2187 q^{2/3}}{6656 \pi^5} \left( \frac{\Gamma(5, 4 \pi y/3)}{q} + \dots \right) + \frac{216513 q^{2/3}}{26624 \pi^5} \left( 1 + \frac{33344 q}{34375} + \dots \right).
\end{align}
\end{itemize}

\section{DM Self-interaction Cross-sections}\label{appen:isidm}

\begin{figure}[h]
\centering
\includegraphics[width=0.4\linewidth]{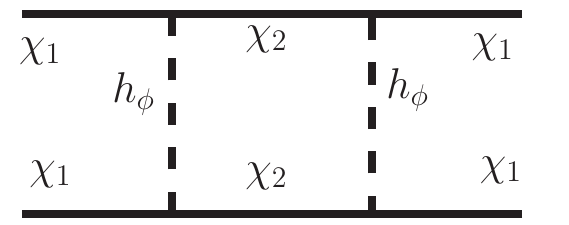}
\includegraphics[width=0.4\linewidth]{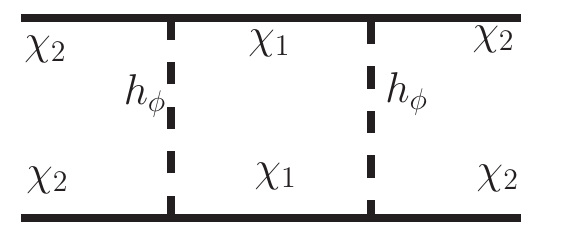}
\includegraphics[width=0.4\linewidth]{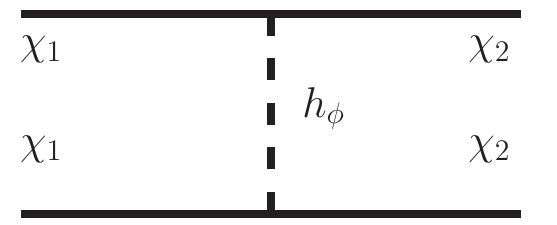}
\includegraphics[width=0.4\linewidth]{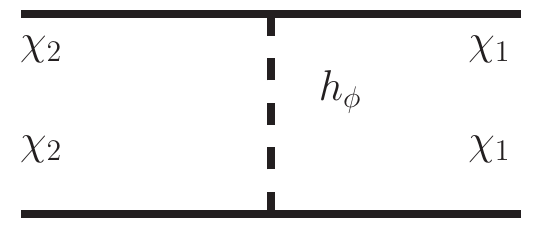}
\caption{Feynman diagrams for ``ground to ground" scattering (top left), ``excited to excited" scattering (top right), ``ground to excited" scattering (bottom left), and ``excited to ground" scattering (bottom right).}
\label{fig:sidmFD}
\end{figure}

The non-perturbative potential is approximated by matching to a Hulthén potential at a matching radius $r_M$, determined by \cite{Schutz:2014nka}
\begin{eqnarray}
    \frac{e^{-\epsilon_\phi r_M}}{r_M}={\rm max}\left[ \frac{\epsilon_\delta^2}{2},\epsilon_\phi^2 \right].
\end{eqnarray}
For $r>r_M$, the potential follows a Yukawa profile $V(r)\sim V_0 e^{-\mu r}$, where the effective screening mass $\mu$ and amplitude $V_0$ are:
\begin{eqnarray}
    \mu=\epsilon_\phi\left( \frac{1}{2}+\frac{1}{2}\sqrt{1+\frac{4}{\epsilon_\phi r_M}} \right), \quad V_0=\frac{e^{\epsilon_\phi r_M\left(-\frac{1}{2}+\frac{1}{2}\sqrt{1+\frac{4}{\epsilon_\phi r_M}}\right)}}{r_M}.
\end{eqnarray}
The scattering amplitudes are governed by the Gamma functions of the kinematic variables:
\begin{eqnarray}
    \Gamma_v&=&\Gamma\left( 1+\frac{i\epsilon_v}{\mu} \right) \Gamma\left( \frac{i\epsilon_v-i\epsilon_\Delta}{2\mu}+\frac{1}{2} \right) \Gamma\left( \frac{i\epsilon_v+i\epsilon_\Delta}{2\mu}+\frac{1}{2} \right),\\
    \Gamma_\Delta&=&\Gamma\left( 1+\frac{i\epsilon_\Delta}{\mu} \right) \Gamma\left( \frac{i\epsilon_\Delta-i\epsilon_v}{2\mu}+\frac{1}{2} \right) \Gamma\left( \frac{i\epsilon_v+i\epsilon_\Delta}{2\mu}+\frac{1}{2} \right),
\end{eqnarray}
where $\epsilon_\Delta=\sqrt{\epsilon_v^2-\epsilon_\delta^2}$. The relevant momentum-transfer cross-sections ($\tilde{\sigma}$) for the ground state (gr) $\chi_1$ and excited state (ex) $\chi_2$ transitions are analytically expressed as:
\begin{eqnarray}
    \tilde{\sigma}^1_{\rm gr\rightarrow gr}&=&\frac{\pi}{\epsilon_v^2}\left| 1+\left( \frac{V_0}{4\mu^2} \right)^{-2i\epsilon_v/\mu} \frac{\Gamma_v}{\Gamma_v^*} \frac{\cosh(a)\sinh(b+i\varphi)}{\cosh(-b)\sinh(a-i\varphi)} \right|^2,\\
    \tilde{\sigma}^2_{\rm ex\rightarrow ex}&=&\frac{\pi}{\epsilon_\Delta^2}\left| 1+\left( \frac{V_0}{4\mu^2} \right)^{-2i\epsilon_\Delta/\mu} \frac{\Gamma_\Delta}{\Gamma_\Delta^*} \frac{\cosh(a)\sinh(-b+i\varphi)}{\cosh(-b)\sinh(a-i\varphi)} \right|^2,\\
    \tilde{\sigma}^3_{\rm gr\rightarrow ex}&=&\frac{2\pi \cos^2\varphi\sinh(\pi\epsilon_v/\mu)\sinh(\pi\epsilon_\Delta/\mu)}{\epsilon_v^2\cosh^2(-b)\left(\cosh(a)-\cosh(2\varphi)\right)},\\
    \tilde{\sigma}^4_{\rm ex\rightarrow gr}&=&\frac{2\pi \cos^2\varphi\sinh(\pi\epsilon_v/\mu)\sinh(\pi\epsilon_\Delta/\mu)}{\epsilon_\Delta^2\cosh^2(-b)\left(\cosh(a)-\cosh(2\varphi)\right)},
\end{eqnarray}
where $a=\pi\left(\frac{\epsilon_\Delta+\epsilon_v}{2\mu}\right)$ and $b=\pi\left(\frac{\epsilon_v-\epsilon_\Delta}{2\mu}\right)$. 

These general expressions simplify in specific kinematic limits. In the low-velocity limit ($\epsilon_v\rightarrow0$):
\begin{eqnarray}
    \tilde{\sigma}^5_{\rm gr\rightarrow gr}=\frac{\pi}{\mu^2}\left( \pi \cot\left( \varphi-\frac{\pi \epsilon_\delta}{2\mu} \right) +2\ln\left( \frac{V_0}{\mu^2} \right) +2\gamma_E-4\ln2-2\psi_0\left( \frac{\epsilon_\delta}{2\mu}+\frac{1}{2} \right) \right),
\end{eqnarray}
where $\gamma_E$ is the Euler-Mascheroni constant and $\psi_0$ is the digamma function. At the kinematic threshold for endothermic up-scattering ($\epsilon_v=\epsilon_\delta$):
\begin{eqnarray}
    \tilde{\sigma}^6_{\rm gr\rightarrow ex}&\approx& \frac{4\pi^2\epsilon_\Delta\cos^2\varphi\tanh\left( \frac{\pi\epsilon_\delta}{2\mu} \right)}{\epsilon_\delta^2\mu\left( \cosh\left( \frac{\mu\epsilon_\delta}{\mu} -\cos(2\varphi) \right) \right)},\\
    \tilde{\sigma}^7_{\rm ex\rightarrow gr}&\approx& \frac{4\pi^2\cos^2\varphi\tanh\left( \frac{\pi\epsilon_\delta}{2\mu} \right)}{\epsilon_\Delta\mu\left( \cosh\left( \frac{\mu\epsilon_\delta}{\mu} -\cos(2\varphi) \right) \right)},\\
    \tilde{\sigma}^8_{\rm ex\rightarrow ex}&\approx& \frac{1}{\mu^2}\left( \zeta^2 +\frac{2\pi\cos\varphi}{\cosh\left( \frac{\pi\epsilon_\delta}{\mu} \right)-\cos(2\varphi)} \left( 2\zeta\sin\varphi+\cos\varphi~{\rm sech}^2\left(\frac{\pi\epsilon_\delta}{2\mu}\right) \right)  \right)^2,
\end{eqnarray}
where $\zeta=2\gamma_E-2\psi_0\left( \frac{i\epsilon_\delta}{2\mu}+\frac{1}{2} \right)+i\pi\tanh\left(\frac{\pi\epsilon_\delta}{2\mu}\right) +\ln\left(\frac{V_0^2}{16\mu^4}\right)$. 

Finally, in the Born regime ($\epsilon_\phi\geq 1$):
\begin{eqnarray}
    \tilde{\sigma}^9_{\rm gr\rightarrow gr} &=& \frac{\pi}{\epsilon^6}\frac{1}{(1+\epsilon_\delta/\epsilon_\phi)^4},\\
    \tilde{\sigma}^{10}_{\rm ex\rightarrow ex}&=&\frac{\pi}{\epsilon_\phi^6}\frac{1}{(1+\epsilon_\delta^2/\epsilon_\phi^2)^2},\\
    \tilde{\sigma}^{11}_{\rm gr\rightarrow ex}&=&\frac{\epsilon_\Delta}{\epsilon_v}\frac{4\pi}{\epsilon_\phi^4(1-\epsilon_\delta^2/\epsilon_\phi^2)^2+4\epsilon_v^2/\epsilon_\phi^2},\\
    \tilde{\sigma}^{12}_{\rm ex\rightarrow gr}&=&\frac{\epsilon_v}{\epsilon_\Delta}\frac{4\pi}{\epsilon_\phi^4(1-\epsilon_\delta^2/\epsilon_\phi^2)^2+4\epsilon_v^2/\epsilon_\phi^2}.
\end{eqnarray}
The physical cross-sections are recovered via $\sigma=\tilde{\sigma} / (\alpha^2 m_\chi^2)$. The unsuppressed off-diagonal coupling permits efficient exothermic down-scattering ($\chi_2 \chi_2 \to \chi_1 \chi_1$) and co-scattering ($\chi_1 \chi_2 \to \chi_1 \chi_2$), providing the necessary heat transfer to flatten galactic density profiles. The Feynman diagrams are shown in Fig. \ref{fig:sidmFD}.

\providecommand{\href}[2]{#2}\begingroup\raggedright\endgroup


\begin{thebibliography}{100}
	
	\bibitem{ParticleDataGroup:2018ovx}
	{\scshape Particle Data Group} collaboration, \emph{{Review of Particle
			Physics}}, \href{https://doi.org/10.1103/PhysRevD.98.030001}{\emph{Phys. Rev.
			D} {\bfseries 98} (2018) 030001}.
	
	\bibitem{Altarelli:2010gt}
	G.~Altarelli and F.~Feruglio, \emph{{Discrete Flavor Symmetries and Models of
			Neutrino Mixing}},
	\href{https://doi.org/10.1103/RevModPhys.82.2701}{\emph{Rev. Mod. Phys.}
		{\bfseries 82} (2010) 2701}
	[\href{https://arxiv.org/abs/1002.0211}{{\ttfamily 1002.0211}}].
	
	\bibitem{King:2013eh}
	S.F.~King and C.~Luhn, \emph{{Neutrino Mass and Mixing with Discrete
			Symmetry}}, \href{https://doi.org/10.1088/0034-4885/76/5/056201}{\emph{Rept.
			Prog. Phys.} {\bfseries 76} (2013) 056201}
	[\href{https://arxiv.org/abs/1301.1340}{{\ttfamily 1301.1340}}].
	
	\bibitem{Petcov:2017ggy}
	S.T.~Petcov, \emph{{Discrete Flavour Symmetries, Neutrino Mixing and Leptonic
			CP Violation}},
	\href{https://doi.org/10.1140/epjc/s10052-018-6158-5}{\emph{Eur. Phys. J. C}
		{\bfseries 78} (2018) 709}
	[\href{https://arxiv.org/abs/1711.10806}{{\ttfamily 1711.10806}}].
	
	\bibitem{Chauhan:2023faf}
	G.~Chauhan, P.S.B.~Dev, I.~Dubovyk, B.~Dziewit, W.~Flieger, K.~Grzanka et~al.,
	\emph{{Phenomenology of lepton masses and mixing with discrete flavor
			symmetries}}, \href{https://doi.org/10.1016/j.ppnp.2024.104126}{\emph{Prog.
			Part. Nucl. Phys.} {\bfseries 138} (2024) 104126}
	[\href{https://arxiv.org/abs/2310.20681}{{\ttfamily 2310.20681}}].
	
	\bibitem{Borah:2024gql}
	D.~Borah, P.~Das, B.~Karmakar and S.~Mahapatra, \emph{{Discrete dark matter
			with light Dirac neutrinos}},
	\href{https://doi.org/10.1103/PhysRevD.111.035032}{\emph{Phys. Rev. D}
		{\bfseries 111} (2025) 035032}
	[\href{https://arxiv.org/abs/2406.17861}{{\ttfamily 2406.17861}}].
	
	\bibitem{Borah:2017dmk}
	D.~Borah and B.~Karmakar, \emph{{$A_4$ flavour model for Dirac neutrinos: Type
			I and inverse seesaw}},
	\href{https://doi.org/10.1016/j.physletb.2018.03.047}{\emph{Phys. Lett. B}
		{\bfseries 780} (2018) 461}
	[\href{https://arxiv.org/abs/1712.06407}{{\ttfamily 1712.06407}}].
	
	\bibitem{Das:2018qyt}
	P.~Das, A.~Mukherjee and M.K.~Das, \emph{{Active and sterile neutrino
			phenomenology with $A_4$ based minimal extended seesaw}},
	\href{https://doi.org/10.1016/j.nuclphysb.2019.02.024}{\emph{Nucl. Phys. B}
		{\bfseries 941} (2019) 755}
	[\href{https://arxiv.org/abs/1805.09231}{{\ttfamily 1805.09231}}].
	
	\bibitem{Feruglio:2017spp}
	F.~Feruglio, \emph{{Are neutrino masses modular forms?}},  in \emph{{From My
			Vast Repertoire ...}: {Guido Altarelli's Legacy}}, A.~Levy, S.~Forte and
	G.~Ridolfi, eds., pp.~227--266 (2019),
	\href{https://doi.org/10.1142/9789813238053_0012}{DOI}
	[\href{https://arxiv.org/abs/1706.08749}{{\ttfamily 1706.08749}}].
	
	\bibitem{Feruglio:2021dte}
	F.~Feruglio, V.~Gherardi, A.~Romanino and A.~Titov, \emph{{Modular invariant
			dynamics and fermion mass hierarchies around $\tau = i$}},
	\href{https://doi.org/10.1007/JHEP05(2021)242}{\emph{JHEP} {\bfseries 05}
		(2021) 242} [\href{https://arxiv.org/abs/2101.08718}{{\ttfamily
			2101.08718}}].
	
	\bibitem{Ding:2023htn}
	G.-J.~Ding and S.F.~King, \emph{{Neutrino mass and mixing with modular
			symmetry}}, \href{https://doi.org/10.1088/1361-6633/ad52a3}{\emph{Rept. Prog.
			Phys.} {\bfseries 87} (2024) 084201}
	[\href{https://arxiv.org/abs/2311.09282}{{\ttfamily 2311.09282}}].
	
	\bibitem{Nomura:2019xsb}
	T.~Nomura, H.~Okada and S.~Patra, \emph{{An inverse seesaw model with $A_4$
			-modular symmetry}},
	\href{https://doi.org/10.1016/j.nuclphysb.2021.115395}{\emph{Nucl. Phys. B}
		{\bfseries 967} (2021) 115395}
	[\href{https://arxiv.org/abs/1912.00379}{{\ttfamily 1912.00379}}].
	
	\bibitem{Asaka:2019vev}
	T.~Asaka, Y.~Heo, T.H.~Tatsuishi and T.~Yoshida, \emph{{Modular $A_4$
			invariance and leptogenesis}},
	\href{https://doi.org/10.1007/JHEP01(2020)144}{\emph{JHEP} {\bfseries 01}
		(2020) 144} [\href{https://arxiv.org/abs/1909.06520}{{\ttfamily
			1909.06520}}].
	
	\bibitem{Nomura:2019lnr}
	T.~Nomura, H.~Okada and O.~Popov, \emph{{A modular $A_4$ symmetric scotogenic
			model}}, \href{https://doi.org/10.1016/j.physletb.2020.135294}{\emph{Phys.
			Lett. B} {\bfseries 803} (2020) 135294}
	[\href{https://arxiv.org/abs/1908.07457}{{\ttfamily 1908.07457}}].
	
	\bibitem{Chen:2025ruj}
	M.-C.~Chen, S.~Perez and M.~Ratz, \emph{{Scale-Independent Relations Between
			Neutrino Mass Parameters}},
	\href{https://doi.org/10.3390/universe12020046}{\emph{Universe} {\bfseries
			12} (2026) 46} [\href{https://arxiv.org/abs/2511.03974}{{\ttfamily
			2511.03974}}].
	
	\bibitem{Granelli:2025lds}
	A.~Granelli, D.~Meloni, M.~Parriciatu, J.T.~Penedo and S.T.~Petcov,
	\emph{{Modular-symmetry-protected seesaw}},
	\href{https://doi.org/10.1007/JHEP12(2025)035}{\emph{JHEP} {\bfseries 12}
		(2025) 035} [\href{https://arxiv.org/abs/2505.21405}{{\ttfamily
			2505.21405}}].
	
	\bibitem{Pathak:2024sei}
	G.~Pathak, P.~Das and M.K.~Das, \emph{{Neutrino mass genesis in scoto-inverse
			seesaw with modular $A_4$}},
	\href{https://doi.org/10.1140/epjc/s10052-025-14263-1}{\emph{Eur. Phys. J. C}
		{\bfseries 85} (2025) 569}
	[\href{https://arxiv.org/abs/2411.13895}{{\ttfamily 2411.13895}}].
	
	\bibitem{Qu:2024rns}
	B.-Y.~Qu and G.-J.~Ding, \emph{{Non-holomorphic modular flavor symmetry}},
	\href{https://doi.org/10.1007/JHEP08(2024)136}{\emph{JHEP} {\bfseries 08}
		(2024) 136} [\href{https://arxiv.org/abs/2406.02527}{{\ttfamily
			2406.02527}}].
	
	\bibitem{Gao:2025jlw}
	X.-Y.~Gao and C.-C.~Li, \emph{{Minimal lepton models with non-holomorphic
			modular A $_{4}$ symmetry*}},
	\href{https://doi.org/10.1088/1674-1137/ae3f0a}{\emph{Chin. Phys. C}
		{\bfseries 50} (2026) 053109}
	[\href{https://arxiv.org/abs/2512.07158}{{\ttfamily 2512.07158}}].
	
	\bibitem{Tapender:2026ets}
	Tapender and S.~Verma, \emph{{Tri-resonant leptogenesis in a non-holomorphic
			modular A$_4$ scotogenic model}},
	\href{https://doi.org/10.1140/epjc/s10052-026-16096-y}{\emph{Eur. Phys. J. C}
		{\bfseries 86} (2026) 860}
	[\href{https://arxiv.org/abs/2602.17243}{{\ttfamily 2602.17243}}].
	
	\bibitem{Nasri:2026nbf}
	S.~Nasri, L.~Singh, Tapender and S.~Verma, \emph{{Dark-portal leptogenesis in a
			nonholomorphic modular scoto-seesaw model}},
	\href{https://doi.org/10.1103/vbtk-561v}{\emph{Phys. Rev. D} {\bfseries 113}
		(2026) 115008} [\href{https://arxiv.org/abs/2601.06435}{{\ttfamily
			2601.06435}}].
	
	\bibitem{Okada:2025nap}
	H.~Okada and S.~Jangid, \emph{{A Radiative Seesaw Model in a Noninvertible
			Selection Rule with the Assistance of a Nonholomorphic Modular A4 Symmetry}},
	\href{https://doi.org/10.1093/ptep/ptag070}{\emph{PTEP} {\bfseries 2026}
		(2026) 053B05} [\href{https://arxiv.org/abs/2510.17292}{{\ttfamily
			2510.17292}}].
	
	\bibitem{Zhang:2025dsa}
	X.~Zhang and Y.~Reyimuaji, \emph{{Inverse seesaw model in nonholomorphic
			modular A4 flavor symmetry}},
	\href{https://doi.org/10.1103/17p3-bw5r}{\emph{Phys. Rev. D} {\bfseries 112}
		(2025) 075050} [\href{https://arxiv.org/abs/2507.06945}{{\ttfamily
			2507.06945}}].
	
	\bibitem{Qu:2025ddz}
	B.-Y.~Qu, J.-N.~Lu and G.-J.~Ding, \emph{{Non-holomorphic modular flavor
			symmetry and odd weight polyharmonic Maa{\ss} form}},
	\href{https://doi.org/10.1007/JHEP11(2025)140}{\emph{JHEP} {\bfseries 11}
		(2025) 140} [\href{https://arxiv.org/abs/2506.19822}{{\ttfamily
			2506.19822}}].
	
	\bibitem{Nomura:2024vus}
	T.~Nomura and H.~Okada, \emph{{A More Novel Approach of Radiative Linear Seesaw
			in a Modular A4 Symmetry}},
	\href{https://doi.org/10.1093/ptep/ptaf044}{\emph{PTEP} {\bfseries 2025}
		(2025) 043B04} [\href{https://arxiv.org/abs/2410.21843}{{\ttfamily
			2410.21843}}].
	
	\bibitem{Kobayashi:2025hnc}
	T.~Kobayashi, H.~Okada and Y.~Orikasa, \emph{{Zee{\textendash}Babu model in a
			Nonholomorphic Modular A4 Symmetry and Modular Stabilization}},
	\href{https://doi.org/10.1093/ptep/ptag039}{\emph{PTEP} {\bfseries 2026}
		(2026) 033B07} [\href{https://arxiv.org/abs/2502.12662}{{\ttfamily
			2502.12662}}].
	
	\bibitem{Cheshta:2026fls}
	Cheshta, Priya, S.~Dutt and B.C.~Chauhan, \emph{{A Type-I seesaw framework with
			non-holomorphic modular symmetry}},
	\href{https://doi.org/10.1140/epjc/s10052-026-15675-3}{\emph{Eur. Phys. J. C}
		{\bfseries 86} (2026) 400}
	[\href{https://arxiv.org/abs/2604.18070}{{\ttfamily 2604.18070}}].
	
	\bibitem{Behera:2026tmo}
	M.K.~Behera, J.~Das and N.~Mondal, \emph{{A Non-Holomorphic Modular $A_4$
			Framework for Resonant Leptogenesis with Gravitational Wave Signatures}},
	\href{https://arxiv.org/abs/2607.18803}{{\ttfamily 2607.18803}}.
	
	\bibitem{CentellesChulia:2023osj}
	S.~Centelles~Chuli{\'a}, R.~Kumar, O.~Popov and R.~Srivastava, \emph{{Neutrino
			mass sum rules from modular A4 symmetry}},
	\href{https://doi.org/10.1103/PhysRevD.109.035016}{\emph{Phys. Rev. D}
		{\bfseries 109} (2024) 035016}
	[\href{https://arxiv.org/abs/2308.08981}{{\ttfamily 2308.08981}}].
	
	\bibitem{Wang:2020dbp}
	X.~Wang, \emph{{Dirac neutrino mass models with a modular $S_4$ symmetry}},
	\href{https://doi.org/10.1016/j.nuclphysb.2020.115247}{\emph{Nucl. Phys. B}
		{\bfseries 962} (2021) 115247}
	[\href{https://arxiv.org/abs/2007.05913}{{\ttfamily 2007.05913}}].
	
	\bibitem{Singh:2024imk}
	L.~Singh, M.~Kashav and S.~Verma, \emph{{Minimal type-I Dirac seesaw and
			leptogenesis under A4 modular invariance}},
	\href{https://doi.org/10.1016/j.nuclphysb.2024.116666}{\emph{Nucl. Phys. B}
		{\bfseries 1007} (2024) 116666}
	[\href{https://arxiv.org/abs/2405.07165}{{\ttfamily 2405.07165}}].
	
	\bibitem{Spergel:1999mh}
	D.N.~Spergel and P.J.~Steinhardt, \emph{{Observational evidence for
			selfinteracting cold dark matter}},
	\href{https://doi.org/10.1103/PhysRevLett.84.3760}{\emph{Phys. Rev. Lett.}
		{\bfseries 84} (2000) 3760}
	[\href{https://arxiv.org/abs/astro-ph/9909386}{{\ttfamily
			astro-ph/9909386}}].
	
	\bibitem{Tulin:2017ara}
	S.~Tulin and H.-B.~Yu, \emph{{Dark Matter Self-interactions and Small Scale
			Structure}}, \href{https://doi.org/10.1016/j.physrep.2017.11.004}{\emph{Phys.
			Rept.} {\bfseries 730} (2018) 1}
	[\href{https://arxiv.org/abs/1705.02358}{{\ttfamily 1705.02358}}].
	
	\bibitem{Bullock:2017xww}
	J.S.~Bullock and M.~Boylan-Kolchin, \emph{{Small-Scale Challenges to the
			$\Lambda$CDM Paradigm}},
	\href{https://doi.org/10.1146/annurev-astro-091916-055313}{\emph{Ann. Rev.
			Astron. Astrophys.} {\bfseries 55} (2017) 343}
	[\href{https://arxiv.org/abs/1707.04256}{{\ttfamily 1707.04256}}].
	
	\bibitem{Buckley:2009in}
	M.R.~Buckley and P.J.~Fox, \emph{{Dark Matter Self-Interactions and Light Force
			Carriers}}, \href{https://doi.org/10.1103/PhysRevD.81.083522}{\emph{Phys.
			Rev. D} {\bfseries 81} (2010) 083522}
	[\href{https://arxiv.org/abs/0911.3898}{{\ttfamily 0911.3898}}].
	
	\bibitem{Feng:2009hw}
	J.L.~Feng, M.~Kaplinghat and H.-B.~Yu, \emph{{Halo Shape and Relic Density
			Exclusions of Sommerfeld-Enhanced Dark Matter Explanations of Cosmic Ray
			Excesses}}, \href{https://doi.org/10.1103/PhysRevLett.104.151301}{\emph{Phys.
			Rev. Lett.} {\bfseries 104} (2010) 151301}
	[\href{https://arxiv.org/abs/0911.0422}{{\ttfamily 0911.0422}}].
	
	\bibitem{Feng:2009mn}
	J.L.~Feng, M.~Kaplinghat, H.~Tu and H.-B.~Yu, \emph{{Hidden Charged Dark
			Matter}}, \href{https://doi.org/10.1088/1475-7516/2009/07/004}{\emph{JCAP}
		{\bfseries 07} (2009) 004} [\href{https://arxiv.org/abs/0905.3039}{{\ttfamily
			0905.3039}}].
	
	\bibitem{Loeb:2010gj}
	A.~Loeb and N.~Weiner, \emph{{Cores in Dwarf Galaxies from Dark Matter with a
			Yukawa Potential}},
	\href{https://doi.org/10.1103/PhysRevLett.106.171302}{\emph{Phys. Rev. Lett.}
		{\bfseries 106} (2011) 171302}
	[\href{https://arxiv.org/abs/1011.6374}{{\ttfamily 1011.6374}}].
	
	\bibitem{Bringmann:2016din}
	T.~Bringmann, F.~Kahlhoefer, K.~Schmidt-Hoberg and P.~Walia, \emph{{Strong
			constraints on self-interacting dark matter with light mediators}},
	\href{https://doi.org/10.1103/PhysRevLett.118.141802}{\emph{Phys. Rev. Lett.}
		{\bfseries 118} (2017) 141802}
	[\href{https://arxiv.org/abs/1612.00845}{{\ttfamily 1612.00845}}].
	
	\bibitem{Kaplinghat:2015aga}
	M.~Kaplinghat, S.~Tulin and H.-B.~Yu, \emph{{Dark Matter Halos as Particle
			Colliders: Unified Solution to Small-Scale Structure Puzzles from Dwarfs to
			Clusters}}, \href{https://doi.org/10.1103/PhysRevLett.116.041302}{\emph{Phys.
			Rev. Lett.} {\bfseries 116} (2016) 041302}
	[\href{https://arxiv.org/abs/1508.03339}{{\ttfamily 1508.03339}}].
	
	\bibitem{vandenAarssen:2012vpm}
	L.G.~van~den Aarssen, T.~Bringmann and C.~Pfrommer, \emph{{Is dark matter with
			long-range interactions a solution to all small-scale problems of
			{\textbackslash}Lambda CDM cosmology?}},
	\href{https://doi.org/10.1103/PhysRevLett.109.231301}{\emph{Phys. Rev. Lett.}
		{\bfseries 109} (2012) 231301}
	[\href{https://arxiv.org/abs/1205.5809}{{\ttfamily 1205.5809}}].
	
	\bibitem{Tulin:2013teo}
	S.~Tulin, H.-B.~Yu and K.M.~Zurek, \emph{{Beyond Collisionless Dark Matter:
			Particle Physics Dynamics for Dark Matter Halo Structure}},
	\href{https://doi.org/10.1103/PhysRevD.87.115007}{\emph{Phys. Rev. D}
		{\bfseries 87} (2013) 115007}
	[\href{https://arxiv.org/abs/1302.3898}{{\ttfamily 1302.3898}}].
	
	\bibitem{Borah:2022ask}
	D.~Borah, S.~Mahapatra and N.~Sahu, \emph{{New realization of light thermal
			self-interacting dark matter and detection prospects}},
	\href{https://doi.org/10.1103/PhysRevD.108.L091702}{\emph{Phys. Rev. D}
		{\bfseries 108} (2023) L091702}
	[\href{https://arxiv.org/abs/2211.15703}{{\ttfamily 2211.15703}}].
	
	\bibitem{Borah:2024wos}
	D.~Borah, S.~Mahapatra, P.K.~Paul, N.~Sahu and P.~Shukla, \emph{{Asymmetric
			self-interacting dark matter with a canonical seesaw model}},
	\href{https://doi.org/10.1103/PhysRevD.110.035033}{\emph{Phys. Rev. D}
		{\bfseries 110} (2024) 035033}
	[\href{https://arxiv.org/abs/2404.14912}{{\ttfamily 2404.14912}}].
	
	\bibitem{LZ:2024zvo}
	{\scshape LZ} collaboration, \emph{{Dark Matter Search Results from
			4.2{\,}{\,}Tonne-Years of Exposure of the LUX-ZEPLIN (LZ) Experiment}},
	\href{https://doi.org/10.1103/4dyc-z8zf}{\emph{Phys. Rev. Lett.} {\bfseries
			135} (2025) 011802} [\href{https://arxiv.org/abs/2410.17036}{{\ttfamily
			2410.17036}}].
	
	\bibitem{XENON:2024wpa}
	{\scshape XENON} collaboration, \emph{{The XENONnT dark matter experiment}},
	\href{https://doi.org/10.1140/epjc/s10052-024-12982-5}{\emph{Eur. Phys. J. C}
		{\bfseries 84} (2024) 784}
	[\href{https://arxiv.org/abs/2402.10446}{{\ttfamily 2402.10446}}].
	
	\bibitem{PandaX:2025rrz}
	{\scshape PandaX} collaboration, \emph{{Search for Light Dark Matter with 259
			Days of Data in PandaX-4T}},
	\href{https://doi.org/10.1103/rtnh-jn8s}{\emph{Phys. Rev. Lett.} {\bfseries
			135} (2025) 211001} [\href{https://arxiv.org/abs/2507.11930}{{\ttfamily
			2507.11930}}].
	
	\bibitem{Choi:2026kxe}
	K.-Y.~Choi, E.~Lkhagvadorj and S.~Mahapatra, \emph{{Scalar Portal Verifiable
			Light Dark Matter and Correlated Gravitational Wave Signatures}},
	\href{https://arxiv.org/abs/2609.02501}{{\ttfamily 2609.02501}}.
	
	\bibitem{Tucker-Smith:2001myb}
	D.~Tucker-Smith and N.~Weiner, \emph{{Inelastic dark matter}},
	\href{https://doi.org/10.1103/PhysRevD.64.043502}{\emph{Phys. Rev. D}
		{\bfseries 64} (2001) 043502}
	[\href{https://arxiv.org/abs/hep-ph/0101138}{{\ttfamily hep-ph/0101138}}].
	
	\bibitem{Cui:2009xq}
	Y.~Cui, D.E.~Morrissey, D.~Poland and L.~Randall, \emph{{Candidates for
			Inelastic Dark Matter}},
	\href{https://doi.org/10.1088/1126-6708/2009/05/076}{\emph{JHEP} {\bfseries
			05} (2009) 076} [\href{https://arxiv.org/abs/0901.0557}{{\ttfamily
			0901.0557}}].
	
	\bibitem{He:2020sat}
	H.-J.~He, Y.-C.~Wang and J.~Zheng, \emph{{GeV-scale inelastic dark matter with
			dark photon mediator via direct detection and cosmological and laboratory
			constraints}}, \href{https://doi.org/10.1103/PhysRevD.104.115033}{\emph{Phys.
			Rev. D} {\bfseries 104} (2021) 115033}
	[\href{https://arxiv.org/abs/2012.05891}{{\ttfamily 2012.05891}}].
	
	\bibitem{Borah:2020smw}
	D.~Borah, S.~Mahapatra and N.~Sahu, \emph{{Connecting Low scale Seesaw for
			Neutrino Mass to Inelastic sub-GeV Dark Matter with Abelian Gauge Symmetry}},
	\href{https://doi.org/10.1016/j.nuclphysb.2021.115407}{\emph{Nucl. Phys. B}
		{\bfseries 968} (2021) 115407}
	[\href{https://arxiv.org/abs/2009.06294}{{\ttfamily 2009.06294}}].
	
	\bibitem{Cho:2024lhp}
	W.~Cho, K.-Y.~Choi and S.~Mahapatra, \emph{{Reconciling cosmological tensions
			with inelastic dark matter and dark radiation in a U(1)D framework}},
	\href{https://doi.org/10.1088/1475-7516/2024/09/065}{\emph{JCAP} {\bfseries
			09} (2024) 065} [\href{https://arxiv.org/abs/2408.03004}{{\ttfamily
			2408.03004}}].
	
	\bibitem{Borah:2020jzi}
	D.~Borah, S.~Mahapatra, D.~Nanda and N.~Sahu, \emph{{Inelastic fermion dark
			matter origin of XENON1T excess with muon $(g- 2)$ and light neutrino
			mass}}, \href{https://doi.org/10.1016/j.physletb.2020.135933}{\emph{Phys.
			Lett. B} {\bfseries 811} (2020) 135933}
	[\href{https://arxiv.org/abs/2007.10754}{{\ttfamily 2007.10754}}].
	
	\bibitem{LZ:2026axp}
	{\scshape LZ} collaboration, \emph{{Search for dark matter particle
			interactions in an extended nuclear recoil energy window with the LUX-ZEPLIN
			(LZ) experiment}},  \href{https://arxiv.org/abs/2609.02823}{{\ttfamily
			2609.02823}}.
	
	\bibitem{Fan:2026kxx}
	J.~Fan and M.~Reece, \emph{{Higgsino Above the Sea of Fog}},
	\href{https://arxiv.org/abs/2609.01504}{{\ttfamily 2609.01504}}.
	
	\bibitem{Lou:2026idn}
	Y.~Lou and C.-T.~Lu, \emph{{Fermionic Dark Matter Absorption and the
			High-Energy Event in LUX-ZEPLIN}},
	\href{https://arxiv.org/abs/2609.01592}{{\ttfamily 2609.01592}}.
	
	\bibitem{Freese:2026sga}
	K.~Freese and D.P.~Theodosopoulos, \emph{{Higgsino Dark Matter Interpretation
			of the LUX-ZEPLIN 248 keV Nuclear-Recoil Event}},
	\href{https://arxiv.org/abs/2609.01583}{{\ttfamily 2609.01583}}.
	
	\bibitem{Su:2026rwz}
	L.~Su, J.M.~Yang and W.-N.~Yang, \emph{{Inelastic Dark Matter Signature at High
			Recoil Energy in LUX-ZEPLIN and CRESST}},
	\href{https://arxiv.org/abs/2609.01475}{{\ttfamily 2609.01475}}.
	
	\bibitem{Chattopadhyay:2026ryw}
	U.~Chattopadhyay, D.~Das, R.~Puri and J.~Roy, \emph{{Sub-TeV Singlino Dark
			Matter in light from Sagittarius A$^\ast$ and LUX-ZEPLIN Nuclear-Recoil
			Event}},  \href{https://arxiv.org/abs/2609.02994}{{\ttfamily 2609.02994}}.
	
	\bibitem{Yamashita:2026ump}
	K.~Yamashita, \emph{{Inelastic Dark Photon Dark Matter for the LUX-ZEPLIN
			High-Recoil Event and the Galactic Halo Gamma-Ray Excess}},
	\href{https://arxiv.org/abs/2609.02868}{{\ttfamily 2609.02868}}.
	
	\bibitem{Visinelli:2026kgt}
	L.~Visinelli, \emph{{A Peccei--Quinn Origin for Inelastic Electroweak Dark
			Matter after LUX-ZEPLIN}},
	\href{https://arxiv.org/abs/2609.02807}{{\ttfamily 2609.02807}}.
	
	\bibitem{DiMauro:2026ldr}
	M.~Di~Mauro, \emph{{Dark Matter at the Kinematic Edge: Interpreting the 248 keV
			LZ Nuclear-Recoil Candidate}},
	\href{https://arxiv.org/abs/2609.02608}{{\ttfamily 2609.02608}}.
	
	\bibitem{Rodd:2026tyn}
	N.L.~Rodd, B.R.~Safdi, T.R.~Slatyer and W.L.~Xu, \emph{{Confronting the
			Higgsino Interpretation of the LZ Event with the High-Energy Sideband}},
	\href{https://arxiv.org/abs/2609.04175}{{\ttfamily 2609.04175}}.
	
	\bibitem{Jeesun:2026vzo}
	S.~Jeesun and A.~Majumdar, \emph{{Atmospheric neutrino up-scattering
			explanation of LZ 2026 excess}},
	\href{https://arxiv.org/abs/2609.04185}{{\ttfamily 2609.04185}}.
	
	\bibitem{McCabe:2026crm}
	C.~McCabe, \emph{{Seasonal dark matter from the LUX-ZEPLIN high-energy event}},
	\href{https://arxiv.org/abs/2609.04181}{{\ttfamily 2609.04181}}.
	
	\bibitem{Unwin:2026rdp}
	J.~Unwin, \emph{{Axion Portal Dark Matter and the LUX-ZEPLIN High-Recoil
			Event}},  \href{https://arxiv.org/abs/2609.04186}{{\ttfamily 2609.04186}}.
	
	\bibitem{Smirnov:2026aqk}
	J.~Smirnov, S.~Griffith and J.F.~Beacom, \emph{{Inelastic Signatures of
			Electroweak Dark Matter}},
	\href{https://arxiv.org/abs/2609.04144}{{\ttfamily 2609.04144}}.
	
	\bibitem{Du:2026guj}
	X.~Du and F.~Wang, \emph{{TeV Higgsino Interpretation of the LZ High-Recoil
			Event with Intermediate-Scale Electroweak Gauginos}},
	\href{https://arxiv.org/abs/2609.04163}{{\ttfamily 2609.04163}}.
	
	\bibitem{Vilenkin:2000jqa}
	A.~Vilenkin and E.P.S.~Shellard, \emph{{Cosmic Strings and Other Topological
			Defects}}, Cambridge University Press (7, 2000).
	
	\bibitem{Gelmini:1988sf}
	G.B.~Gelmini, M.~Gleiser and E.W.~Kolb, \emph{{Cosmology of Biased Discrete
			Symmetry Breaking}},
	\href{https://doi.org/10.1103/PhysRevD.39.1558}{\emph{Phys. Rev. D}
		{\bfseries 39} (1989) 1558}.
	
	\bibitem{Larsson:1996sp}
	S.E.~Larsson, S.~Sarkar and P.L.~White, \emph{{Evading the cosmological domain
			wall problem}}, \href{https://doi.org/10.1103/PhysRevD.55.5129}{\emph{Phys.
			Rev. D} {\bfseries 55} (1997) 5129}
	[\href{https://arxiv.org/abs/hep-ph/9608319}{{\ttfamily hep-ph/9608319}}].
	
	\bibitem{Saikawa:2017hiv}
	K.~Saikawa, \emph{{A review of gravitational waves from cosmic domain walls}},
	\href{https://doi.org/10.3390/universe3020040}{\emph{Universe} {\bfseries 3}
		(2017) 40} [\href{https://arxiv.org/abs/1703.02576}{{\ttfamily 1703.02576}}].
	
	\bibitem{Nakayama:2016gxi}
	K.~Nakayama, F.~Takahashi and N.~Yokozaki, \emph{{Gravitational waves from
			domain walls and their implications}},
	\href{https://doi.org/10.1016/j.physletb.2017.05.010}{\emph{Phys. Lett. B}
		{\bfseries 770} (2017) 500}
	[\href{https://arxiv.org/abs/1612.08327}{{\ttfamily 1612.08327}}].
	
	\bibitem{Paul:2024iie}
	P.K.~Paul, N.~Sahu and P.~Shukla, \emph{{Thermal leptogenesis, dark matter, and
			gravitational waves from an extended canonical seesaw scenario}},
	\href{https://doi.org/10.1103/w8gl-wbjd}{\emph{Phys. Rev. D} {\bfseries 112}
		(2025) 015032} [\href{https://arxiv.org/abs/2409.08828}{{\ttfamily
			2409.08828}}].
	
	\bibitem{Ma:2025bjf}
	E.~Ma, P.K.~Paul and N.~Sahu, \emph{{Lepton parity dark matter and naturally
			unstable domain walls}}, \href{https://doi.org/10.1103/tj6t-dyqn}{\emph{Phys.
			Rev. D} {\bfseries 112} (2025) 095020}
	[\href{https://arxiv.org/abs/2508.02642}{{\ttfamily 2508.02642}}].
	
	\bibitem{Borah:2026kfo}
	D.~Borah, P.K.~Paul and N.~Sahu, \emph{{Can Dirac neutrinos destabilize
			$\mathcal{Z}_2$ domain wall network?}},
	\href{https://arxiv.org/abs/2602.07380}{{\ttfamily 2602.07380}}.
	
	\bibitem{CentellesChulia:2020dfh}
	S.~Centelles~Chuli{\'a}, R.~Srivastava and A.~Vicente, \emph{{The inverse
			seesaw family: Dirac and Majorana}},
	\href{https://doi.org/10.1007/JHEP03(2021)248}{\emph{JHEP} {\bfseries 03}
		(2021) 248} [\href{https://arxiv.org/abs/2011.06609}{{\ttfamily
			2011.06609}}].
	
	\bibitem{Esteban:2024eli}
	I.~Esteban, M.C.~Gonzalez-Garcia, M.~Maltoni, I.~Martinez-Soler, J.P.~Pinheiro
	and T.~Schwetz, \emph{{NuFit-6.0: updated global analysis of three-flavor
			neutrino oscillations}},
	\href{https://doi.org/10.1007/JHEP12(2024)216}{\emph{JHEP} {\bfseries 12}
		(2024) 216} [\href{https://arxiv.org/abs/2410.05380}{{\ttfamily
			2410.05380}}].
	
	\bibitem{NuFIT6.1}
	``{NuFIT} 6.1 (2025).'' \url{https://www.nu-fit.org}, 2025.
	
	\bibitem{Luo:2020sho}
	X.~Luo, W.~Rodejohann and X.-J.~Xu, \emph{{Dirac neutrinos and $N_{{\rm
					eff}}$}}, \href{https://doi.org/10.1088/1475-7516/2020/06/058}{\emph{JCAP}
		{\bfseries 06} (2020) 058}
	[\href{https://arxiv.org/abs/2005.01629}{{\ttfamily 2005.01629}}].
	
	\bibitem{Borah:2025fkd}
	D.~Borah, S.~Mahapatra, D.~Nanda, S.K.~Sahoo and N.~Sahu, \emph{{Effective
			theory of light Dirac neutrino portal dark matter with observable
			{\ensuremath{\Delta}}Neff}},
	\href{https://doi.org/10.1103/m7my-1cjy}{\emph{Phys. Rev. D} {\bfseries 112}
		(2025) 055010} [\href{https://arxiv.org/abs/2502.10318}{{\ttfamily
			2502.10318}}].
	
	\bibitem{Mahapatra:2023oyh}
	S.~Mahapatra, S.K.~Sahoo, N.~Sahu and V.S.~Thounaojam, \emph{{Self-interacting
			dark matter and Dirac neutrinos via lepton quarticity}},
	\href{https://doi.org/10.1103/PhysRevD.109.055036}{\emph{Phys. Rev. D}
		{\bfseries 109} (2024) 055036}
	[\href{https://arxiv.org/abs/2312.12322}{{\ttfamily 2312.12322}}].
	
	\bibitem{Abazajian:2019eic}
	K.~Abazajian et~al., \emph{{CMB-S4 Science Case, Reference Design, and Project
			Plan}},  \href{https://arxiv.org/abs/1907.04473}{{\ttfamily 1907.04473}}.
	
	\bibitem{CMB-HD:2022bsz}
	{\scshape CMB-HD} collaboration, \emph{{Snowmass2021 CMB-HD White Paper}},
	\href{https://arxiv.org/abs/2203.05728}{{\ttfamily 2203.05728}}.
	
	\bibitem{Schutz:2014nka}
	K.~Schutz and T.R.~Slatyer, \emph{{Self-Scattering for Dark Matter with an
			Excited State}},
	\href{https://doi.org/10.1088/1475-7516/2015/01/021}{\emph{JCAP} {\bfseries
			01} (2015) 021} [\href{https://arxiv.org/abs/1409.2867}{{\ttfamily
			1409.2867}}].
	
	\bibitem{Blennow:2016gde}
	M.~Blennow, S.~Clementz and J.~Herrero-Garcia, \emph{{Self-interacting
			inelastic dark matter: A viable solution to the small scale structure
			problems}}, \href{https://doi.org/10.1088/1475-7516/2017/03/048}{\emph{JCAP}
		{\bfseries 03} (2017) 048}
	[\href{https://arxiv.org/abs/1612.06681}{{\ttfamily 1612.06681}}].
	
	\bibitem{Zhang:2016dck}
	Y.~Zhang, \emph{{Self-interacting Dark Matter Without Direct Detection
			Constraints}}, \href{https://doi.org/10.1016/j.dark.2016.12.003}{\emph{Phys.
			Dark Univ.} {\bfseries 15} (2017) 82}
	[\href{https://arxiv.org/abs/1611.03492}{{\ttfamily 1611.03492}}].
	
	\bibitem{Dutta:2021wbn}
	M.~Dutta, S.~Mahapatra, D.~Borah and N.~Sahu, \emph{{Self-interacting Inelastic
			Dark Matter in the light of XENON1T excess}},
	\href{https://doi.org/10.1103/PhysRevD.103.095018}{\emph{Phys. Rev. D}
		{\bfseries 103} (2021) 095018}
	[\href{https://arxiv.org/abs/2101.06472}{{\ttfamily 2101.06472}}].
	
	\bibitem{Arkani-Hamed:2008hhe}
	N.~Arkani-Hamed, D.P.~Finkbeiner, T.R.~Slatyer and N.~Weiner, \emph{{A Theory
			of Dark Matter}},
	\href{https://doi.org/10.1103/PhysRevD.79.015014}{\emph{Phys. Rev. D}
		{\bfseries 79} (2009) 015014}
	[\href{https://arxiv.org/abs/0810.0713}{{\ttfamily 0810.0713}}].
	
	\bibitem{Madhavacheril:2013cna}
	M.S.~Madhavacheril, N.~Sehgal and T.R.~Slatyer, \emph{{Current Dark Matter
			Annihilation Constraints from CMB and Low-Redshift Data}},
	\href{https://doi.org/10.1103/PhysRevD.89.103508}{\emph{Phys. Rev. D}
		{\bfseries 89} (2014) 103508}
	[\href{https://arxiv.org/abs/1310.3815}{{\ttfamily 1310.3815}}].
	
	\bibitem{Slatyer:2015jla}
	T.R.~Slatyer, \emph{{Indirect dark matter signatures in the cosmic dark ages.
			I. Generalizing the bound on s-wave dark matter annihilation from Planck
			results}}, \href{https://doi.org/10.1103/PhysRevD.93.023527}{\emph{Phys. Rev.
			D} {\bfseries 93} (2016) 023527}
	[\href{https://arxiv.org/abs/1506.03811}{{\ttfamily 1506.03811}}].
	
	\bibitem{Planck:2018vyg}
	{\scshape Planck} collaboration, \emph{{Planck 2018 results. VI. Cosmological
			parameters}},
	\href{https://doi.org/10.1051/0004-6361/201833910}{\emph{Astron. Astrophys.}
		{\bfseries 641} (2020) A6}
	[\href{https://arxiv.org/abs/1807.06209}{{\ttfamily 1807.06209}}].
	
	\bibitem{Elor:2015bho}
	G.~Elor, N.L.~Rodd, T.R.~Slatyer and W.~Xue, \emph{{Model-Independent Indirect
			Detection Constraints on Hidden Sector Dark Matter}},
	\href{https://doi.org/10.1088/1475-7516/2016/06/024}{\emph{JCAP} {\bfseries
			06} (2016) 024} [\href{https://arxiv.org/abs/1511.08787}{{\ttfamily
			1511.08787}}].
	
	\bibitem{Profumo:2017obk}
	S.~Profumo, F.S.~Queiroz, J.~Silk and C.~Siqueira, \emph{{Searching for
			Secluded Dark Matter with H.E.S.S., Fermi-LAT, and Planck}},
	\href{https://doi.org/10.1088/1475-7516/2018/03/010}{\emph{JCAP} {\bfseries
			03} (2018) 010} [\href{https://arxiv.org/abs/1711.03133}{{\ttfamily
			1711.03133}}].
	
	\bibitem{Fermi-LAT:2015att}
	{\scshape Fermi-LAT} collaboration, \emph{{Searching for Dark Matter
			Annihilation from Milky Way Dwarf Spheroidal Galaxies with Six Years of Fermi
			Large Area Telescope Data}},
	\href{https://doi.org/10.1103/PhysRevLett.115.231301}{\emph{Phys. Rev. Lett.}
		{\bfseries 115} (2015) 231301}
	[\href{https://arxiv.org/abs/1503.02641}{{\ttfamily 1503.02641}}].
	
	\bibitem{HESS:2018cbt}
	{\scshape HESS} collaboration, \emph{{Search for $\gamma$-Ray Line Signals from
			Dark Matter Annihilations in the Inner Galactic Halo from 10 Years of
			Observations with H.E.S.S.}},
	\href{https://doi.org/10.1103/PhysRevLett.120.201101}{\emph{Phys. Rev. Lett.}
		{\bfseries 120} (2018) 201101}
	[\href{https://arxiv.org/abs/1805.05741}{{\ttfamily 1805.05741}}].
	
	\bibitem{Choquette:2016xsw}
	J.~Choquette, J.M.~Cline and J.M.~Cornell, \emph{{p-wave Annihilating Dark
			Matter from a Decaying Predecessor and the Galactic Center Excess}},
	\href{https://doi.org/10.1103/PhysRevD.94.015018}{\emph{Phys. Rev. D}
		{\bfseries 94} (2016) 015018}
	[\href{https://arxiv.org/abs/1604.01039}{{\ttfamily 1604.01039}}].
	
	\bibitem{An:2016kie}
	H.~An, M.B.~Wise and Y.~Zhang, \emph{{Strong CMB Constraint On P-Wave
			Annihilating Dark Matter}},
	\href{https://doi.org/10.1016/j.physletb.2017.08.010}{\emph{Phys. Lett. B}
		{\bfseries 773} (2017) 121}
	[\href{https://arxiv.org/abs/1606.02305}{{\ttfamily 1606.02305}}].
	
	\bibitem{Fuss:2025xwe}
	L.~Fu{\ss}, M.~Garny and A.~Ibarra, \emph{{Connecting cosmologically decaying
			dark matter to neutrino physics}},
	\href{https://doi.org/10.1088/1475-7516/2026/05/075}{\emph{JCAP} {\bfseries
			05} (2026) 075} [\href{https://arxiv.org/abs/2509.19596}{{\ttfamily
			2509.19596}}].
	
	\bibitem{Krnjaic:2025zjl}
	G.~Krnjaic, D.~McKeen, R.~Mizuta, G.~Mohlabeng, D.E.~Morrissey and D.~Tuckler,
	\emph{{X-rays from inelastic dark matter freeze-in}},
	\href{https://doi.org/10.1103/99z7-kz4s}{\emph{Phys. Rev. D} {\bfseries 112}
		(2025) 115039} [\href{https://arxiv.org/abs/2509.19428}{{\ttfamily
			2509.19428}}].
	
	\bibitem{Slatyer:2016qyl}
	T.R.~Slatyer and C.-L.~Wu, \emph{{General Constraints on Dark Matter Decay from
			the Cosmic Microwave Background}},
	\href{https://doi.org/10.1103/PhysRevD.95.023010}{\emph{Phys. Rev. D}
		{\bfseries 95} (2017) 023010}
	[\href{https://arxiv.org/abs/1610.06933}{{\ttfamily 1610.06933}}].
	
	\bibitem{Essig:2013goa}
	R.~Essig, E.~Kuflik, S.D.~McDermott, T.~Volansky and K.M.~Zurek,
	\emph{{Constraining Light Dark Matter with Diffuse X-Ray and Gamma-Ray
			Observations}}, \href{https://doi.org/10.1007/JHEP11(2013)193}{\emph{JHEP}
		{\bfseries 11} (2013) 193} [\href{https://arxiv.org/abs/1309.4091}{{\ttfamily
			1309.4091}}].
	
	\bibitem{Barlow:1990vc}
	R.J.~Barlow, \emph{{Extended maximum likelihood}},
	\href{https://doi.org/10.1016/0168-9002(90)91334-8}{\emph{Nucl. Instrum.
			Meth. A} {\bfseries 297} (1990) 496}.
	
	\bibitem{Hiramatsu:2013qaa}
	T.~Hiramatsu, M.~Kawasaki and K.~Saikawa, \emph{{On the estimation of
			gravitational wave spectrum from cosmic domain walls}},
	\href{https://doi.org/10.1088/1475-7516/2014/02/031}{\emph{JCAP} {\bfseries
			02} (2014) 031} [\href{https://arxiv.org/abs/1309.5001}{{\ttfamily
			1309.5001}}].
	
	\bibitem{Novichkov:2019sqv}
	P.P.~Novichkov, J.T.~Penedo, S.T.~Petcov and A.V.~Titov, \emph{{Generalised CP
			Symmetry in Modular-Invariant Models of Flavour}},
	\href{https://doi.org/10.1007/JHEP07(2019)165}{\emph{JHEP} {\bfseries 07}
		(2019) 165} [\href{https://arxiv.org/abs/1905.11970}{{\ttfamily
			1905.11970}}].
	
	\bibitem{Liu:2019khw}
	X.-G.~Liu and G.-J.~Ding, \emph{{Neutrino Masses and Mixing from Double
			Covering of Finite Modular Groups}},
	\href{https://doi.org/10.1007/JHEP08(2019)134}{\emph{JHEP} {\bfseries 08}
		(2019) 134} [\href{https://arxiv.org/abs/1907.01488}{{\ttfamily
			1907.01488}}].
	
	\bibitem{Lu:2019vgm}
	J.-N.~Lu, X.-G.~Liu and G.-J.~Ding, \emph{{Modular symmetry origin of texture
			zeros and quark lepton unification}},
	\href{https://doi.org/10.1103/PhysRevD.101.115020}{\emph{Phys. Rev. D}
		{\bfseries 101} (2020) 115020}
	[\href{https://arxiv.org/abs/1912.07573}{{\ttfamily 1912.07573}}].
	
	\bibitem{Bringmann:2017}
	K.~Bringmann, A.~Folsom, K.~Ono and L.~Rolen, \emph{Harmonic Maass Forms and
		Mock Modular Forms: Theory and Applications}, vol.~64 of \emph{Colloquium
		Publications}, American Mathematical Society, Providence, RI (2017).
	
	\bibitem{He:2006dk}
	X.-G.~He, Y.-Y.~Keum and R.R.~Volkas, \emph{{A(4) flavor symmetry breaking
			scheme for understanding quark and neutrino mixing angles}},
	\href{https://doi.org/10.1088/1126-6708/2006/04/039}{\emph{JHEP} {\bfseries
			04} (2006) 039} [\href{https://arxiv.org/abs/hep-ph/0601001}{{\ttfamily
			hep-ph/0601001}}].
	
	\bibitem{Ishimori:2010au}
	H.~Ishimori, T.~Kobayashi, H.~Ohki, Y.~Shimizu, H.~Okada and M.~Tanimoto,
	\emph{{Non-Abelian Discrete Symmetries in Particle Physics}},
	\href{https://doi.org/10.1143/PTPS.183.1}{\emph{Prog. Theor. Phys. Suppl.}
		{\bfseries 183} (2010) 1} [\href{https://arxiv.org/abs/1003.3552}{{\ttfamily
			1003.3552}}].
	
\end{thebibliography}
\end{document}